\documentclass[11pt,a4paper]{article}
\usepackage{jcappub}
\usepackage[compat=1.1.0]{tikz-feynhand}
\usepackage{bm}
\usepackage{amsmath}
\usepackage{mathrsfs}
\usepackage[mathscr]{euscript}
\usepackage{feynmp}
\usepackage{comment}
\usepackage{natbib}
\usepackage{ulem}
\def\dd{\mathrm{d}}

\def\mcP{\mathcal{P}}

\def\Mpl{M_{\rm Pl}}

\def\0{{(0)}}
\def\sig0{\dot{\sigma}_0}
\def\dsig{\delta \sigma}

\def\ph0{\dot{\phi}_0}

\def\dn{\Delta n}

\title{
Statistically-Anisotropic Tensor Bispectrum from Inflation}

\author[1]{Takashi Hiramatsu,}
\author[2,3]{Kai Murai,}
\author[4,5]{Ippei Obata,}
\author[3,6]{and Shuichiro Yokoyama}
\affiliation[1]{Department of Physics, Rikkyo University, Toshima, Tokyo 171-8501, Japan}
\affiliation[2]{ Institute for Cosmic Ray Research,
  The University of Tokyo, Kashiwa 277-8582, Japan}
\affiliation[3]{ Kavli Institute for the Physics and Mathematics of the
  Universe (WPI), The University of Tokyo,
  Kashiwa 277-8583, Japan}
\affiliation[4]{Department of Physics, Kyoto University, Kyoto 606-8502, Japan}
\affiliation[5]{Max-Planck-Institut f{\"u}r Astrophysik, Karl-Schwarzschild-Str. 1, 85741 Garching, Germany}
\affiliation[6]{Kobayashi-Maskawa Institute, Nagoya University, Aichi 464-8602, Japan}
\emailAdd{hiramatz@rikkyo.ac.jp}
\emailAdd{kmurai@icrr.u-tokyo.ac.jp}
\emailAdd{obata@tap.scphys.kyoto-u.ac.jp}
\emailAdd{shu@kmi.nagoya-u.ac.jp}

\abstract{
We develop a possibility of generating tensor non-Gaussianity in a kind of anisotropic inflation, where a $U(1)$ gauge field is kinetically coupled to a spectator scalar field.
Owing to this coupling, the coherent mode of the electric field appears and softly breaks the isotropy of the Universe. 
We compute the bispectrum of linearly-polarized tensor perturbations sourced by the gauge field and find that it is strongly red-tilted
and has distinctive statistical anisotropies including higher-order multipole moments.
Interestingly, the tensor bispectra with the specific combinations of linear polarization modes are dominant, and their amplitudes depend on the different sets of multipole moments.
This new type of statistically-anisotropic tensor non-Gaussianity can be potentially testable with the upcoming cosmic microwave background B-mode polarization experiments.
}

\keywords{inflation, primordial gravitational waves (theory)}
\arxivnumber{2008.XXXXX}

\usepackage{amsfonts}

\begin{document}

\begin{flushright}
KUNS-2831, RUP-20-26
\end{flushright}

\maketitle


\section{Introduction}

Over the next decades, the measurement of B-mode polarization in cosmic microwave background (CMB) will probe the gravitational waves from an inflationary universe with more substantial improvement in accuracy.
According to the standard lore, primordial tensor perturbations are provided by the quantum fluctuations of quasi-de-Sitter space-time with a nearly flat inflaton's potential.
The resultant power spectrum of the gravitational waves is then nearly scale-invariant (but slightly red-tilted), statistically isotropic, parity symmetric, and almost Gaussian.
However, these predictions are not necessarily true if the tensor perturbations were provided by the matter sector in the early universe.

The particle production of a gauge field coupled to a scalar field during inflation is a possible mechanism generating
primordial gravitational waves, which has been widely investigated.
For instance, the gauge field axially coupled to a rolling axionic field experiences a tachyonic instability in one of two helicity modes when it leaves the horizon, and such an amplified gauge field predicts a scale-dependent helical gravitational wave power spectrum, in both of Abelian ($U(1)$) model \cite{Sorbo:2011rz,Barnaby:2011vw,Cook:2011hg,Mukohyama:2014gba,Domcke:2016bkh,Garcia-Bellido:2017aan,Ozsoy:2017blg,Ozsoy:2020ccy} and non-Abelian ($SU(2)$) model \cite{Dimastrogiovanni:2012ew,Adshead:2013qp,Obata:2016tmo,Maleknejad:2016qjz,Obata:2016xcr,Dimastrogiovanni:2016fuu,Fujita:2017jwq,Thorne:2017jft,Domcke:2018rvv,Maleknejad:2018nxz,Mirzagholi:2020irt,Watanabe:2020ctz} (or more references therein).
These sourced primordial gravitational waves are typically non-Gaussian and yield a non-zero bispectrum (three-point correlation function) of the tensor perturbations \cite{Cook:2013xea,Namba:2015gja,Agrawal:2017awz,Agrawal:2018mrg,Dimastrogiovanni:2018xnn,Fujita:2018vmv}.
The resultant bispectrum is enhanced at the equilateral configuration since the production of tensor perturbations is efficient at around the horizon scale.
Moreover, the nonzero parity-odd signals are produced in CMB bispectrum, which never appear in the standard parity-invariant scenario \cite{Kamionkowski:2010rb,Shiraishi:2013kxa,Shiraishi:2016yun,Shiraishi:2019yux}.
Such information would be a useful tool for high energy physics in the early universe and open a new window to discriminate between different inflationary models.

On the other hand, the gauge field coupled to a dilatonic field can also induce the particle production in the inflationary period.
The rolling dilatonic field generically violates the conformal invariance of gauge field via the time variation of gauge kinetic function, and an instability occurs in both helicity modes of gauge field on the super-horizon scales.
Since the longer wavelength modes are more enhanced while the instability persists, the coherent vector field naturally appears and it breaks the rotational invariance in space \cite{Watanabe:2009ct,Kanno:2010nr,Do:2011zza,Ohashi:2013pca,Naruko:2014bxa,Ito:2017bnn}.
Then, the scalar and tensor perturbations acquire the directional dependence, which induces a quadrupole anisotropy in their power spectra \cite{Watanabe:2010fh,Watanabe:2010bu,Ohashi:2013qba}.
This inflationary model is called anisotropic inflation and the above phenomenological consequence was firstly motivated to explain the quadrupolar anomaly in WMAP data \cite{Groeneboom:2008fz,Groeneboom:2009cb}.
However, the anomaly was later found to be due to the WMAP's asymmetric beam effect \cite{2010PhRvD..81j3003H,Bennett:2012zja},
and unfortunately no evidence for the violation of rotational symmetry has been found by the latest Planck observations \cite{Kim:2013gka, Akrami:2018odb}. Therefore the original anisotropic inflation model, where a $U(1)$ gauge field is directly coupled to an inflaton, is severely constrained, or a strong fine-tuning in the target space of the model parameters is needed \cite{Bartolo:2012sd,Fujita:2017lfu}.

The above difficulty can be evaded if the gauge field couples to a spectator field instead of the inflaton \cite{Fujita:2018zbr}.
In this case, 
the generation of quadrupole anisotropy in the curvature perturbation is suppressed due to the gravitational coupling between the gauge field and the inflaton.
In addition, interestingly, a sizable amount of statistically anisotropic gravitational waves are produced by the growth of background gauge field on large scales.
Ref. \cite{Fujita:2018zbr} has evaluated the power spectrum (two-point function) of tensor perturbations and 
Ref. \cite{Hiramatsu:2018vfw} has shown that its amplitude and the statistical anisotropies are potentially testable with the upcoming CMB observations.
Inspired by these results, in this work, we develop the generation of tensor non-Gaussianity possessing the statistical anisotropies from this scenario, and consider the possibility to test them in future CMB measurements.

The outline of this paper is as follows.
In section \ref{model setup}, we give a short review of the $U(1)$-spectator model
following Ref. \cite{Fujita:2018zbr}, and present the statistically anisotropic tensor perturbations in section \ref{twopoint}.
In section \ref{hhh}, we calculate 3-point function of tensor perturbations and discuss its detectability in upcoming CMB observations.
Finally, we summarize our work and discuss the outlook in section \ref{sum}.
In this paper, we use the natural unit $\hbar = c = 1$.

\section{Model Setup}
\label{model setup}

In this section,  
we review our model setup in which 
a $U(1)$ gauge field couples to a spectator field.
This model was originally proposed in Ref.~\cite{Fujita:2018zbr}.
The total action is
\begin{align}
S &= \int dtd\bm{x}\sqrt{-g}\left[
\frac{\Mpl^2}{2}R
+ \mathcal{L}_{\rm inflaton} + 
\mathcal{L}_{U(1)\mathchar`-{\rm spectator}}\right] \ , \\
\mathcal{L}_{\rm inflaton} &= -\dfrac{1}{2}(\partial_\mu \phi)^2 -U(\phi) \ , \\
\mathcal{L}_{U(1)\mathchar`-{\rm spectator}} &= -\dfrac{1}{2}(\partial_\mu \sigma)^2 -V(\sigma)
-\dfrac{1}{4}I^2(\sigma)F_{\mu\nu}F^{\mu\nu},
\label{model action}
\end{align}
where $\Mpl$ is the reduced Planck mass, $R$ is the Ricci scalar, and we do not specify the inflaton's potential $U(\phi)$.
In $\mathcal{L}_{U(1)\mathchar`-{\rm spectator}}$,
$\sigma$ is a spectator field with a potential $V(\sigma)$, and $F_{\mu\nu} = \partial_\mu A_\nu - \partial_\nu A_\mu$ is a field strength of $U(1)$ field, $A_\mu$, which couples to the spectator field through a kinetic coupling function, $I(\sigma).$

\subsection{background dynamics}

For the metric, we assume the usual flat Friedmann-Lema\^{i}tre-Robertson-Walker metric as
\begin{equation}
    ds^2 = -dt^2 + a(t)^2d\bm{x}^2 \ .
\end{equation}
Strictly speaking, the spatial isotropy is slightly broken due to the presence of spatially homogeneous U(1) vector field.
However, as we will briefly see later, its degree is at most the order of slow-roll parameters, and 
hence we neglect the spatial anisotropy in the background metric.

Regarding the kinetic function, $I(\sigma)$, we adopt a simple monotonic function
\begin{equation}
    I(\sigma) = \exp(\sigma/\Lambda) \ , \qquad H \ll \Lambda \ll \Mpl \ .
\end{equation}
As for the $U(1)$ gauge field, we choose the Coulomb gauge $\partial_i A_i = 0$ and get the Hamiltonian constraint as $\bar{A}_0(t) = 0$. Hereinafter, we denote the spatially homogeneous background value by using a bar.
We introduce the homogeneous electric field, $\bar{\bm{E}} \equiv -\bar{I}\dot{\bar{\bm{A}}}/a = \bar{E}_z(t){\bm e}_z$,
which is assumed to be oriented along the $z$-axis.
For $\sigma$ and $\bar{A}_z$, the equations read
\begin{align}
\ddot{\bar{\sigma}} + 3H\dot{\bar{\sigma}} + \bar{V}_\sigma = \frac{2 \bar{\rho}_E}{\Lambda} \ , \quad
\dfrac{d}{dt}\left(a\bar{I}^2\dot{\bar{A}}_z \right) = 0 \label{eq: Bg1} \ .
\end{align}
The latter equation indicates $\bar{E}_z \propto I^{-1}a^{-2}$.
Hence its evolution is determined by the background motion of $I(\sigma)$.
We define the time variation of the kinetic function as
\begin{equation}
    n(t) \equiv -\dfrac{\dot{I}}{HI} = -\dfrac{\dot{\bar{\sigma}}}{H\Lambda} \ . \label{eq: n}
\end{equation}
For the initial condition, we assume that the energy density of background electric field 
$\bar{\rho}_E \equiv \bar{E}_z^2/2$ is negligibly small, and
then $\bar{V}_{\sigma} \simeq - 3 H \dot{\bar{\sigma}} =  3n H^2\Lambda$ is satisfied.
However, since $\bar{\rho}_E \propto a^{2(n-2)}$, the electric field grows up in time if $n > 2$ holds.
We assume that this condition is realized in the early inflationary stage,
and denote the value of $n$ at this stage by $n_{\rm ini}$.
For simplicity, here we consider the case where $n_{\rm ini} \gtrsim 2$ is constant in time (see Ref.~\cite{Fujita:2018zbr} for a concrete model realizing such a case).
Then, at a certain time, the enhanced electric field backreacts on the motion of $\sigma(t)$, and $\bar{\rho}_{E}$ gets balanced to the kinetic energy of the spectator field.
In this phase, $\bar{\rho}_E$ becomes constant in time and this means
$n(t)$ is fixed to be $2$.
That is, the slow-roll solutions are given by
\begin{equation}
\dot{\bar{\sigma}}_{\rm att} \simeq -2H\Lambda \ , \qquad \bar{\rho}_{E, \text{att}} \simeq \dfrac{3}{2}\Delta nH^2\Lambda^2 \qquad (\Delta n \equiv n_{\rm ini} - 2) \ . \label{eq: att}
\end{equation}
This is an attractor solution of {\it anisotropic inflation} in the case of the spectator field coupled with 
the U(1) gauge field.
Eventually, the spectator field settles into a potential minimum and stops enhancing the gauge field.
This is the end of the attractor phase.

As we have mentioned, the existence of the background electric field,
in general, breaks the spatial isotropy, but the rotational symmetry in the $xy$-plane remains. 
In such a case, the background metric can be expressed as
\begin{equation}
    ds^2 = -dt^2 + e^{2\alpha(t) + 2\beta(t)}(dx^2 + dy^2) + e^{2\alpha(t)-4\beta(t)}dz^2~,
\end{equation}
with the isotropic scale factor, $a(t) = e^{\alpha(t)}$, and the spatial shear, $\beta(t)$, respectively.
The isotropic Hubble parameter is introduced as $H \equiv \dot{\alpha}$ and the equations of motions for $\beta$ is given by
\begin{align}
\ddot{\beta} &= -3H\dot{\beta} + e^{4\beta}\dfrac{2}{3\Mpl^2}\bar{\rho}_E \ .
\end{align}
During the attractor phase in our setup, $\ddot{\beta}$ becomes negligible and $\dot{\beta}$ converges to a (nearly) constant value
\begin{equation}
    \dfrac{\dot{\beta}}{H} \simeq \dfrac{2\bar{\rho}_E}{9\Mpl^2H^2} = \mathcal{O}((\Lambda / \Mpl)^2) \ .
\end{equation}
Therefore, 
for $\Lambda / \Mpl \ll 1$,
the spatial anisotropy can be neglected on the background.

\subsection{perturbation dynamics}

Let us discuss the perturbation dynamics of $\delta A_i$ and $\delta\sigma$ in the presence of background electric field.
The mode decomposition of $\delta\sigma$ and $\delta A_i$ in the Fourier space is given by
\begin{align}
    \delta\sigma(t, \bm{x}) &= \int\dfrac{d\bm{k}}{(2\pi)^3} ~\delta\hat{\sigma}_{\bm{k}}(t) ~e^{i\bm{k}\cdot\bm{x}} \ , \\
\delta A_i(t, \bm{x}) &= \int\dfrac{d\bm{k}}{(2\pi)^3}\left[ ie^X_i(\hat{\bm{k}})\delta \hat{A}^X_{\bm{k}}(t) + e^Y_i(\hat{\bm{k}})\delta \hat{A}^Y_{\bm{k}}(t) \right] e^{i\bm{k}\cdot\bm{x}} \ .
\end{align}
Since $\bm{\bar{E}}$ is assumed to be oriented along the $z$-axis, we can use the rotational symmetry and restrict the wave vector of the fluctuations to lying on the $zx$-plane $\bm{k} = k(\sin\theta_{\hat{\bm{k}}}, 0, \cos\theta_{\hat{\bm{k}}}) \ (0 \leq \theta_{\hat{\bm{k}}} \leq 2\pi)$.
Thus, the linear polarization vectors, $e^X_i(\hat{\bm{k}})$ and $e^Y_i(\hat{\bm{k}})$, are respectively taken to be
\begin{equation}
e^X_i(\hat{\bm{k}}) = \left(
\begin{array}{c}
\cos\theta_{\hat{\bm{k}}} \\
0 \\
-\sin\theta_{\hat{\bm{k}}} \\
\end{array} \right) \ , \qquad
e^Y_i(\hat{\bm{k}}) = \left(
\begin{array}{c}
0 \\
1 \\
0 \\
\end{array} \right)~, \label{eq: polvec}
\end{equation}
with an angle $\cos\theta_{\hat{\bm{k}}} = \bm{k}\cdot\bar{\bm{E}}/(|\bm{k}||\bar{\bm{E}}|)$, obeying the transverse and orthonormal conditions\footnote{
We have changed the choice of polar coordinates in our previous work \cite{Fujita:2018zbr}, where we used 3-dimensional polar coordinates.}.
The point is that one polarization vector lies on the $zx$-plane, while the other is always orthogonal to it.
Therefore, the inner products between the background electric field and the polarization vectors are given by
\begin{equation}
\sum_i\dot{\bar{A}}_ie^X_i(\hat{\bm{k}}) = -\sin\theta_{\hat{\bm{k}}}\dfrac{a}{\bar{I}}\sqrt{2\bar{\rho}_E} \ , \qquad \sum_i\dot{\bar{A}}_ie^Y_i(\hat{\bm{k}}) = 0 \ , \label{eq: pro}
\end{equation}
where we choose $\dot{\bar{A}}_z > 0$ without loss of generality.
Since $\delta A_i$ is linearly coupled to $\delta \sigma$ through the term such as $\delta \sigma \dot{\bar{A}}_i \dot{\delta A}_i$,
the above property implies that only the $X$ mode, $\delta A^X$, can linearly couple to $\delta\sigma$.
Note that, as the fluctuation of temporal component $\delta A_0(t, \bm{x}) (\equiv A_0(t, {\bm x})) - \bar{A}_0(t))$ is provided by the scalar field, we eliminate it by solving the gauge constraint equation.

We firstly solve the coupled system of $\delta A^X$ and $\delta\sigma$.
By introducing a vector notation, $\Delta
=(a^{3/2}\delta\hat{\sigma}_{\bm k}, \ a^{1/2}\bar{I} \delta \hat{A}_{\bm k}^X)^T$,
we obtain a relevant quadratic action of $\delta A^X$ and $\delta\sigma$ with a matrix form
\begin{equation}
S^{(2)}_\Delta=\dfrac{1}{2}\int dt\dfrac{d\bm{k}}{(2\pi)^3}\left[
\dot{\Delta}^\dag \dot{\Delta}+ \dot{\Delta}^\dag K \Delta- \Delta^\dag K \dot{\Delta}
-\Delta^\dag \Omega^2 \Delta\right], \label{eq: Qua}
\end{equation}
where
\begin{align}
&K=\dfrac{\sqrt{2\bar{\rho}_E}}{\Lambda}\sin\theta_{\hat{\bm{k}}}\begin{pmatrix}0 & i \\
i & 0 \\
\end{pmatrix},
\notag\\
&\Omega^2 = \begin{pmatrix}\dfrac{k^2}{a^2}-\left(\dfrac{9}{4}-\dfrac{\epsilon_H}{2}\right)H^2+\mu_\sigma^2 &
-i\dfrac{\sqrt{2\bar{\rho}_E}}{\Lambda}\sin\theta_{\hat{\bm{k}}} \dfrac{d}{dt}\left(\ln\left(\bar{I}/a\right)\right) \\
i\dfrac{\sqrt{2\bar{\rho}_E}}{\Lambda}\sin\theta_{\hat{\bm{k}}} \dfrac{d}{dt}\left(\ln\left(\bar{I}/a\right)\right)
& \quad \dfrac{k^2}{a^2} -\left(\dfrac{1}{4}-\dfrac{\epsilon_H}{2}\right)H^2 -\left(\dfrac{\ddot{\bar{I}}}{\bar{I}}+H\dfrac{\dot{\bar{I}}}{\bar{I}}\right) \\
\end{pmatrix} \ , \label{eq: Coupled action}
\end{align}
with $\mu_\sigma^2 \equiv \bar{V}_{\sigma\sigma} +(4\bar{I}^2_\sigma\cos^2\theta - \bar{I}_\sigma^2 - \bar{I}\bar{I}_{\sigma\sigma})\dot{\bar{A}}^2/a^2$.
This action yields the equations of motion for $\dsig$ and $\delta A$ as
\begin{equation}
\ddot{\Delta} + 2K \dot{\Delta} + (\Omega^2 + \dot{K})\Delta =0 \ .
\label{matrix EoM}
\end{equation}
The non-diagonal terms represent the mixing effect between $\delta A^X$ and $\delta\sigma$ via the background vector field\footnote{We notice that the interactions of $h_{ij}$ are Planck-suppressed in comparison with them, so that the backreaction of $h_{ij}$ to $\delta A_i$ and $\delta\sigma$ are negligible.}.
As is mentioned in the previous section, 
the energy densities of the spectator field and gauge field are balanced in the attractor phase due to the mutual backreaction, which means that the non-diagonal terms in Eq.~\eqref{eq: Coupled action} are not negligible.
Hence, we need to quantize $\delta\hat{A}^X$ and $\delta\hat{\sigma}$ as a coupled system where the mixing effect is treated non-perturbatively \cite{Nilles:2001fg,Gumrukcuoglu:2010yc}.
In order to do so, we introduce two sets of ``intrinsic" and ``sourced" modes originating from the vacuum fluctuations of each field as
\begin{equation}
\Delta
=\begin{pmatrix} a^{3/2}\delta\sigma_{\text{int},\bm{k}} & a^{3/2}\delta\sigma_{\text{src},\bm{k}} \\
a^{1/2}\bar{I}\delta A^{X}_{\text{src},\bm{k}} & a^{1/2}\bar{I}\delta A^{X}_{\text{int},\bm{k}} \\
\end{pmatrix}
\begin{pmatrix} \hat{a}_{\bm k} \\
\hat{b}^X_{\bm k} \\
\end{pmatrix}+{\rm h.c.} \ ,
\label{Matrix quantization}
\end{equation}
where $\{ \hat{a}_{\bm k}, \ \hat{a}^\dagger_{-\bm k} \}$ and $\{ \hat{b}^X_{\bm k}, \ \hat{b}^{X\dagger}_{-\bm k} \}$ are creation/annihilation operators obeying the standard commutation relations.
We find that $\delta\sigma_{\rm src} \ (\delta A^X_{\text{src}})$ is a mode function sourced by $\delta A^X_{\text{int}} \ (\delta\sigma_{\text{int}})$ via the non-diagonal interactions.
We assume that the mode functions satisfy the usual Bunch-Davies initial condition which is expressed as
\begin{equation}
\lim_{k/(aH) \to \infty}
\Delta = \dfrac{e^{-i\int \omega dt}}{\sqrt{2\omega}}\begin{pmatrix}1 & 0 \\
0 & 1 \\
\end{pmatrix} \ , \qquad \omega = k/a \ .
\label{Bunch-Davies condition}
\end{equation}
Then, the equations of motion Eq. (\ref{matrix EoM}) are reduced to the coupled equations of motion for $\{\delta\sigma_{\text{int}}, \ \delta A^X_{\text{src}}\}$ and $\{\delta\sigma_{\text{src}}, \ \delta A^X_{\text{int}}\}$.

Let us show the solutions for $\delta\sigma_{\bm{k}}$ and $\delta A^X_{\bm{k}}$ on the super-horizon scale 
in the anisotropic attractor phase.
As has been shown in \cite{Fujita:2018zbr}, the contributions of $\{\delta\sigma_{\text{src}}, \ \delta A^X_{\text{int}}\}$ are dominant in the generation of tensor power spectrum, while $\{\delta\sigma_{\text{int}}, \ \delta A^X_{\text{src}}\}$ are sub-leading.
Hence, we neglect the latter contributions to the spectrum.
With this treatment, 
the equations of motion Eq. (\ref{matrix EoM})
on super-horizon scales ($k \ll aH$)
are reduced to
\begin{align}
\ddot{\delta\sigma}_{\bm{k}}+3H\dot{\delta\sigma}_{\bm{k}}+\left(4\dfrac{\bar{\rho}_E}{\Lambda^2}\cos2\theta_{\hat{\bm{k}}}\right)\delta\sigma_{\bm{k}}
&\simeq -2i\frac{\sqrt{2\bar{\rho}_E}}{\Lambda}\sin\theta_{\hat{\bm{k}}}\ \frac{\bar{I}\delta\dot{A}_{\bm{k}}^X}{a},
\label{dsig EoM t}
\\
\partial_t \left(\frac{\bar{I}\delta\dot{A}_{\bm{k}}^X}{a}\right)+
a^{-3}\partial_t \left(a^2\bar{I}\right) \delta \dot{A}_k^X
&\simeq
-2i\frac{\sqrt{2\bar{\rho}_E}}{\Lambda}\sin\theta_{\hat{\bm{k}}} \ \delta\dot{\sigma}_{\bm{k}} \ ,
\label{dA EoM t}
\end{align}
where we have omitted the index ``int/src" since both systems $\{\delta\sigma_{\text{int}}, \ \delta A^X_{\text{src}}\}$ and $\{\delta\sigma_{\text{src}}, \ \delta A^X_{\text{int}}\}$ obey the same equation.
During the attractor phase, 
as we have mentioned, $\bar{\rho}_E = \text{const.}$
is realized
and this means 
$\partial_t \left(a^2\bar{I}\right) =0$.
Then, 
the second term in the left hand side
of the equation of motion for $\delta A^X$ \eqref{dA EoM t}
vanishes, and hence we can easily obtain a
solution for $\delta A^X$ on super-horizon scales as
\begin{equation}
\frac{\bar{I}\delta\dot{A}_{\bm{k}}^X}{a}=
-2i\frac{\sqrt{2\bar{\rho}_E}}{\Lambda} \sin\theta_{\hat{\bm{k}}} \ \delta\sigma_{\bm{k}} +
C(k) 
\ ,
\end{equation}
where $C(k)$ is an integration constant.
Substituting it into Eq. \eqref{dsig EoM t}, we have
\begin{equation}
\ddot{\delta\sigma}_{\bm{k}}+3H\dot{\delta\sigma}_{\bm{k}}+\dfrac{4\bar{\rho}_E}{\Lambda^2}\delta\sigma_{\bm{k}}
=
-2i\dfrac{\sqrt{2\bar{\rho}_E}}{\Lambda}\sin\theta_{\hat{\bm{k}}}~C(k)
\ .
\end{equation}
As has been shown in \cite{Fujita:2018zbr}, $\delta \sigma_{\bm{k}}$ and $\bar{I}\delta \dot{A}_{\bm{k}}^X/a$ have a constant solution while the others are decaying solutions during the attractor phase. Therefore, we focus on the constant solution.
Assuming $\delta \sigma_{\bm{k}} \simeq {\rm const.}$, the solution for the sourced mode on super-horizon scales is given by
\begin{align}
\delta\sigma_{\bm{k}}=& -i \dfrac{H}{\sqrt{2k}k}D(k)\sin\theta_{\hat{\bm{k}}} 
\ ,
\label{eq: deltasigmaatt}
\end{align}
where we have redefined a normalization constant $D(k)$ as $D(k) \equiv C(k)\Lambda k^{3/2}/(H\sqrt{\bar{\rho}_E})$.
Accordingly, we also obtain the solution for the gauge field perturbation as
\begin{align}
\frac{\bar{I}\delta \dot{A}_{\bm{k}}^X}{a} &= \dfrac{H}{\sqrt{2k}k}\frac{\sqrt{2\bar{\rho}_E}}{\Lambda}D(k)\cos2\theta_{\hat{\bm{k}}} \ .
\label{delta A D1D2}
\end{align}
On the other hand, $\delta A^Y$ does not couple to $\delta\sigma$ 
at linear level implied by Eq. \eqref{eq: pro}. That is, it only possesses an intrinsic operator $\delta \hat{A}^Y = \delta \hat{A}^Y_{\text{int}}$.
Hence, we can get the solution of $\delta A^Y = \delta A^Y_{\text{int}}$ in the attractor phase by taking the limit, $\theta_{\bm k} \rightarrow 0$, 
\begin{equation}
\dfrac{\bar{I}\delta \dot{A}^{Y}_k}{a}= \dfrac{H}{\sqrt{2k}k}\frac{\sqrt{2\bar{\rho}_E}}{\Lambda}D(k) \ .
\label{eq: appatt2}
\end{equation}
The absolute value of $D(k)$ in $\{\delta\sigma_{\text{src}}, \ \delta A^X_{\text{int}}\}$
is determined by the Bunch-Davies initial condition and 
is found to be\footnote{We have corrected a typo in \cite{Fujita:2018zbr} and changed the notation of $\gamma(n)$.}
\begin{equation}
    |D(k)| = \gamma(n_{\rm ini})\left(\dfrac{k_A}{k}\right)^{\Delta n} \ , \qquad \gamma(n_{\rm ini})\equiv \dfrac{2^{n_{\rm ini}-1}\Gamma(n_{\rm ini}+1/2)}{\sqrt{3\pi}\Delta n^{3/2}} \ ,
\label{eq: D}
\end{equation}
where $k_A$ is the comoving wavenumber which exits the horizon when the attractor phase starts.
The scale-dependence of $D(k)$ comes from the fact
that the system $\{\delta\sigma_{\text{src}}, \ \delta A^X_{\text{int}}\}$ 
grows even on super-horizon scales in the phase where the electric field grows up in time before the attractor phase, as shown in \cite{Fujita:2018zbr}. Here, we take $k_A$ to be larger than $k_{\rm CMB}$ corresponding to the typical comoving wavenumber that we can observe in CMB experiments. Thus, as we will see later, the sourced tensor perturbations have also red-tilted spectrum.
In addition, we note that an angular factor 
$\cos2\theta_{\hat{\bm{k}}}$
in Eq.~\eqref{delta A D1D2} provides the higher multipoles in the tensor spectrum.

\section{Generation of statistically-anisotropic tensor power spectrum}
\label{twopoint}

In this section, we show the amplification of gravitational waves, $h_{ij}$, sourced by $U(1)$-spectator dynamics, which has been also
discussed in Ref.~\cite{Fujita:2018zbr}.
While in Ref.~\cite{Fujita:2018zbr} the calculation is based on the Green's function method, we employ the so-called in-in formalism \cite{Weinberg:2005vy}.

The leading second-order interaction between $h_{ij} \ , \delta A_i$, and $\delta\sigma$ comes from the $I^2FF$ term,
which is given by
\begin{align}
\sqrt{-g}\mathcal{L}_2(t, \bm{x}) &= -a h_{ij}\left(\bar{I}^2\dot{\bar{A}}_i\delta\dot{A}_j + \bar{I}\bar{I}_\sigma\dot{\bar{A}}_i\dot{\bar{A}}_j\delta\sigma \right) \label{eq: tenlag2} \ .
\end{align}
The gravitational waves, $h_{ij}$, can be decomposed into Fourier modes as
\begin{align}
h_{ij}(t,\bm x) &=
\int \dfrac{\dd^3 k}{(2\pi)^3} e^{i\bm{k}\cdot\bm{x}}\, \left[e^+_{ij}(\hat{\bm k}) \hat{h}_{\bm k}^+(t)+i e^\times_{ij}(\hat{\bm k}) \hat{h}_{\bm k}^\times (t)\right] \ , \\
\hat{h}_{\bm k}^s(t) &= h^s_{\bm k}(t)\hat{c}^s_{\bm k} + h^{s*}_{-\bm k}(t)\hat{c}^{s\dagger}_{-\bm k} \qquad (s = +, \ \times) \ .
\end{align}
Polarization tensors, $e^+_{ij}(\hat{\bm k})$ and $e^\times_{ij}(\hat{\bm k})$, are defined as the combination of linear polarization vectors given by Eq. (\ref{eq: polvec}), as
\begin{align}
e^+_{ij}(\hat{\bm k}) &\equiv \dfrac{1}{\sqrt{2}}\left( e^X_i(\hat{\bm{k}})e^X_j(\hat{\bm{k}}) - e^Y_i(\hat{\bm{k}})e^Y_j(\hat{\bm{k}}) \right) \notag \\
&=\dfrac{1}{\sqrt{2}}\begin{pmatrix}
\cos^2\theta_{\hat{\bm{k}}} \ & 0 \ & -\sin\theta_{\hat{\bm{k}}}\cos\theta_{\hat{\bm{k}}} \\
0 \ &  - 1 \ & 0 \\
-\sin\theta_{\hat{\bm{k}}}\cos\theta_{\hat{\bm{k}}} \ & 0 & \sin^2\theta_{\hat{\bm{k}}} \\
\end{pmatrix} \ , \\
e^\times_{ij}(\hat{\bm k}) &\equiv \dfrac{1}{\sqrt{2}}\left( e^X_i(\hat{\bm{k}})e^Y_j(\hat{\bm{k}}) + e^Y_i(\hat{\bm{k}})e^X_j(\hat{\bm{k}}) \right) \notag \\
&=\dfrac{1}{\sqrt{2}}\begin{pmatrix}
0 \ & \cos\theta_{\hat{\bm{k}}} \ & 0 \\
\cos\theta_{\hat{\bm{k}}} \ & 0 \ & -\sin\theta_{\hat{\bm{k}}} \\
0 & -\sin\theta_{\hat{\bm{k}}} \ & 0 \\
\end{pmatrix} \ ,
\end{align}
which obey the transverse-traceless and orthonormal conditions.
Due to the identities~\eqref{eq: pro}, the plus mode, $h^+$, linearly couples to $\delta A^X$ and $\delta\sigma$, while the cross mode $h^\times$ couples only to $\delta A^Y$ via the background vector field.
Contrary to the perturbations $\delta A_i$ and $\delta\sigma$, which are strongly coupled to each other, the interactions of $h_{ij}$ are Planck-suppressed.
Therefore, we can quantize the gravitational waves by introducing the intrinsic creation/annihilation operators $\{ c^s_{\bm k}, \ c^{\dagger s}_{-\bm k} \}$ and calculate
their correlation functions with the perturbative method.

\begin{figure}[htpb]    
\begin{center}     
\begin{tikzpicture} 
\begin{feynhand}    
    \vertex [particle] (i1) at (-5,1) {$h^+$};
    \vertex [particle] (f1) at (-1,1) {$h^+$};
    \vertex [dot] (w1) at (-4,1) {};
    \vertex [dot] (w2) at (-2,1) {};
    \propag [plain] (i1) to (w1);
    \propag [photon] (w1) to (w2);
    \propag [plain] (w2) to (f1);
    \node at (-3,0.5) {$\delta A^X_{\text{int}}$};
    \vertex [particle] (i3) at (1,1) {$h^+$};
    \vertex [particle] (f3) at (5,1) {$h^+$};
    \vertex [dot] (w7) at (2,1) {};
    \vertex [dot] (w8) at (4,1) {};
    \propag [plain] (i3) to (w7);
    \propag [photon] (w7) to (w8);
    \propag [plain] (w8) to (f3);
    \node at (3,0.5) {$\delta \sigma_{\text{src}}$};
    \vertex [particle] (i3) at (-5,-1) {$h^+$};
    \vertex [particle] (f3) at (-1,-1) {$h^+$};
    \vertex [dot] (w5) at (-4,-1) {};
    \vertex [crossdot] (w5m) at (-3,-1) {};
    \vertex [dot] (w6) at (-2,-1) {};
    \propag [plain] (i3) to (w5);
    \propag [photon] (w5) to (w5m);
    \propag [photon] (w5m) to (w6);
    \propag [plain] (w6) to (f3);
    \node at (-3,-1.5) {$\delta A^X_{\text{int}} \ \delta\sigma_{\text{src}}$};
    \vertex [particle] (i2) at (1,-1) {$h^\times$};
    \vertex [particle] (f2) at (5,-1) {$h^\times$};
    \vertex [dot] (w3) at (2,-1) {};
    \vertex [dot] (w4) at (4,-1) {};
    \propag [plain] (i2) to (w3);
    \propag [photon] (w3) to (w4);
    \node at (3,-1.5) {$\delta A^Y$};
    \propag [plain] (w4) to (f2);
    \end{feynhand}
\end{tikzpicture}
\caption{Tree-level contributions to the tensor power spectrum sourced by gauge field. The black dots represent the vertices of quadratic interactions $H_2$ in Eq.~\eqref{eq: H2}. The external solid lines and the internal wavy lines represent $h^{+/\times}$ and $\delta A^{X/Y}_{\rm int}, \ \delta\sigma_{\rm src}$, respectively. As to the propagator of $\delta\sigma_{\text{src}}$, we use the common notation as that of $\delta A_{\rm int}$ since $\delta\sigma_{\text{src}}$ is originated from the gauge boson. The crossed circle represents the mixing effect between $\delta A_{\rm int}$ and $\delta\sigma_{\rm src}$.}
\label{fig: twopoint}
\end{center}
\end{figure}
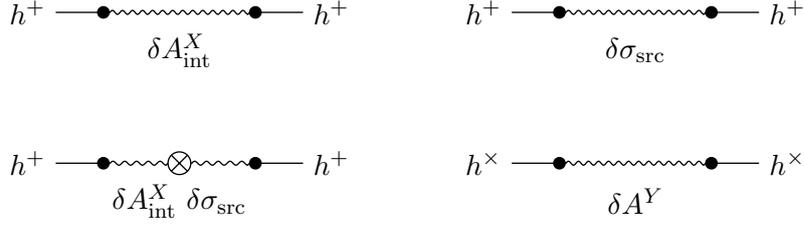

Let us compute the power spectrum of tensor mode:
\begin{align}
(2\pi)^3\delta^{(3)}(\bm{k} + \bm{k}')P_h^{s_1s_2}(\bm{k}) &= \lim_{\tau \to 0}\langle \hat{h}^{s_1}_{\bm{k}}(\tau)\hat{h}^{s_2}_{\bm{k}'}(\tau) \rangle \\
&= \lim_{\tau \to 0}\left(\dfrac{2}{a(\tau)\Mpl}\right)^2\langle \hat{\psi}^{s_1}_{\bm{k}}(\tau)\hat{\psi}^{s_2}_{\bm{k}'}(\tau) \rangle \quad (s_i = +, \ \times)\ ,
\end{align}
where $\tau$ is a conformal time and we have taken the canonical variable, $\psi_{ij} \equiv a\Mpl h_{ij}/2$.
The vacuum expectation value of an operator $\mathcal{O}(t)$ in the ``in" state is given by
\begin{align}
\langle\mathcal{O}(t)\rangle &= \left\langle\left[ T^*\exp\left(i\int^t_{-\infty}H_I(t')dt'\right)\right]\mathcal{O}_I(t)\left[ T\exp\left(-i\int^t_{-\infty}H_I(t'')dt''\right)\right]\right\rangle \\
&= \sum_{N=0}^\infty i^N\int_{-\infty}^tdt_N\int_{-\infty}^{t_N}dt_{N-1}\cdot\cdot\cdot\int_{-\infty}^{t_2}dt_1\left\langle \left[ H_I(t_1), \left[H_I(t_2), \cdot\cdot\cdot \left[H_I(t_N), \mathcal{O}_I(t)\right] \cdot\cdot\cdot \right] \right] \right\rangle \ , \notag
\end{align}
where $T$ and $T^*$ are the time-ordering and anti-time-ordering operators, respectively, and the subscript $I$ denotes operators in the interaction picture.
In order to compute them, we have to find the interaction Hamiltonian, $H_I = -\int d\bm{x}\mathcal{L}_I$, of tensor perturbations.
From Eq.~\eqref{eq: tenlag2}, the leading quadratic interaction Hamiltonian, $H_2$, is given by
\begin{align}
H_2(t) &= H_{2+}(t) + H_{2\times}(t) \ , \label{eq: H2} \\
H_{2+}(t) &= a^2\dfrac{\Lambda}{\Mpl}\int\dfrac{d\bm{p}d\bm{q}}{(2\pi)^3}\hat{\psi}^+_{\bm{p}}\hat{\mathcal{F}}^+_{\bm{q}}(\delta \hat{A}^X_{\rm int}, \delta\hat{\sigma}_{\rm src}, t)\delta^{(3)}(\bm{p}+\bm{q}) \ , \\
H_{2\times}(t) &= a^2\dfrac{\Lambda}{\Mpl}\int\dfrac{d\bm{p}d\bm{q}}{(2\pi)^3}\hat{\psi}^\times_{\bm{p}}\hat{\mathcal{F}}^\times_{\bm{q}}(\delta \hat{A}^Y_{\rm int}, t)\delta^{(3)}(\bm{p}+\bm{q}) \ ,
\end{align}
where
\begin{align}
\hat{\mathcal{F}}^+_{\bm{q}}(\delta \hat{A}^X_{\rm int}, \delta\hat{\sigma}_{\rm src}, t) &\equiv -\sqrt{2}\dfrac{\sqrt{2\bar{\rho}_E}}{\Lambda}\sin\theta_{\hat{\bm{q}}}\left( i\dfrac{\bar{I}\delta\dot{\hat{A}}^X_{\text{int},\bm{q}}}{a} - \dfrac{\sqrt{2\bar{\rho}_E}}{\Lambda}\sin\theta_{\hat{\bm{q}}}\delta\hat{\sigma}_{\text{src},\bm{q}} \right) \label{eq: Fp} \ , \\
\hat{\mathcal{F}}^\times_{\bm{q}}(\delta \hat{A}^Y_{\rm int}, t) &\equiv i\sqrt{2}\dfrac{\sqrt{2\bar{\rho}_E}}{\Lambda}\sin\theta_{\hat{\bm{q}}}\dfrac{\bar{I}\delta\dot{\hat{A}}^Y_{\text{int},\bm{q}}}{a} \label{eq: Fc} \ .
\end{align}
Then, up to the leading order the two-point function
of tensor perturbations generated from $\delta A_i$ and $\delta \sigma$ is given by
\begin{align}
\langle \hat{\psi}^{s_1}_{\bm{k}}(t)\hat{\psi}^{s_2}_{\bm{k}'}(t) \rangle &= -\int_{-\infty}^tdt_2\int_{-\infty}^{t_2}dt_{1}\left\langle \left[ H_{2}(t_1), \left[H_{2}(t_2), \hat{\psi}^{s_1}_{\bm{k}}(t)\hat{\psi}^{s_2}_{\bm{k}'}(t) \right] \right] \right\rangle \ ,
\end{align}
where
\begin{align}
&\left\langle \left[ H_{2}(t_1), \left[H_{2}(t_2), \hat{\psi}^{s_1}_{\bm{k}}(t)\hat{\psi}^{s_2}_{\bm{k}'}(t) \right] \right] \right\rangle \notag \\
&= -\dfrac{4\Lambda^2}{\Mpl^2}a(t_1)^2a(t_2)^2\langle \hat{\mathcal{F}}^{s_1}_{\bm{k}}(t_1)\hat{\mathcal{F}}^{s_2}_{\bm{k}'}(t_2) \rangle \text{Im}[\psi_{k'}(t_2)\psi^{*}_{k'}(t)]\text{Im}[\psi_{k}(t_1)\psi^{*}_{k}(t)] + (t_1 \leftrightarrow t_2) \notag \\
 &+i\dfrac{2\Lambda^2}{\Mpl^2}a(t_1)^2a(t_2)^2\left[\hat{\mathcal{F}}^{s_1}_{\bm{k}}(t_1), \ \hat{\mathcal{F}}^{s_2}_{\bm{k}'}(t_2)\right] \text{Im}[\psi_{k'}(t_2)\psi^{*}_{k'}(t)]\langle\hat{\psi}_{k}(t_1)\hat{\psi}_{k}(t)\rangle + (t_1 \leftrightarrow t_2) \label{eq: com} \ .
\end{align}
We can ignore the contribution from the second term since the commutators of $\hat{\mathcal{F}}^{+/\times}$ vanish on the super-horizon scale (see the discussion in Appendix \ref{appendix negligible}).
In order to perform the time integral, we rewrite the cosmic time into the conformal time $d\tau = dt/a$ and use a dimensionless time variable $x_i \equiv -k\tau_i$.
On the super-horizon regime, we have
\begin{equation}
\text{Im}[\psi_{k}(x_i)\psi^{*}_{k}(x)] = \dfrac{1}{2k}\dfrac{x_i^3 - x^3}{3x_ix} \ .
\end{equation}
The expectation value of $\mathcal{F}$ reads
\begin{equation}
\langle \hat{\mathcal{F}}^{s_1}_{\bm{k}}(t_1)\hat{\mathcal{F}}^{s_2}_{\bm{k}'}(t_2) \rangle = \delta_{s_1s_2}(2\pi)^3\delta^{(3)}(\bm{k}+\bm{k}')\mathcal{F}^{s_1}_{\bm{k}}(t_1)\mathcal{F}^{s_2*}_{\bm{k}}(t_2) \ . \label{eq: FF}
\end{equation}
We illustrate the diagrams of sourced tensor power spectrum in Figure \ref{fig: twopoint}.
As has been shown in the previous section, not only the intrinsic mode of gauge field, but the sourced spectator field also provides the tensor perturbations.
%
%
The dominant contribution of time integration comes from the mode functions $\mathcal{F}$ settling in a constant value on the attractor phase.
Substituting from Eq.~\eqref{eq: deltasigmaatt} to Eq.~\eqref{eq: D} into the expressions of $\mathcal{F}^{+/\times}$, we obtain
\begin{align}
|\mathcal{F}^+_{\bm{k}}| &\simeq \dfrac{H^3}{k^{3/2}}\gamma(n_{\rm ini})\left(\dfrac{k_A}{k}\right)^{\Delta n}\left(\dfrac{\sqrt{2\bar{\rho}_E}}{H\Lambda}\right)^2\sin\theta_{\hat{\bm{k}}}\cos^2\theta_{\hat{\bm{k}}} \ , \\
|\mathcal{F}^\times_{\bm{k}}| &\simeq \dfrac{H^3}{k^{3/2}}\gamma(n_{\rm ini})\left(\dfrac{k_A}{k}\right)^{\Delta n}\left(\dfrac{\sqrt{2\bar{\rho}_E}}{H\Lambda}\right)^2\sin\theta_{\hat{\bm{k}}} \ .
\end{align}
We notice that the additional complex phases, $\mathcal{F}^{s_i} = |\mathcal{F}^{s_i}|e^{i\delta_{s_i}}$, are cancelled out in Eq.~\eqref{eq: FF}.
Finally, the obtained dimensionless power spectrum of the sourced tensor mode for $k \lesssim k_A$ is
\begin{align}
\mcP_h(\bm{k})&=  \mcP_h^{++}(\bm{k}) + \mcP_h^{\times\times}(\bm{k})
\notag\\
&\simeq \mcP_h^{\rm (vac)}\left(1-\cos^2\theta_{\hat{\bm{k}}}+\cos^4\theta_{\hat{\bm{k}}}-\cos^6\theta_{\hat{\bm{k}}}\right)
\left[\dn\,\gamma(n_{\rm ini})N_A\dfrac{\Lambda}{\Mpl}\left(\dfrac{k_A}{k}\right)^{\dn}
 \right]^2 \ ,
\label{sourced Ph}
\end{align}
where we have used Eq.~\eqref{eq: att} and $\mcP_h^{\rm (vac)}=2H^2/(\pi^2 \Mpl^2)$, and $N_A$ is the number of e-foldings of the attractor phase.
It is interesting to note that the statistical anisotropy of $\mcP_h$ characterized by $(1-\cos^2\theta_{\hat{\bm{k}}}+\cos^4\theta_{\hat{\bm{k}}}-\cos^6\theta_{\hat{\bm{k}}})$ does not depend on any model parameters and thus it is a unique and robust prediction of this model.
It should be also noted that the tensor power spectra of the two linear polarizations have different angular dependencies,
\begin{equation}
\mcP_h^{++} \propto \cos^4\theta_{\hat{\bm{k}}}(1-\cos^2\theta_{\hat{\bm{k}}}),
\qquad
\mcP_h^{\times\times} \propto 1-\cos^2\theta_{\hat{\bm{k}}},
\label{different SA}
\end{equation}
and therefore the spectrum is linearly polarized at the same order
\begin{equation}
\mcP^{\rm linear}_h \equiv \dfrac{1}{2}\left(\mcP_h^{++} - \mcP_h^{\times\times}\right) \propto \sin^4\theta_{\hat{\bm{k}}}\left(1+\cos^2\theta_{\hat{\bm{k}}}\right) \ .
\end{equation}
This is another fascinating observational signature of this scenario.
We notice that, since $\mathcal{P}^{+\times}_h = \mathcal{P}^{\times+}_h = 0$, this scenario would not predict the parity-violating signals.


\section{Statistically anisotropic tensor bispectrum}
\label{hhh}

In a similar manner to the power spectrum, by making use of the in-in formalism, we calculate the tensor bispectrum $B_h^{s_1s_2s_3} \ (s_{1,2,3} = \{+, \ \times\})$ and investigate the statistical anisotropy due to the existence of the homogeneous $U(1)$ gauge field.
The bispectrum of tensor mode is defined as
\begin{align}
(2\pi)^3\delta^{(3)}(\bm{k}_1 + \bm{k}_2 + \bm{k}_3)B_h^{s_1s_2s_3}(\bm{k}_1, \bm{k}_2, \bm{k}_3) &= \lim_{\tau \to 0}\langle \hat{h}^{s_1}_{\bm{k_1}}(\tau)\hat{h}^{s_2}_{\bm{k_2}}(\tau)\hat{h}^{s_3}_{\bm{k_3}}(\tau) \rangle \notag \\
&= \lim_{\tau \to 0}\left(\dfrac{2}{a(\tau)\Mpl}\right)^3\langle \hat{\psi}^{s_1}_{\bm{k_1}}(\tau)\hat{\psi}^{s_2}_{\bm{k_2}}(\tau)\hat{\psi}^{s_3}_{\bm{k_3}}(\tau) \rangle \ .
\end{align}

\subsection{Cubic interaction Hamiltonian}

The considerable main three-point vertices come from the gauge kinetic term $I^2FF$ as well as in the case of the power spectrum, which leads to the following cubic interaction Hamiltonian
\begin{equation}
H_3 = H_{h A A} + H_{h A \sigma} + H_{h \sigma \sigma} \ .
\end{equation}
Firstly, the interaction $h \delta A\delta A$ is given by
\begin{align}
\delta\left( -\dfrac{1}{4}I(\sigma)^2F_{\mu\nu}F^{\mu\nu} \right)_{h\delta A\delta A}
&= -\dfrac{\bar{I}^2}{2a^2}h_{ij}\left( \delta\dot{A}_i\delta\dot{A}_j - \dfrac{1}{a^2}\delta F_{ki}\delta F_{kj} \right) \ .
\end{align}
As we have mentioned, we can eliminate the fluctuation of the temporal component $\delta A_0$ by solving the gauge constraint equation and, up to the leading order in slow-roll parameters, it
is provided by $\delta \sigma$ as
\begin{equation}
\bar{I}\partial_i^2\delta A_{0} = \dfrac{2\bar{I}_\sigma}{\bar{I}}\bar{I}\dot{\bar{A}}_i\partial_i\delta\sigma + \mathcal{O}(\epsilon_H)(\delta A_i, \ \delta\varphi, \ \delta\sigma, \ h_{ij}) \ .
\label{gauge constraint}
\end{equation}
Thus, in the above expression, the contribution from $\delta A_0$ is slow-roll suppressed and can be neglected.
On the other hand, however, we have to take into account this contribution to $\delta\sigma$.
Then, the interaction $h\delta A\delta\sigma$ is given by
\begin{align}
\delta\left( -\dfrac{1}{4}I(\sigma)^2F_{\mu\nu}F^{\mu\nu} \right)_{h\delta A\delta \sigma}
&= -\dfrac{\bar{I}^2}{a^2}h_{ij}\left( 2\dfrac{\bar{I}_\sigma}{\bar{I}}\delta\sigma\dot{\bar{A}}_i\delta\dot{A}_j - \delta\dot{A}_i \partial_j\delta A_0 \right) \simeq 0 \ .
\end{align}
Remarkably, this contribution vanishes at leading order in the slow-roll approximation.
Finally, we obtain the interaction $h \delta\sigma \delta\sigma$ as
\begin{align}
&\delta\left( -\dfrac{1}{4}I(\sigma)^2F_{\mu\nu}F^{\mu\nu} \right)_{h\delta \sigma \delta \sigma} \notag\\
&= -\dfrac{1}{2a^2}h_{ij}\left( \dot{\bar{A}}_i\dot{\bar{A}}_j\left( \bar{I}_\sigma^2 + \bar{I}\bar{I}_{\sigma\sigma} \right)\delta\sigma^2 -4\bar{I}\bar{I}_\sigma\delta\sigma\dot{\bar{A}}_i\partial_j\delta A_0  + \bar{I}^2\partial_i\delta A_0\partial_j\delta A_0 \right) \ ,
\end{align}
with Eq. (\ref{gauge constraint}).
Therefore, the relevant cubic interaction Hamiltonians are as follows:
\begin{align}
H_{hAA}(t) &= \dfrac{1}{\Mpl}a^2\int \dfrac{d\bm{k}~d\bm{p}~d\bm{q}}{(2\pi)^6}\delta^{(3)}(\bm{k}+\bm{p}+\bm{q})\left(\hat{\psi}^+_{\bm{k}}e^+_{ij}(\hat{\bm{k}})+i\hat{\psi}^\times_{\bm{k}}e^\times_{ij}(\hat{\bm{k}})\right)
\notag \\
 &\quad \times \left[ -\dfrac{\bar{I}^2}{a^2}\delta\dot{\hat{A}}^X_{\bm{p}}\delta\dot{\hat{A}}^X_{\bm{q}}e^X_i(\hat{\bm{p}})e^X_j(\hat{\bm{q}}) 
+\dfrac{\bar{I}^2}{a^2}\delta\dot{\hat{A}}^Y_{\bm{p}}\delta\dot{\hat{A}}^Y_{\bm{q}}e^Y_i(\hat{\bm{p}})e^Y_j(\hat{\bm{q}}) + 2i\dfrac{\bar{I}^2}{a^2}\delta\dot{\hat{A}}^X_{\bm{p}}\delta\dot{\hat{A}}^Y_{\bm{q}}e^X_i(\hat{\bm{p}})e^Y_j(\hat{\bm{q}}) \right] \ , \label{eq: AAh} \\
H_{h\sigma\sigma}(t) &= -\dfrac{2}{\Mpl}a^2\int \dfrac{d\bm{k}~d\bm{p}~d\bm{q}}{(2\pi)^6}\delta^{(3)}(\bm{k}+\bm{p}+\bm{q})\left(\hat{\psi}^+_{\bm{k}}e^+_{ij}(\hat{\bm{k}})+i\hat{\psi}^\times_{\bm{k}}e^\times_{ij}(\hat{\bm{k}})\right)\dfrac{\bar{I}^2}{a^2\Lambda^2}\dot{\bar{A}}_i\dot{\bar{A}}_j\delta\hat{\sigma}_{\bm{p}}\delta\hat{\sigma}_{\bm{q}} \notag \\
&= -\dfrac{2\sqrt{2}}{\Mpl}\dfrac{\bar{\rho}_E}{\Lambda^2}a^2\int \dfrac{d\bm{k}~d\bm{p}~d\bm{q}}{(2\pi)^6}\delta^{(3)}(\bm{k}+\bm{p}+\bm{q})\hat{\psi}^+_{\bm{k}}\sin^2\theta_{\hat{\bm{k}}}\delta\hat{\sigma}_{\bm{p}}\delta\hat{\sigma}_{\bm{q}} \label{eq: sigsigh} \ .
\end{align}
Note that the contribution of the gradient energy in Eq.~\eqref{eq: AAh} was ignored.

\subsection{Computation of bispectrum based on in-in formalism}

Let us compute the bispectrum of tensor perturbations.
By using the in-in formalism, it is given by
\begin{align}
&\langle \hat{\psi}^{s_1}_{\bm{k_1}}(t)\hat{\psi}^{s_2}_{\bm{k_2}}(t)\hat{\psi}^{s_3}_{\bm{k_3}}(t) \rangle \notag \\
&= i^3\int_{-\infty}^tdt_3\int_{-\infty}^{t_3}dt_{2}\int_{-\infty}^{t_2}dt_{1}\left\langle \left[ H_I(t_1), \left[ H_I(t_2), \left[H_I(t_3), \hat{\psi}^{s_1}_{\bm{k_1}}\hat{\psi}^{s_2}_{\bm{k_2}}\hat{\psi}^{s_3}_{\bm{k_3}}(t)\right] \right] \right] \right\rangle \notag \\
&= i^3\int_{-\infty}^tdt_3\int_{-\infty}^{t_3}dt_{2}\int_{-\infty}^{t_2}dt_{1}\left\langle \left[ H_3(t_1), \left[ H_2(t_2), \left[H_2(t_3), \hat{\psi}^{s_1}_{\bm{k_1}}\hat{\psi}^{s_2}_{\bm{k_2}}\hat{\psi}^{s_3}_{\bm{k_3}}(t)\right] \right] \right] \right\rangle + ... \label{eq: hhhad} \ ,
\end{align}
where $...$ denotes additional two terms obtained by the permutation of the position of $H_3$.
For our interest, 
we consider the case where $\bm{k}_1,\bm{k}_2$, and $\bm{k}_3 $ all lie in the $zx$-plane
since the statistical anisotropy is expected to appear along the direction of background electric field $\bm{\bar{E}} \propto \hat{z}$.
Then, we can use
\begin{equation}
e^X_i(\hat{\bm{k}}_m)e^X_i(\hat{\bm{k}}_n) = \cos\theta_{\hat{\bm{k}}_m\cdot\hat{\bm{k}}_n} \ , \quad e^Y_i(\hat{\bm{k}}_m)e^Y_i(\hat{\bm{k}}_n) = 1 \ , \ \quad e^X_i(\hat{\bm{k}}_m)e^Y_i(\hat{\bm{k}}_n) = 0 \ , \label{eq: xz}
\end{equation}
where $\cos \theta_{\hat{\bm k}_m \cdot \hat{\bm k}_n} \equiv \hat{\bm k}_m \cdot \hat{\bm k}_n$.
Then, we can rewrite Eq.~\eqref{eq: AAh} as
\begin{align}
H_{hAA}(t) &= \dfrac{1}{\sqrt{2}\Mpl}a^2\int \dfrac{d\bm{k}~d\bm{p}~d\bm{q}}{(2\pi)^6}\delta^{(3)}(\bm{k}+\bm{p}+\bm{q})
\label{eq: AAh2} \\
&\times\left[ -\hat{\psi}^+_{\bm{k}}\left(\dfrac{\bar{I}^2}{a^2}\delta\dot{\hat{A}}^X_{\bm{p}}\delta\dot{\hat{A}}^X_{\bm{q}}\cos\theta_{\hat{\bm{k}}\cdot\hat{\bm{p}}}\cos\theta_{\hat{\bm{k}}\cdot\hat{\bm{q}}} + \dfrac{\bar{I}^2}{a^2}\delta\dot{\hat{A}}^Y_{\bm{p}}\delta\dot{\hat{A}}^Y_{\bm{q}} \right) - 2\hat{\psi}^\times_{\bm{k}}\dfrac{\bar{I}^2}{a^2}\delta\dot{\hat{A}}^X_{\bm{p}}\delta\dot{\hat{A}}^Y_{\bm{q}}\cos\theta_{\hat{\bm{k}}\cdot\hat{\bm{p}}} \right] \ , \notag
\end{align}
where we have used the fact that
$\{\bm{k}, \bm{p}, \bm{q}\}$ corresponds to a permutation of $\{\bm{k}_1, \bm{k}_2, \bm{k}_3\}$ and then lies in the $zx$-plane.

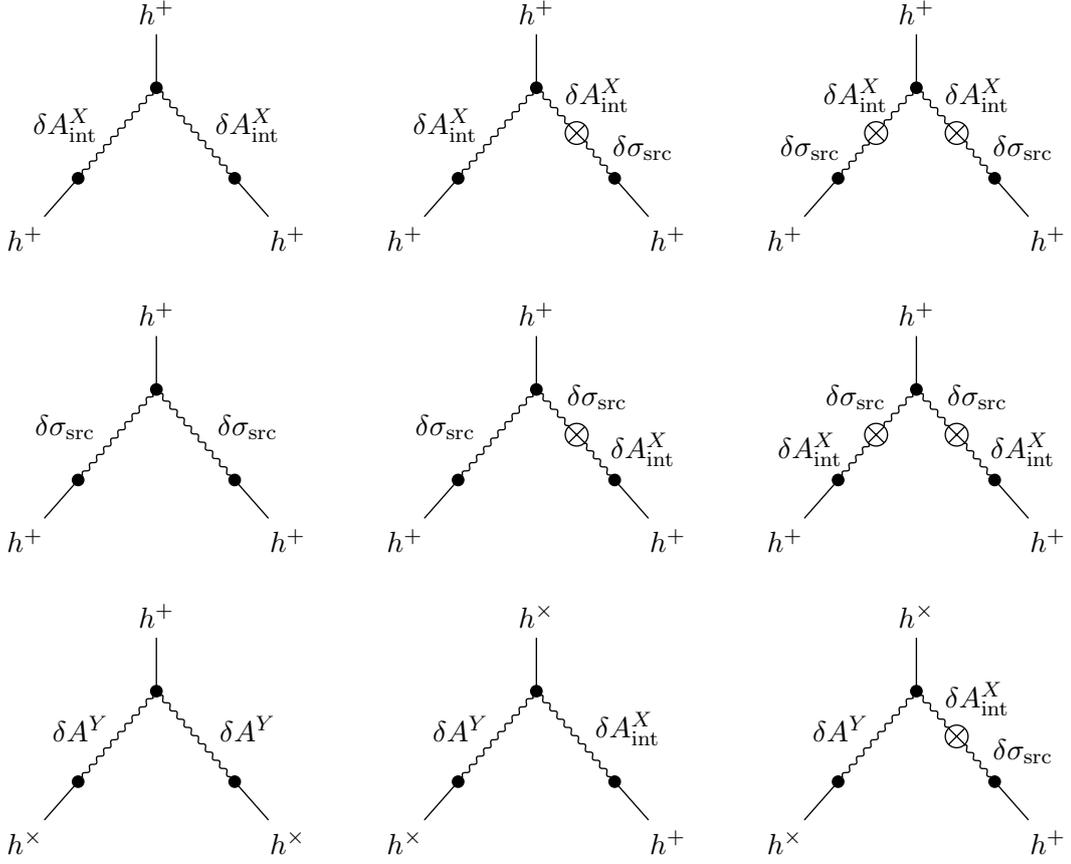
\begin{figure}[htpb]    
\begin{center}     
\begin{tikzpicture} 
\begin{feynhand}    
    \vertex [particle] (p1) at (-7-1.73,-1) {$h^+$};
    \vertex [particle] (p2) at (-7+1.73,-1) {$h^+$};
    \vertex [particle] (p3) at (-7,2) {$h^+$};
    \vertex [dot] (wp1) at (-7-1.73+0.7,-1+0.8) {};
    \vertex [dot] (wp2) at (-7+1.73-0.7,-1+0.8) {};
    \vertex [dot] (wp3) at (-7,1) {};
    \node at (-7-1.2,1-0.5) {$\delta A^X_{\text{int}}$};
    \node at (-7+1.2,1-0.5) {$\delta A^X_{\text{int}}$};
    \propag [plain] (p1) to (wp1);
    \propag [plain] (p2) to (wp2);
    \propag [plain] (p3) to (wp3);
    \propag [photon] (wp1) to (wp3);
    \propag [photon] (wp2) to (wp3);
    \vertex [particle] (p21) at (-2-1.73,-1) {$h^+$};
    \vertex [particle] (p22) at (-2+1.73,-1) {$h^+$};
    \vertex [particle] (p23) at (-2,2) {$h^+$};
    \vertex [dot] (wp21) at (-2-1.73+0.7,-1+0.8) {};
    \vertex [dot] (wp22) at (-2+1.73-0.7,-1+0.8) {};
    \vertex [dot] (wp23) at (-2,1) {};
    \vertex [crossdot] (wp2m2) at (-2+1.73-0.7-0.5,-1+0.8+0.6) {};
    \node at (-3.2,0.5) {$\delta A^X_{\text{int}}$};
    \node at (-1.2,0.9) {$\delta A^X_{\text{int}}$}; 
    \node at (-0.6,0.2) {$\delta \sigma_{\text{src}}$};    
    \propag [plain] (p21) to (wp21);
    \propag [plain] (p22) to (wp22);
    \propag [plain] (p23) to (wp23);
    \propag [photon] (wp21) to (wp23);
    \propag [photon] (wp22) to (wp2m2);
    \propag [photon] (wp2m2) to (wp23);
    \vertex [particle] (p31) at (3-1.73,-1) {$h^+$};
    \vertex [particle] (p32) at (3+1.73,-1) {$h^+$};
    \vertex [particle] (p33) at (3,2) {$h^+$};
    \vertex [dot] (wp31) at (3-1.73+0.7,-1+0.8) {};
    \vertex [dot] (wp32) at (3+1.73-0.7,-1+0.8) {};
    \vertex [dot] (wp33) at (3,1) {};
    \vertex [crossdot] (wp3m1) at (3-1.73+0.7+0.5,-1+0.8+0.6) {};
    \vertex [crossdot] (wp3m2) at (3+1.73-0.7-0.5,-1+0.8+0.6) {};
    \node at (3-0.8,0.9) {$\delta A^X_{\text{int}}$}; 
    \node at (3-1.4,0.2) {$\delta \sigma_{\text{src}}$};    
    \node at (3+0.8,0.9) {$\delta A^X_{\text{int}}$}; 
    \node at (3+1.4,0.2) {$\delta \sigma_{\text{src}}$};    
    \propag [plain] (p31) to (wp31);
    \propag [plain] (p32) to (wp32);
    \propag [plain] (p33) to (wp33);
    \propag [photon] (wp31) to (wp3m1);
    \propag [photon] (wp3m1) to (wp33);
    \propag [photon] (wp32) to (wp3m2);
    \propag [photon] (wp3m2) to (wp33);
    \vertex [particle] (p41) at (-7-1.73,-5) {$h^+$};
    \vertex [particle] (p42) at (-7+1.73,-5) {$h^+$};
    \vertex [particle] (p43) at (-7,-2) {$h^+$};
    \vertex [dot] (wp41) at (-7-1.73+0.7,-5+0.8) {};
    \vertex [dot] (wp42) at (-7+1.73-0.7,-5+0.8) {};
    \vertex [dot] (wp43) at (-7,-3) {};
    \node at (-7-1.2,-3-0.5) {$\delta \sigma_{\text{src}}$};
    \node at (-7+1.2,-3-0.5) {$\delta \sigma_{\text{src}}$};
    \propag [plain] (p41) to (wp41);
    \propag [plain] (p42) to (wp42);
    \propag [plain] (p43) to (wp43);
    \propag [photon] (wp41) to (wp43);
    \propag [photon] (wp42) to (wp43);
    \vertex [particle] (p51) at (-2-1.73,-5) {$h^+$};
    \vertex [particle] (p52) at (-2+1.73,-5) {$h^+$};
    \vertex [particle] (p53) at (-2,-2) {$h^+$};
    \vertex [dot] (wp51) at (-2-1.73+0.7,-5+0.8) {};
    \vertex [dot] (wp52) at (-2+1.73-0.7,-5+0.8) {};
    \vertex [dot] (wp53) at (-2,-3) {};
    \vertex [crossdot] (wp5m2) at (-2+1.73-0.7-0.5,-5+0.8+0.6) {};
    \node at (-3.2,0.5-4) {$\delta \sigma_{\text{src}}$};
    \node at (-1.2,0.9-4) {$\delta \sigma_{\text{src}}$}; 
    \node at (-0.6,0.2-4) {$\delta A^X_{\text{int}}$};    
    \propag [plain] (p51) to (wp51);
    \propag [plain] (p52) to (wp52);
    \propag [plain] (p53) to (wp53);
    \propag [photon] (wp51) to (wp53);
    \propag [photon] (wp52) to (wp5m2);
    \propag [photon] (wp5m2) to (wp53);
    \vertex [particle] (p61) at (3-1.73,-5) {$h^+$};
    \vertex [particle] (p62) at (3+1.73,-5) {$h^+$};
    \vertex [particle] (p63) at (3,-2) {$h^+$};
    \vertex [dot] (wp61) at (3-1.73+0.7,-5+0.8) {};
    \vertex [dot] (wp62) at (3+1.73-0.7,-5+0.8) {};
    \vertex [dot] (wp63) at (3,-3) {};
    \vertex [crossdot] (wp6m1) at (3-1.73+0.7+0.5,-5+0.8+0.6) {};
    \vertex [crossdot] (wp6m2) at (3+1.73-0.7-0.5,-5+0.8+0.6) {};
    \node at (3-0.8,0.9-4) {$\delta \sigma_{\text{src}}$}; 
    \node at (3-1.4,0.2-4) {$\delta A^X_{\text{int}}$};    
    \node at (3+0.8,0.9-4) {$\delta \sigma_{\text{src}}$}; 
    \node at (3+1.4,0.2-4) {$\delta A^X_{\text{int}}$};    
    \propag [plain] (p61) to (wp61);
    \propag [plain] (p62) to (wp62);
    \propag [plain] (p63) to (wp63);
    \propag [photon] (wp61) to (wp6m1);
    \propag [photon] (wp6m1) to (wp63);
    \propag [photon] (wp62) to (wp6m2);
    \propag [photon] (wp6m2) to (wp63);
    \vertex [particle] (c11) at (-7-1.73,-9) {$h^\times$};
    \vertex [particle] (c12) at (-7+1.73,-9) {$h^\times$};
    \vertex [particle] (c13) at (-7,-6) {$h^+$};
    \vertex [dot] (wc11) at (-7-1.73+0.7,-9+0.8) {};
    \vertex [dot] (wc12) at (-7+1.73-0.7,-9+0.8) {};
    \vertex [dot] (wc13) at (-7,-7) {};
    \node at (-8,-7.5) {$\delta A^Y$};
    \node at (-5.8,-7.5) {$\delta A^Y$};
    \propag [plain] (c11) to (wc11);
    \propag [plain] (c12) to (wc12);
    \propag [plain] (c13) to (wc13);
    \propag [photon] (wc11) to (wc13);
    \propag [photon] (wc12) to (wc13);
    \vertex [particle] (pc21) at (-2-1.73,-9) {$h^\times$};
    \vertex [particle] (pc22) at (-2+1.73,-9) {$h^+$};
    \vertex [particle] (pc23) at (-2,-6) {$h^\times$};
    \vertex [dot] (wc21) at (-2-1.73+0.7,-9+0.8) {};
    \vertex [dot] (wc22) at (-2+1.73-0.7,-9+0.8) {};
    \vertex [dot] (wc23) at (-2,-7) {};
    \node at (-8+5,-7.5) {$\delta A^Y$}; 
    \node at (-5.8+5,-7.5) {$\delta A^X_{\text{int}}$};    
    \propag [plain] (pc21) to (wc21);
    \propag [plain] (pc22) to (wc22);
    \propag [plain] (pc23) to (wc23);
    \propag [photon] (wc21) to (wc23);
    \propag [photon] (wc22) to (wc23);
    \vertex [particle] (c31) at (3-1.73,-9) {$h^\times$};
    \vertex [particle] (c32) at (3+1.73,-9) {$h^+$};
    \vertex [particle] (c33) at (3,-6) {$h^\times$};
    \vertex [dot] (wc31) at (3-1.73+0.7,-9+0.8) {};
    \vertex [dot] (wc32) at (3+1.73-0.7,-9+0.8) {};
    \vertex [dot] (wc33) at (3,-7) {};
    \vertex [crossdot] (wc3m2) at (3+1.73-0.7-0.5,-9+0.8+0.6) {};
    \node at (-8+5+5,-7.5) {$\delta A^Y$};  
    \node at (3+0.8,0.9-4-4) {$\delta A^X_{\text{int}}$}; 
    \node at (3+1.4,0.2-4-4) {$\delta \sigma_{\text{src}}$};    
    \propag [plain] (c31) to (wc31);
    \propag [plain] (c32) to (wc32);
    \propag [plain] (c33) to (wc33);
    \propag [photon] (wc31) to (wc33);
    \propag [photon] (wc32) to (wc3m2);
    \propag [photon] (wc3m2) to (wc33);
    \end{feynhand}
\end{tikzpicture}
\caption{Main diagrams which create tensor bispectra in our model. The black dots represent the quadratic/cubic interactions $H_2/H_3$.}
\label{fig: threepo}
\end{center}
\end{figure}

We illustrate the tree-level contributions from the cubic interactions \eqref{eq: sigsigh} and \eqref{eq: AAh2} to the bispectrum of sourced tensor perturbations in Figure \ref{fig: threepo}.
One can find that there are only two combinations: $B^{+++}_h$ and $B_h^{+\times\times}$.
This would be a natural consequence from the parity-conserving scenario.
Regarding the other possible diagrams, such as the diagram with self-interactions between $\delta A^X_{\rm int}$ and $\delta\sigma_{\rm src}$ or loop diagrams, we have discussed in Appendix \ref{con} and shown that these contributions can be neglected for our parameter space.

%
%

\subsubsection*{{\bf (i) $B^{+++}_h$}}

First of all, let us compute $B^{+++}_h$.
As to the cubic interaction Hamiltonian contributing to $B^{+++}_h$, we denote it as
\begin{align}
H_{3++} &= \dfrac{a^2}{\Mpl}\int\dfrac{d\bm{k}d\bm{p}d\bm{q}}{(2\pi)^6}\delta^{(3)}(\bm{k}+\bm{p}+\bm{q})\psi^+_{\bm{k}}\hat{\mathcal{F}}^{++}_{\bm{k} \ \bm{p} \ \bm{q}}(\delta\hat{A}^X_{\rm int}, \delta\hat{\sigma}_{\rm src}, t) \ , \\
\hat{\mathcal{F}}^{++}_{\bm{k} \ \bm{p} \ \bm{q}} &\equiv 
-\dfrac{1}{\sqrt{2}}\dfrac{\bar{I}^2}{a^2}\delta\dot{\hat{A}}^X_{\text{int},\bm{p}}\delta\dot{\hat{A}}^X_{\text{int},\bm{q}}\cos\theta_{\hat{\bm{k}}\cdot\hat{\bm{p}}}\cos\theta_{\hat{\bm{k}}\cdot\hat{\bm{q}}} - \sqrt{2}\dfrac{2\bar{\rho}_E}{\Lambda^2}\sin^2\theta_{\hat{\bm{k}}}\delta\hat{\sigma}_{\text{src},\bm{p}}\delta\hat{\sigma}_{\text{src},\bm{q}} \ . \label{eq: F++}
\end{align}
Note that the vertex from $\delta \hat{A}^Y$ in $H_{hAA}$ does not provide the diagram of $h^+h^+h^+$.
As to the commutation form in the bispectrum, disregarding the disconnected pieces, we get
\begin{align}
&\left\langle\left[ H_{3++}(t_1), \left[ H_{2+}(t_2), \left[H_{2+}(t_3),\hat{\psi}^+_{\bm{k_1}}(t)\hat{\psi}^{+}_{\bm{k_2}}(t)\hat{\psi}^{+}_{\bm{k_3}}(t)\right] \right] \right] \right\rangle \notag \\
&= -\Lambda^2\Mpl^{-3}a(t_1)^2a(t_2)^2a(t_3)^2\int\dfrac{d\bm{p}d\bm{q}}{(2\pi)^6}(2\pi)^3\delta^{(3)}(-\bm{k}_3+\bm{p}+\bm{q})~(2i)^3  \notag \\
 &\times \langle \hat{\mathcal{F}}^{++}_{-\bm{k}_3 \ \bm{p} \ \bm{q}}(t_1)\hat{\mathcal{F}}^+_{\bm{k}_2}(t_2)\hat{\mathcal{F}}^+_{\bm{k}_1}(t_3) \rangle \text{Im}[\psi^{+}_{k_1}(t_3)\psi^{+*}_{k_1}(t)]\text{Im}[\psi^{+}_{k_2}(t_2)\psi^{+*}_{k_2}(t)]\text{Im}[\psi^{+}_{k_3}(t_1)\psi^{+*}_{k_3}(t)] + ... \notag \\
 &= -\Lambda^2\Mpl^{-3}a(t_1)^2a(t_2)^2a(t_3)^2(2\pi)^3\delta^{(3)}(\bm{k}_1+\bm{k}_2+\bm{k}_3)~(2i)^3\mathcal{F}^{+*}_{\bm{k}_1}(t_3)\mathcal{F}^{+*}_{\bm{k}_2}(t_2) \\
 & \times  \left( \mathcal{F}^{++}_{\bm{k}_3 \ \bm{k}_1 \ \bm{k}_2}(t_1) + \mathcal{F}^{++}_{\bm{k}_3 \ \bm{k}_2 \ \bm{k}_1}(t_1) \right) \text{Im}[\psi^{+}_{k_1}(t_3)\psi^{+*}_{k_1}(t)]\text{Im}[\psi^{+}_{k_2}(t_2)\psi^{+*}_{k_2}(t)]\text{Im}[\psi^{+}_{k_3}(t_1)\psi^{+*}_{k_3}(t)] + ... \notag  \ ,
\end{align}
where the dots indicate additional terms obtained by the permutation of momenta $\{\bm{k}_1, \ \bm{k}_2, \ \bm{k}_3\}$.
Note that we have neglected the contributions from the commutators of $\hat{\mathcal{F}}^{+/++}$ due to the same reason as
in Eq.~\eqref{eq: com}.
In terms of a dimensionless time variable $x_i \equiv -k_3\tau_i$, we get
\begin{align}
B^{+++}_h
= &-\dfrac{2^3\Lambda^2}{H^6\Mpl^6}
\int_{1}^x\dfrac{dx_3}{x_3}\int_{1}^{x_3}\dfrac{dx_2}{x_2}\int_{1}^{x_2}\dfrac{dx_1}{x_1}\prod_{i=1}^3\dfrac{x_i^3 - x^3}{3x_i^3}  \notag \\
&\times\left[2\mathcal{F}^{+*}_{\bm{k}_1}(x_3)\mathcal{F}^{+*}_{\bm{k}_2}(x_2)\mathcal{F}^{++}_{\bm{k}_3 \ \bm{k}_1 \ \bm{k}_2}(x_1) + (\bm{k}_1\leftrightarrow\bm{k}_2\leftrightarrow\bm{k}_3) \right] + (\text{permutation of $H_3$}) \ , \label{eq: Bppp}
\end{align}
where we cut off the UV effect in the time integration $(x_i > 1)$.
Hereafter we assume that all momenta are in the following window: $k_{\text{CMB}} \leq k_i \leq k_A$,
where $k_\text{CMB}$ represents a typical scale observed in CMB experiments.
Then, the dominant contribution of the time integration comes from the mode functions $\mathcal{F}$ settling in a constant value on the attractor phase.
Substituting Eqs.~\eqref{eq: deltasigmaatt} - \eqref{eq: D} into Eq.~\eqref{eq: F++}, we obtain
\begin{align}
\mathcal{F}^{++}_{\bm{k} \ \bm{p} \ \bm{q}} &\simeq -\sqrt{2}\dfrac{H^2}{\sqrt{2p}p}\dfrac{H^2}{\sqrt{2q}q}\left(\dfrac{k_A}{p}\right)^{\Delta n}\left(\dfrac{k_A}{q}\right)^{\Delta n}\gamma(n_{\rm ini})^2\left(\dfrac{\sqrt{2\bar{\rho}_E}}{H\Lambda}\right)^{2} \notag \\
& \times\left[
\dfrac{1}{2}\cos2\theta_{\hat{\bm{p}}}\cos2\theta_{\hat{\bm{q}}}\cos\theta_{\hat{\bm{k}}\cdot\hat{\bm{p}}}\cos\theta_{\hat{\bm{k}}\cdot\hat{\bm{q}}} - \sin^2\theta_{\hat{\bm{k}}}\sin\theta_{\hat{\bm{p}}}\sin\theta_{\hat{\bm{q}}} \right] \label{eq: F3} \ .
\end{align}
Here we have dropped the complex phase of $\mathcal{F}^{++}$ since it is finally cancelled out from the contribution of $\mathcal{F}^{+*}$ in Eq.~\eqref{eq: Bppp}.
Since the dominant integrand becomes constant, we can find
\begin{equation}
\int_{1}^x\dfrac{dx_3}{x_3}\int_{1}^{x_3}\dfrac{dx_2}{x_2}\int_{1}^{x_2}\dfrac{dx_1}{x_1}[...] \simeq \dfrac{1}{3!}\int_{1}^x\dfrac{dx_3}{x_3}\int_{1}^{x}\dfrac{dx_2}{x_2}\int_{1}^{x}\dfrac{dx_1}{x_1}[...] \ .
\end{equation}
At this time, the additional two terms coming from the permutation in terms of $H_3$ in Eq.~\eqref{eq: hhhad} become the same as the first term.
Then $B^{+++}_h$ is simply approximated by
\begin{align}
B^{+++}_h \simeq&-\dfrac{2^3\Lambda^2}{3!3^3H^6\Mpl^6}\int_{1}^x\dfrac{dx_3}{x_3}\int_{1}^{x}\dfrac{dx_2}{x_2}\int_{1}^{x}\dfrac{dx_1}{x_1}\prod_{i=1}^3\dfrac{x_i^3 - x^3}{x_i^3}  \notag \\
&\times\left[ 2\mathcal{F}^{+*}_{\bm{k}_1}(x_3)\mathcal{F}^{+*}_{\bm{k}_2}(x_2)\mathcal{F}^{++}_{\bm{k}_3 \ \bm{k}_1 \ \bm{k}_2}(x_1) + (\bm{k}_1\leftrightarrow\bm{k}_2\leftrightarrow\bm{k}_3) \right]\times3 \ .
\end{align}
Finally, we obtain
\begin{align}
&B^{+++}_h \simeq 2^{7/2}\dfrac{H^4}{\Mpl^4}\dfrac{\Lambda^2}{\Mpl^2}\gamma(n_{\rm ini})^4\Delta n^{3} N_A^3 ~\mathcal{G}^{+++}(\bm{k}_1, \bm{k}_2, \bm{k}_3) \ , \notag \\
&\mathcal{G}^{+++}(\bm{k}_1, \bm{k}_2, \bm{k}_3) \equiv \dfrac{1}{k_1^3k_2^3}\left(\dfrac{k_A}{k_1}\right)^{2\Delta n}\left(\dfrac{k_A}{k_{2}}\right)^{2\Delta n}\sin\theta_{\hat{\bm{k}}_1}\cos^2\theta_{\hat{\bm{k}}_1}\sin\theta_{\hat{\bm{k}}_2}\cos^2\theta_{\hat{\bm{k}}_2} \notag \\
&\times\left(
\dfrac{1}{2}\cos2\theta_{\hat{\bm{k}}_1}\cos2\theta_{\hat{\bm{k}}_2}\cos\theta_{\hat{\bm{k}}_3\cdot\hat{\bm{k}}_1}\cos\theta_{\hat{\bm{k}}_3\cdot\hat{\bm{k}}_2} - \sin^2\theta_{\hat{\bm{k}}_3}\sin\theta_{\hat{\bm{k}}_1}\sin\theta_{\hat{\bm{k}}_2} \right) \notag \\
&+ (\bm{k}_3\leftrightarrow\bm{k}_{1}, \bm{k}_{2}) \ .
\label{eq: fnl+++}
\end{align}
We can see that Eq.~\eqref{eq: fnl+++} exhibits a rich dependence on angles between the background vector field and the wave vectors caused by the presence of polarization tensor, which does not appear in the anisotropic scalar bispectrum from primordial vector field \cite{Yokoyama:2008xw,Karciauskas:2008bc,Bartolo:2009pa,Bartolo:2011ee,Bartolo:2012sd} (see {\it e.g.}, (A1) in Ref.~\cite{Bartolo:2011ee}).
Also, as discussed in Ref. \cite{Fujita:2018zbr}, 
both the amplitude and the scale-dependence of the tensor power spectrum
are dependent on
the parameter $\Delta n$ and, 
observationally, an interesting parameter region is $\Delta n = \mathcal{O} (1)$.

First, let us evaluate Eq. \eqref{eq: fnl+++} in the squeezed momentum configuration.
Following Refs.~\cite{Akrami:2018odb,Shiraishi:2019yux},
we introduce a non-linearity parameter for tensor perturbations in the squeezed limit as the relative size of bispectrum to the usual scalar local bispectrum template:
\begin{align}
f^{s_1s_2s_3}_{\text{NL,sq}} &\equiv  \dfrac{B^{s_1s_2s_3}_h}{S^{\rm loc}_{k_1k_2k_3}} \ , \qquad S^{\rm loc}_{k_1k_2k_3} \equiv \dfrac{6(2\pi^2\mathcal{P}_\zeta)^2}{5}\dfrac{\sum_ik_i^3}{\prod_ik_i^3} \ . \label{eq: fNLsq}
\end{align}
Since $f_{\text{NL}}^{+++}$ is symmetric between the three momenta $\bm{k}_{1,2,3}$, we consider the following limit $k_1 \simeq k_2 \gg k_3$.
One can find that this squeezed triangle is characterized by $\theta_{(-\hat{\bm{k}}_1)\cdot\hat{\bm{k}}_2} \simeq 0$ and $\theta_{\hat{\bm{k}}_2\cdot\hat{\bm{k}}_3} \simeq \theta_{\hat{\bm{k}}_1\cdot\hat{\bm{k}}_3} + \pi$.
Then, Eq.~\eqref{eq: fNLsq} is given by
\begin{equation}
f^{+++}_{\text{NL}, k_1\simeq k_2 \gg k_3} \simeq \mathcal{A}^{+++}_{\rm sq}(k_1, \ k_3)\mathcal{G}^{+++}_{\rm sq}(\hat{\bm{k}}_1, \ \hat{\bm{k}}_3) \ , \label{eq: fnlsq+++}
\end{equation}
where
\begin{align}
\mathcal{A}^{+++}_{\rm sq}(k_1, \ k_3) &= -\dfrac{5}{2^{3/2}3} r_{\text{vac}}^2\dfrac{\Lambda^2}{\Mpl^2}\gamma(n_{\rm ini})^4\Delta n^{3} N_A^3 \left(\dfrac{k_A}{k_1}\right)^{2\Delta n}\left(\dfrac{k_A}{k_3}\right)^{2\Delta n} \ , \\
\mathcal{G}^{+++}_{\rm sq}(\hat{\bm{k}}_1, \ \hat{\bm{k}}_3) &= \sin\theta_{\hat{\bm{k}}_1}\sin\theta_{\hat{\bm{k}}_3}\cos^2\theta_{\hat{\bm{k}}_1}\cos^2\theta_{\hat{\bm{k}}_3}\left(\sin^3\theta_{\hat{\bm{k}}_1}\sin\theta_{\hat{\bm{k}}_3}-\dfrac{1}{2}\cos2\theta_{\hat{\bm{k}}_1}\cos2\theta_{\hat{\bm{k}}_3}\cos\theta_{\hat{\bm{k}}_1\cdot\hat{\bm{k}}_3}\right) \ .
\end{align}

We also evaluate the non-linearity parameter in the equilateral limit $k_1 = k_2 = k_3 = k$ \cite{Akrami:2018odb,Shiraishi:2019yux}
\begin{equation}
f^{s_1s_2s_3}_{\rm NL, eq} \equiv \dfrac{B^{s_1s_2s_3}_h(k)}{S^{\rm eq}_k} \ , \qquad S^{\rm eq}_k \equiv \dfrac{18}{5}\dfrac{(2\pi^2\mathcal{P}_\zeta)^2}{k^6} \ , \label{eq: fNLeq}
\end{equation}
where $S^{\rm eq}(k)$ is a usual scalar bispectrum template.
Using $\theta_{\hat{\bm{k}}_1\cdot\hat{\bm{k}}_2} = \theta_{\hat{\bm{k}}_2\cdot\hat{\bm{k}}_3} = \theta_{\hat{\bm{k}}_3\cdot\hat{\bm{k}}_1} = 2\pi/3$, we obtain
\begin{equation}
f^{+++}_{\rm NL, eq} = \mathcal{A}^{+++}_{\rm eq}(k)\mathcal{G}^{+++}_{\rm eq}(\hat{\bm{k}}_1) \ , \label{eq: fnleqppp}
\end{equation}
where
\begin{align}
\mathcal{A}^{+++}_{\rm eq}(k) &= -\dfrac{35}{2^{25/2}}r_{\text{vac}}^2\dfrac{\Lambda^2}{\Mpl^2}\gamma(n_{\rm ini})^4\Delta n^{3} N_A^3 \left(\dfrac{k_A}{k}\right)^{4\Delta n} \ , \\
\mathcal{G}^{+++}_{\rm eq}(\hat{\bm{k}}_1) &= 1-\dfrac{60}{7}\cos^2\theta_{\hat{\bm{k}}_1} +\dfrac{160}{7}\cos^4\theta_{\hat{\bm{k}}_1} -\dfrac{320}{21} \cos^6\theta_{\hat{\bm{k}}_1}
\end{align}
and find that it also possesses the angular dependence including higher order multipole moments.

\subsubsection*{{\bf (ii) $B^{+\times\times}_h$}}

Next, the cubic interaction Hamiltonian, which contributes to the bispectrum $B^{+\times\times}_h$, can be given by
\begin{align}
H_{3\times\times} &= \dfrac{a^2}{\Mpl}\int\dfrac{d\bm{k}d\bm{p}d\bm{q}}{(2\pi)^6}\delta^{(3)}(\bm{k}+\bm{p}+\bm{q})\psi^+_{\bm{k}}\hat{\mathcal{F}}^{\times\times}_{\bm{k} \ \bm{p} \ \bm{q}}(\delta\hat{A}^Y_{\rm int}, t) \ , \\
H_{3+\times} &= \dfrac{a^2}{\Mpl}\int\dfrac{d\bm{k}d\bm{p}d\bm{q}}{(2\pi)^6}\delta^{(3)}(\bm{k}+\bm{p}+\bm{q})\psi^\times_{\bm{k}}\hat{\mathcal{F}}^{+\times}_{\bm{k} \ \bm{p} \ \bm{q}}(\delta\hat{A}^X_{\rm int}, \delta\hat{A}^Y_{\rm int}, t) \ ,
\end{align}
where
\begin{equation}
\hat{\mathcal{F}}^{\times\times}_{\bm{k} \ \bm{p} \ \bm{q}} \equiv 
-\dfrac{1}{\sqrt{2}}\dfrac{\bar{I}^2}{a^2}\delta\dot{\hat{A}}^Y_{\text{int},\bm{p}}\delta\dot{\hat{A}}^Y_{\text{int},\bm{q}} \ , \qquad \hat{\mathcal{F}}^{+\times}_{\bm{k} \ \bm{p} \ \bm{q}} \equiv -
\sqrt{2}\dfrac{\bar{I}^2}{a^2}\delta\dot{\hat{A}}^X_{\text{int},\bm{p}}\delta\dot{\hat{A}}^Y_{\text{int},\bm{q}}\cos\theta_{\hat{\bm{k}}\cdot\hat{\bm{p}}} \ .
\end{equation}
We calculate
\begin{align}
&
\left\langle
\left[ H_{3}(t_1), \left[ H_{2}(t_2), \left[H_{2}(t_3),\hat{\psi}^+_{\bm{k_1}}(t)\hat{\psi}^{\times}_{\bm{k_2}}(t)\hat{\psi}^{\times}_{\bm{k_3}}(t)\right] \right] \right]
\right\rangle
\notag \\
=&
\left\langle
\left[ H_{3\times\times}(t_1), \left[ H_{2\times}(t_2), \left[H_{2\times}(t_3),\hat{\psi}^+_{\bm{k_1}}(t)\hat{\psi}^{\times}_{\bm{k_2}}(t)\hat{\psi}^{\times}_{\bm{k_3}}(t)\right] \right] \right]
\right.
\notag \\
 + &\left[ H_{3+\times}(t_1), \left[ H_{2+}(t_2), \left[H_{2\times}(t_3),\hat{\psi}^+_{\bm{k_1}}(t)\hat{\psi}^{\times}_{\bm{k_2}}(t)\hat{\psi}^{\times}_{\bm{k_3}}(t)\right] \right] \right] \notag \\
 + &
 \left.
 \left[ H_{3+\times}(t_1), \left[ H_{2\times}(t_2), \left[H_{2+}(t_3),\hat{\psi}^+_{\bm{k_1}}(t)\hat{\psi}^{\times}_{\bm{k_2}}(t)\hat{\psi}^{\times}_{\bm{k_3}}(t)\right] \right] \right]
 \right\rangle
 \notag \\
 =& \;-\Lambda^2\Mpl^{-3}a(t_1)^2a(t_2)^2a(t_3)^2(2\pi)^3\delta^{(3)}(\bm{k}_1+\bm{k}_2+\bm{k}_3)~(2i)^3  \\
 & \times\left[2\mathcal{F}^{\times\times*}_{\bm{k}_1 \ \bm{k}_2 \ \bm{k}_3}(t_1)\mathcal{F}^\times_{\bm{k}_2}(t_2)\mathcal{F}^\times_{\bm{k}_3}(t_3)\text{Im}[\psi^{+}_{k_1}(t_1)\psi^{+*}_{k_1}(t)]\text{Im}[\psi^{\times}_{k_2}(t_2)\psi^{\times*}_{k_2}(t)]\text{Im}[\psi^{\times}_{k_3}(t_3)\psi^{\times*}_{k_3}(t)] \right. \notag \\
 &\left. +\mathcal{F}^{+\times*}_{\bm{k}_2 \ \bm{k}_1 \ \bm{k}_3}(t_1)\mathcal{F}^+_{\bm{k}_1}(t_2)\mathcal{F}^\times_{\bm{k}_3}(t_3)\text{Im}[\psi^{+}_{k_1}(t_2)\psi^{+*}_{k_1}(t)]\text{Im}[\psi^{\times}_{k_2}(t_1)\psi^{\times*}_{k_2}(t)]\text{Im}[\psi^{\times}_{k_3}(t_3)\psi^{\times*}_{k_3}(t)] \right. \notag \\
 &\left. +\mathcal{F}^{+\times*}_{\bm{k}_3 \ \bm{k}_1 \ \bm{k}_2}(t_1)\mathcal{F}^\times_{\bm{k}_2}(t_2)\mathcal{F}^+_{\bm{k}_1}(t_3)\text{Im}[\psi^{+}_{k_1}(t_3)\psi^{+*}_{k_1}(t)]\text{Im}[\psi^{\times}_{k_2}(t_2)\psi^{\times*}_{k_2}(t)]\text{Im}[\psi^{\times}_{k_3}(t_1)\psi^{\times*}_{k_3}(t)] \right] + (\bm{k}_2 \leftrightarrow \bm{k}_3) \notag \ .
\end{align}
As in the same analysis before, we assume that the commutators including $\delta \dot{\hat{A}}$ or $\delta \hat{\sigma}$ are negligible from the discussion in Appendix \ref{appendix negligible} and that all momenta are smaller than $k_A$.
$\mathcal{F}$ on the attractor phase can be approximated as
\begin{align}
\mathcal{F}^{\times\times}_{\bm{k} \ \bm{p} \ \bm{q}} &\simeq 
-\dfrac{1}{\sqrt{2}}\dfrac{H^2}{\sqrt{2p}p}\dfrac{H^2}{\sqrt{2q}q}\left(\dfrac{k_A}{p}\right)^{\Delta n}\left(\dfrac{k_A}{q}\right)^{\Delta n}\gamma(n_{\rm ini})^2\left(\dfrac{\sqrt{2\bar{\rho}_E}}{H\Lambda}\right)^{2} \label{eq: F32} \ , \\
\mathcal{F}^{+\times}_{\bm{k} \ \bm{p} \ \bm{q}} &\simeq -\sqrt{2}
\dfrac{H^2}{\sqrt{2p}p}\dfrac{H^2}{\sqrt{2q}q}\left(\dfrac{k_A}{p}\right)^{\Delta n}\left(\dfrac{k_A}{q}\right)^{\Delta n}\gamma(n_{\rm ini})^2\left(\dfrac{\sqrt{2\bar{\rho}_E}}{H\Lambda}\right)^{2}\cos2\theta_{\hat{\bm{p}}}\cos\theta_{\hat{\bm{k}}\cdot\hat{\bm{p}}} \label{eq: F33} \ ,
\end{align}
where we have dropped their complex phase factors for the same reason as in $\mathcal{F}^{++}$.
Then, we finally obtain
\begin{align}
&B^{+\times\times}_h
\sim
2^{5/2}\dfrac{H^4}{\Mpl^4}\dfrac{\Lambda^2}{\Mpl^2}\gamma(n_{\rm ini})^4\Delta n^{3} N_A^3 \sum_{i = 1}^3 \mathcal{G}^{+\times\times}_i(\bm{k}_j) \ , \notag \\
&\mathcal{G}^{+\times\times}_1(\bm{k}_j) \equiv
\dfrac{1}{k_2^3k_3^3}
\left( \dfrac{k_A}{k_2} \right)^{2\Delta n}
\left( \dfrac{k_A}{k_3} \right)^{2\Delta n}
\sin\theta_{\hat{\bm{k}}_2}\sin\theta_{\hat{\bm{k}}_3}, \notag \\
&\mathcal{G}^{+\times\times}_2(\bm{k}_j) \equiv
-\dfrac{1}{k_1^3k_3^3}
\left( \dfrac{k_A}{k_1} \right)^{2\Delta n}
\left( \dfrac{k_A}{k_3} \right)^{2\Delta n}
\cos 2\theta_{\hat{\bm{k}}_1}\sin\theta_{\hat{\bm{k}}_1}\cos^2\theta_{\hat{\bm{k}}_1}
\sin\theta_{\hat{\bm{k}}_3}\cos\theta_{\hat{\bm{k}}_1 \cdot \hat{\bm{k}}_2}, \notag \\
&\mathcal{G}^{+\times\times}_3(\bm{k}_j) \equiv
-\dfrac{1}{k_1^3k_2^3}
\left( \dfrac{k_A}{k_1} \right)^{2\Delta n}
\left( \dfrac{k_A}{k_2} \right)^{2\Delta n}
\cos 2\theta_{\hat{\bm{k}}_1}\sin\theta_{\hat{\bm{k}}_1}\cos^2\theta_{\hat{\bm{k}}_1}
\sin\theta_{\hat{\bm{k}}_2}\cos\theta_{\hat{\bm{k}}_1 \cdot \hat{\bm{k}}_3}.
\end{align}
As in the previous case with $B^{+++}_h$, we consider the squeezed limit also here.
There exist two cases in taking limit: $k_2 \simeq k_3 \gg k_1$ and $k_1 \simeq k_2 \gg k_3$ ~(or equivalently $k_1 \simeq k_3 \gg k_2$ under the symmetry between $\bm{k}_2 \leftrightarrow \bm{k}_3$), respectively.
Each amplitude of $f_{\rm NL}$ in Eq.~\eqref{eq: fNLsq} is given by
\begin{align}
f^{+\times\times}_{\text{NL}, k_2 \simeq k_3 \gg k_1}
&\simeq \mathcal{A}_{\rm sq}^{+\times\times}(k_2, k_1)\mathcal{G}^{+\times\times}_{\rm{sq},1}(\hat{\bm{k}}_2, \ \hat{\bm{k}}_1) \ , \quad f^{+\times\times}_{\text{NL}, k_1 \simeq k_2 \gg k_3}
\simeq \mathcal{A}_{\rm sq}^{+\times\times}(k_1, k_3)\mathcal{G}^{+\times\times}_{\rm{sq},3}(\hat{\bm{k}}_1, \ \hat{\bm{k}}_3) \ , \label{eq: fnlsqpcc}
\end{align}
where
\begin{align}
\mathcal{A}_{\rm sq}^{+\times\times}(k_i, k_j) &= -\dfrac{5}{2^{7/2}3}r_{\text{vac}}^2\dfrac{\Lambda^2}{\Mpl^2}\gamma(n_{\rm ini})^4 \Delta n^3 N_A^3\left(\dfrac{k_A}{k_i}\right)^{2\Delta n}\left(\dfrac{k_A}{k_j}\right)^{2\Delta n} \ , \\
\mathcal{G}^{+\times\times}_{\rm{sq},1}(\hat{\bm{k}}_2, \ \hat{\bm{k}}_1) &= 2\cos^2\theta_{\hat{\bm{k}}_1}\cos2\theta_{\hat{\bm{k}}_1}\sin\theta_{\hat{\bm{k}}_1}\sin\theta_{\hat{\bm{k}}_2}\cos\theta_{\hat{\bm{k}}_2\cdot\hat{\bm{k}}_1} \ , \\
\mathcal{G}^{+\times\times}_{\rm{sq},3}(\hat{\bm{k}}_1, \ \hat{\bm{k}}_3) &= \sin^3\theta_{\hat{\bm{k}}_1}\sin\theta_{\hat{\bm{k}}_3}(2+\cos2\theta_{\hat{\bm{k}}_1}) \ .
\end{align}
Also, regarding the non-Gaussianity in the equilateral limit \eqref{eq: fNLeq}, we obtain
\begin{align}
f^{+\times\times}_{\rm NL, eq} &= \mathcal{A}^{+\times\times}_{\rm eq}(k)\mathcal{G}^{+\times\times}_{\rm eq}(\hat{\bm{k}}_1) \label{eq: fnleqpcc} \ , \\
\mathcal{A}^{+\times\times}_{\rm eq}(k) &= \dfrac{5}{2^{9/2}9}r_{\text{vac}}^2\dfrac{\Lambda^2}{\Mpl^2}\gamma(n_{\rm ini})^4\Delta n^{3} N_A^3 \left(\dfrac{k_A}{k}\right)^{4\Delta n} \ , \\
\mathcal{G}^{+\times\times}_{\rm eq}(\hat{\bm{k}}_1)
& = 1 - 2 \cos^2 \theta_{\hat{\bm{k}}_1} -6\cos^4 \theta_{\hat{\bm{k}}_1} +\cos^6 \theta_{\hat{\bm{k}}_1} \ . 
\end{align}

\subsection{Detectability}

Here we investigate the detectability of the above tensor bispectra in the squeezed limit (\eqref{eq: fnlsq+++}, \eqref{eq: fnlsqpcc}) and the equilateral limit (\eqref{eq: fnleqppp}, \eqref{eq: fnleqpcc}).
To begin with, we discuss several constraints in this scenario.
First, the amplified gauge field also affects the curvature perturbations on the uniform energy density slice, $\zeta$.
In the flat slicing, it is given by
\begin{equation}
    \zeta = \dfrac{\sum_i\delta\rho_i}{3\sum_i(\bar{\rho}_i + \bar{P}_i)} \simeq \dfrac{\Omega_\phi\delta\rho_\phi/\bar{\rho}_\phi + \Omega_\sigma\delta\rho_\sigma/\bar{\rho}_\sigma + \Omega_A\delta\rho_A/\bar{\rho}_A }{2\epsilon_H} \ ,
\end{equation}
where $\Omega_i \equiv \bar{\rho}_i/(3\Mpl^2H^2)$ and we have ignored the contribution of the spectator and gauge field in the denominator.
The contributions of the second term and third term are negligible since they have sub-dominant energy and finally become massive during inflation.
Therefore, $\zeta$ is provided by the first term, where $\delta A_{\rm int}$ and $\delta\sigma_{\rm src}$ can source $\delta\phi$ via the gravitational interactions.
In our previous work, we have already estimated the sourced curvature power spectrum, and the ratio of power spectrum of the sourced mode to the vacuum mode $\mathcal{R}_\zeta \equiv \mathcal{P}_\zeta^{(s)}/\mathcal{P}_\zeta^{\rm (vac)}$ is given by \cite{Fujita:2018zbr}
\begin{align}
    \mathcal{R}_\zeta &= \left[n_{\rm ini}\gamma(n_{\rm ini}) \, \sqrt{2\epsilon_H} \frac{\Lambda}{\Mpl}
\left(\frac{k_A}{k}\right)^{\dn}(N_A-1/3)\right]^2
\notag\\
&\quad\times
\left(1-\frac{3n_{\rm ini}-4}{n_{\rm ini}}\cos^2\theta_{\hat{\bm{k}}}+\frac{\dn(3n_{\rm ini}-2)}{n_{\rm ini}^2}\cos^4\theta_{\hat{\bm{k}}}-\frac{\dn^2}{n_{\rm ini}^2}\cos^6\theta_{\hat{\bm{k}}}\right) \ .
\label{sourced Pz}
\end{align}
Then, the strongly red-tilted scale-dependence of the sourced curvature perturbation may affect the scalar spectral index $n_s \equiv 1 + d\ln\mathcal{P}_\zeta/d\ln k$ or its running $dn_s/d\ln k$.
Assuming that $\mathcal{R}_\zeta$ is much smaller than unity, they are approximately estimated as
\begin{align}
n_s = n^{(\rm vac)}_{s} + \dfrac{d\mathcal{R}_\zeta/d\ln k}{1+\mathcal{R}_\zeta} \simeq n^{(\rm vac)}_{s} - 2\Delta n \mathcal{R}_\zeta \ , \quad \dfrac{dn_s}{d\ln k} \simeq \dfrac{dn_s^{(\rm vac)}}{d\ln k} + (2\Delta n)^2\mathcal{R}_\zeta \ ,
\end{align}
where the superscript ``(vac)" means the contribution from the vacuum mode.
We require that these corrections should be smaller than the $1\sigma$ uncertainty
of measured values
\begin{equation}
2\Delta n\mathcal{R}_\zeta \lesssim 4\times10^{-3} \ , \qquad (2\Delta n)^2\mathcal{R}_\zeta \lesssim 7\times10^{-3} \ , \label{eq: ns}
\end{equation}
at the pivot scale $k = 0.05 \text{Mpc}^{-1}$ \cite{Akrami:2018odb}.

Next, for the validity of perturbative approach of the gauge field, $|\bar{A}_i| \gg |\delta A_i|$ should be required
and this means that the background electric field $\bar{\rho}_E$ should be much larger than $\mathcal{O}(H^4)$.
Requiring the condition $\bar{\rho}_E > 10^{2}H^4$, we get an upper bound on the duration of the early growing phase of the gauge field $N_G \equiv \ln(k_A/k)$ \cite{Fujita:2018zbr}:
\begin{equation}
N_G < N_G^{\rm max} \equiv \frac{1}{2\dn}
\ln\left[\frac{3\dn }{10^2\pi^2r_{\rm vac}\mcP_\zeta^{\rm obs}}\frac{\Lambda^2}{\Mpl^2}\right] \ .
\label{bound NG}
\end{equation}

\begin{figure}[htbp]
 \begin{minipage}{0.5\hsize}
  \begin{center}
   \includegraphics[width=75mm]{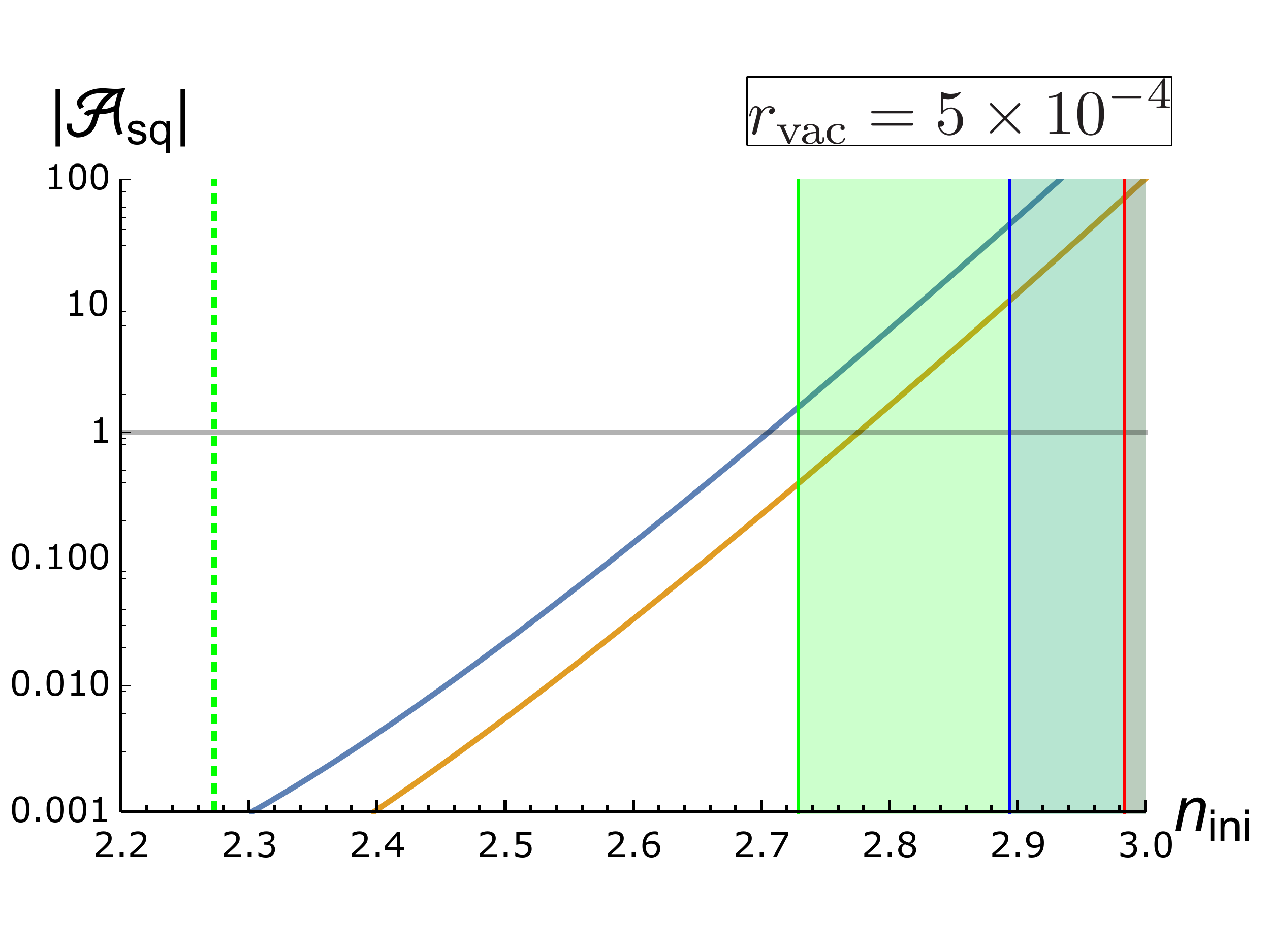}
  \end{center}
 \end{minipage}
  \begin{minipage}{0.5\hsize}
  \begin{center}
   \includegraphics[width=75mm]{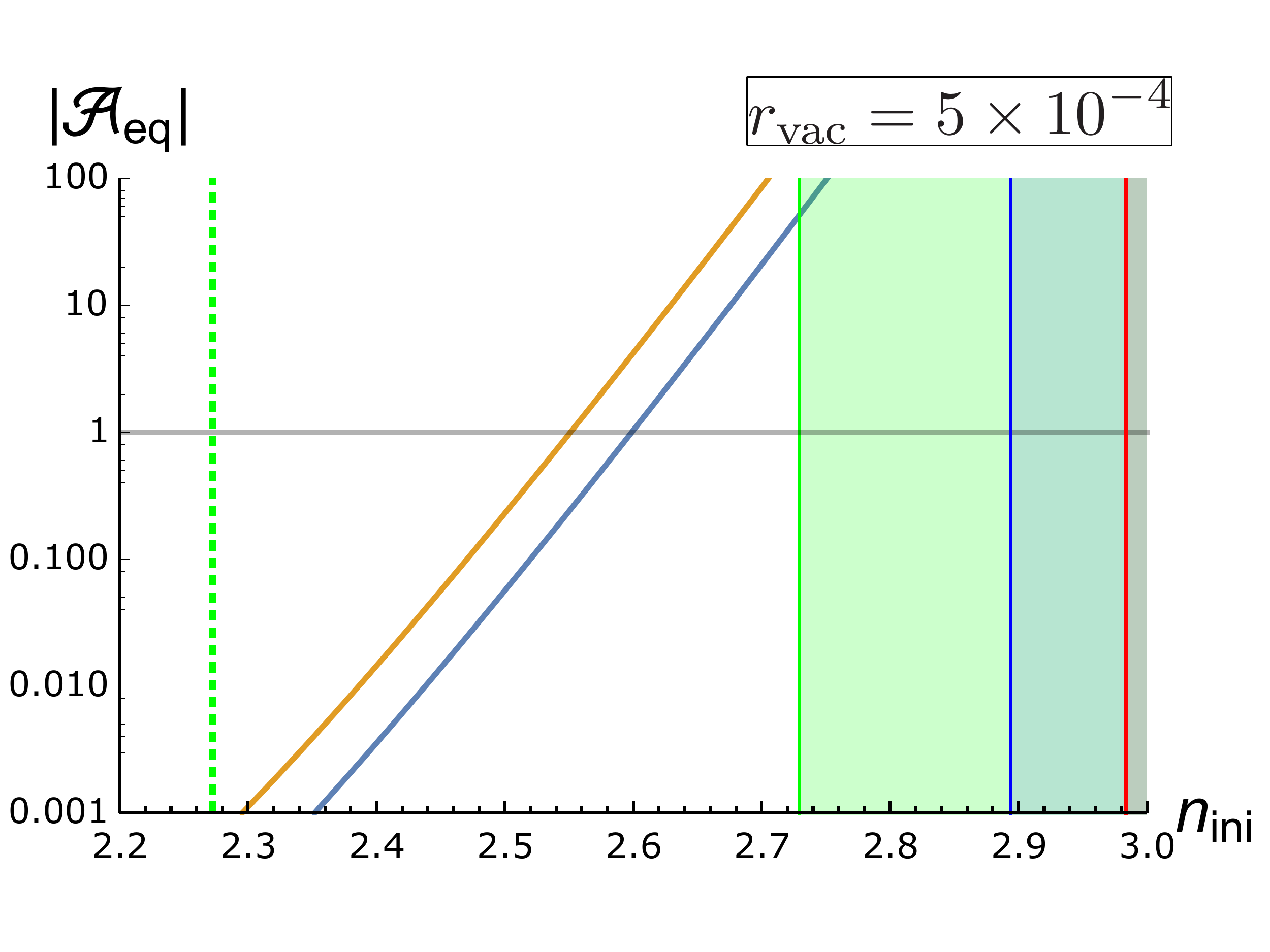}
  \end{center}
 \end{minipage}
  \begin{minipage}{0.5\hsize}
  \begin{center}
   \includegraphics[width=75mm]{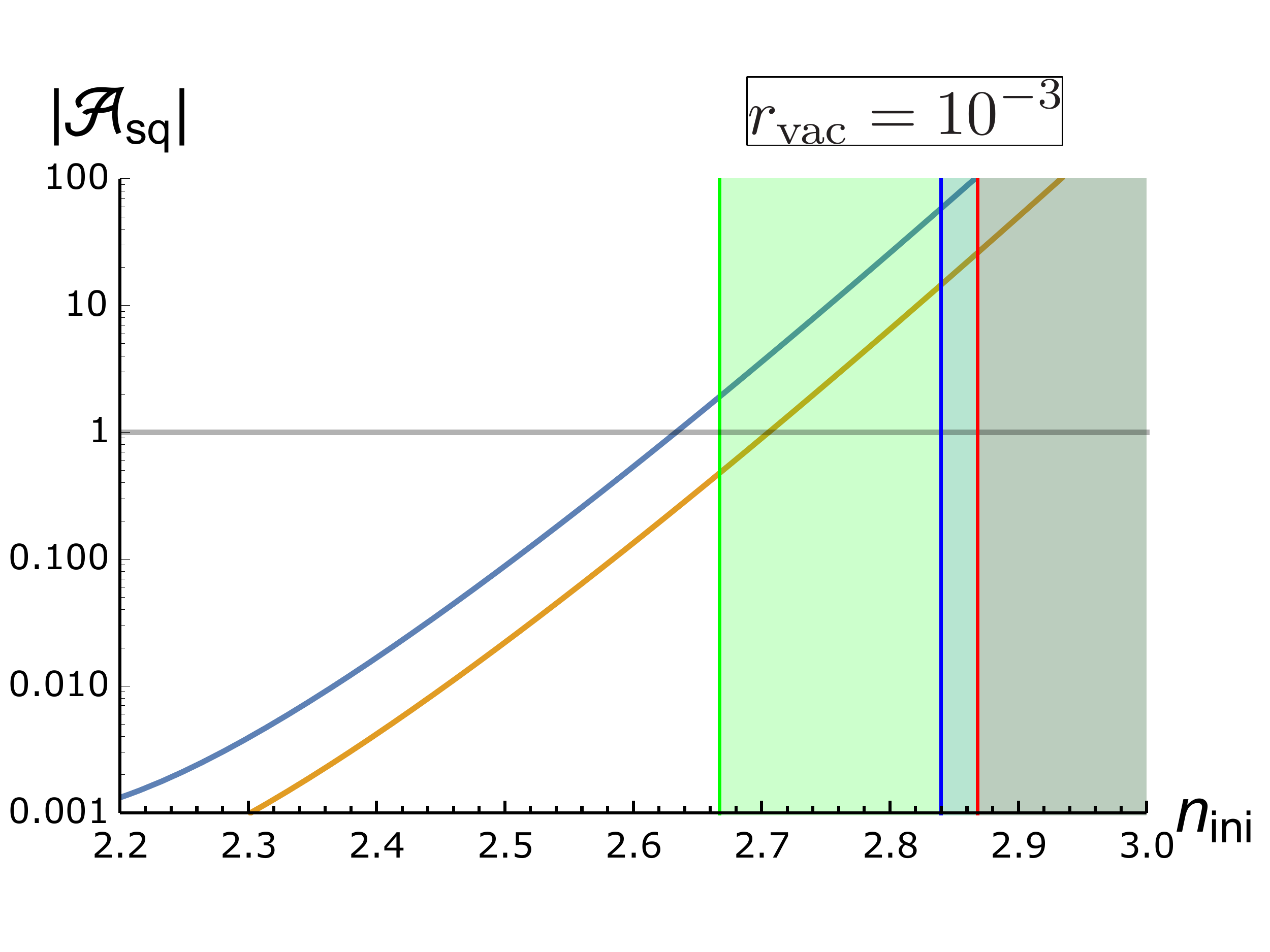}
  \end{center}
 \end{minipage}
  \begin{minipage}{0.5\hsize}
  \begin{center}
   \includegraphics[width=75mm]{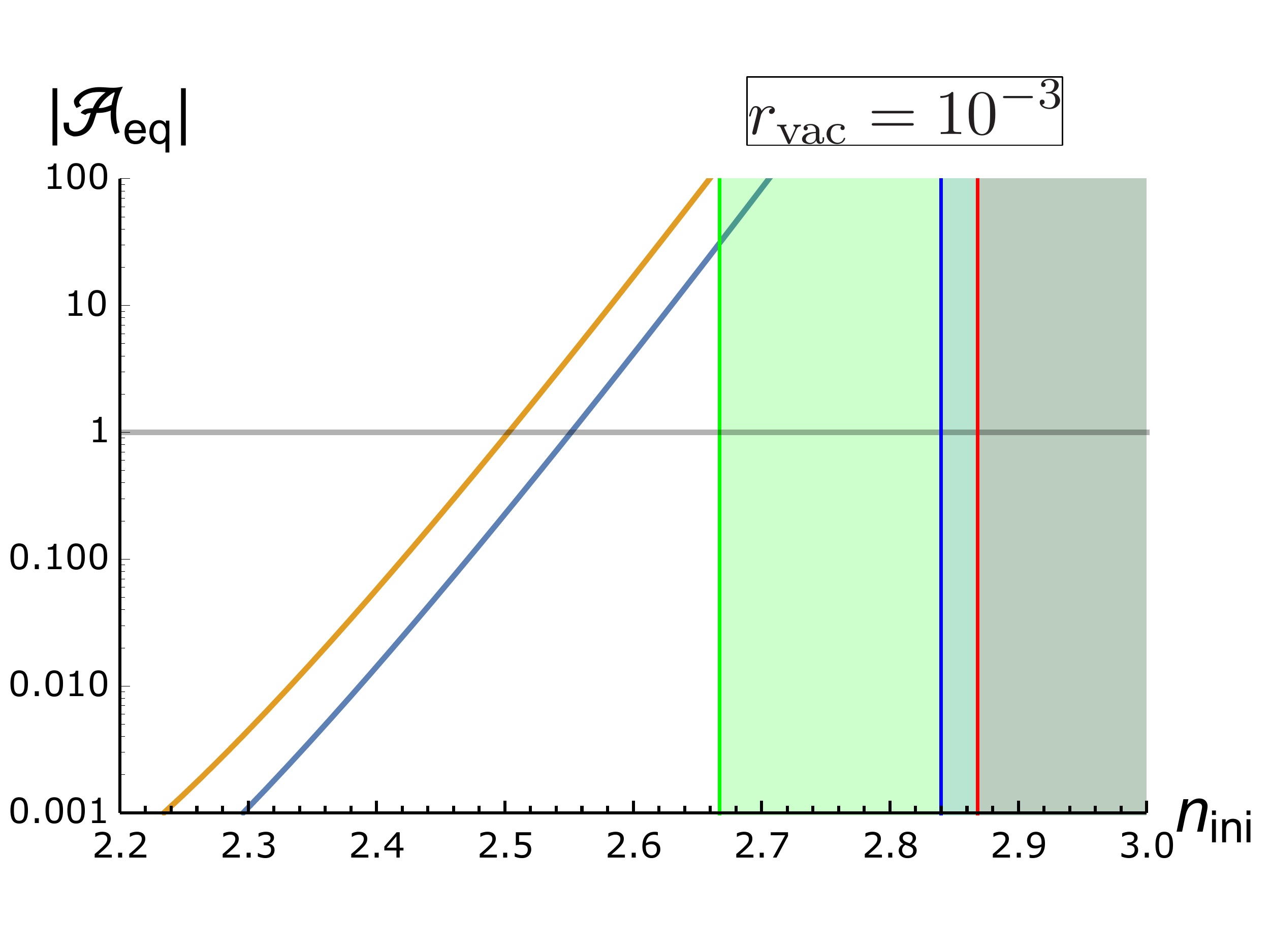}
  \end{center}
 \end{minipage}
  \begin{minipage}{0.5\hsize}
  \begin{center}
   \includegraphics[width=75mm]{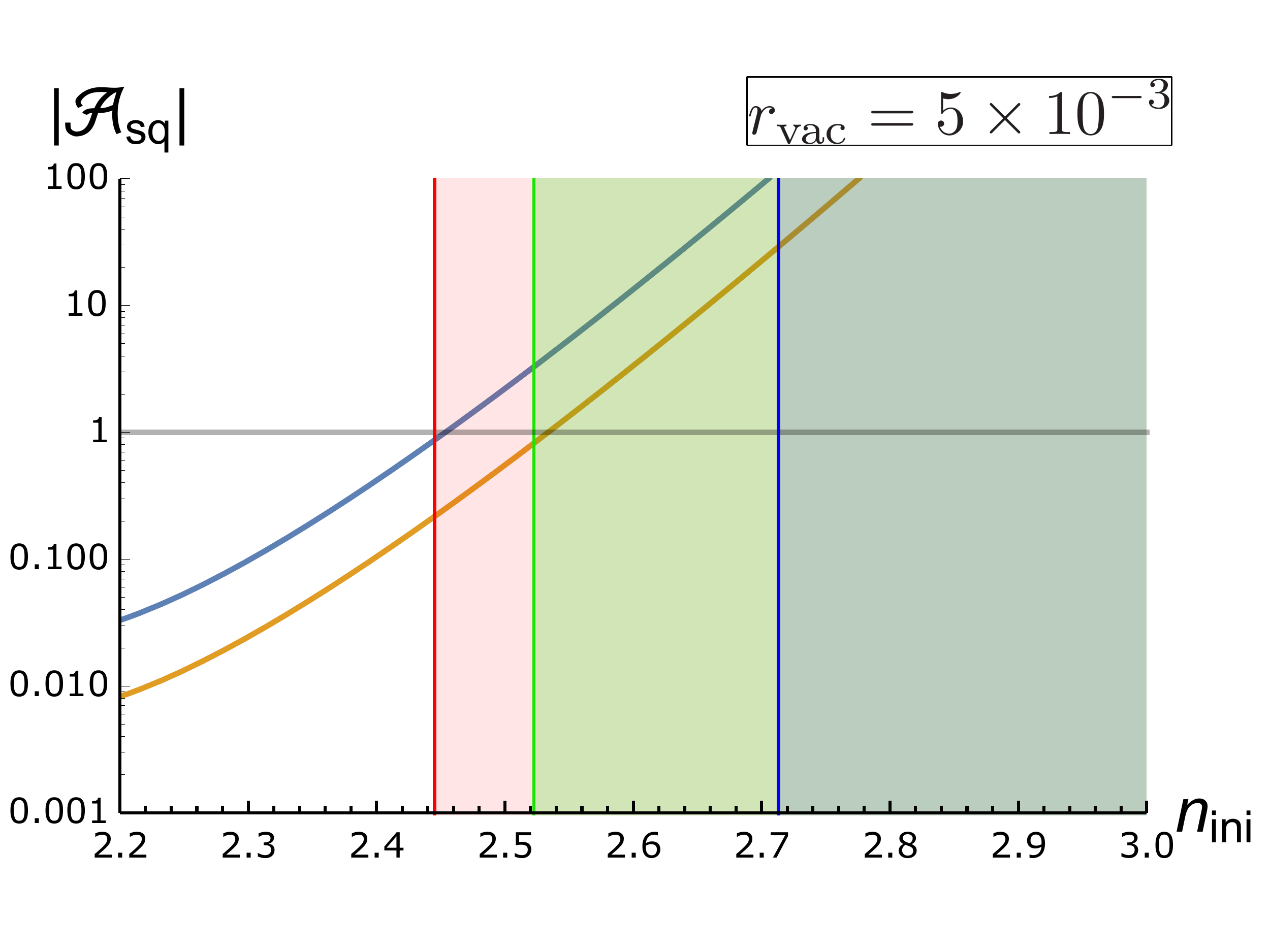}
  \end{center}
 \end{minipage}
  \begin{minipage}{0.5\hsize}
  \begin{center}
   \includegraphics[width=75mm]{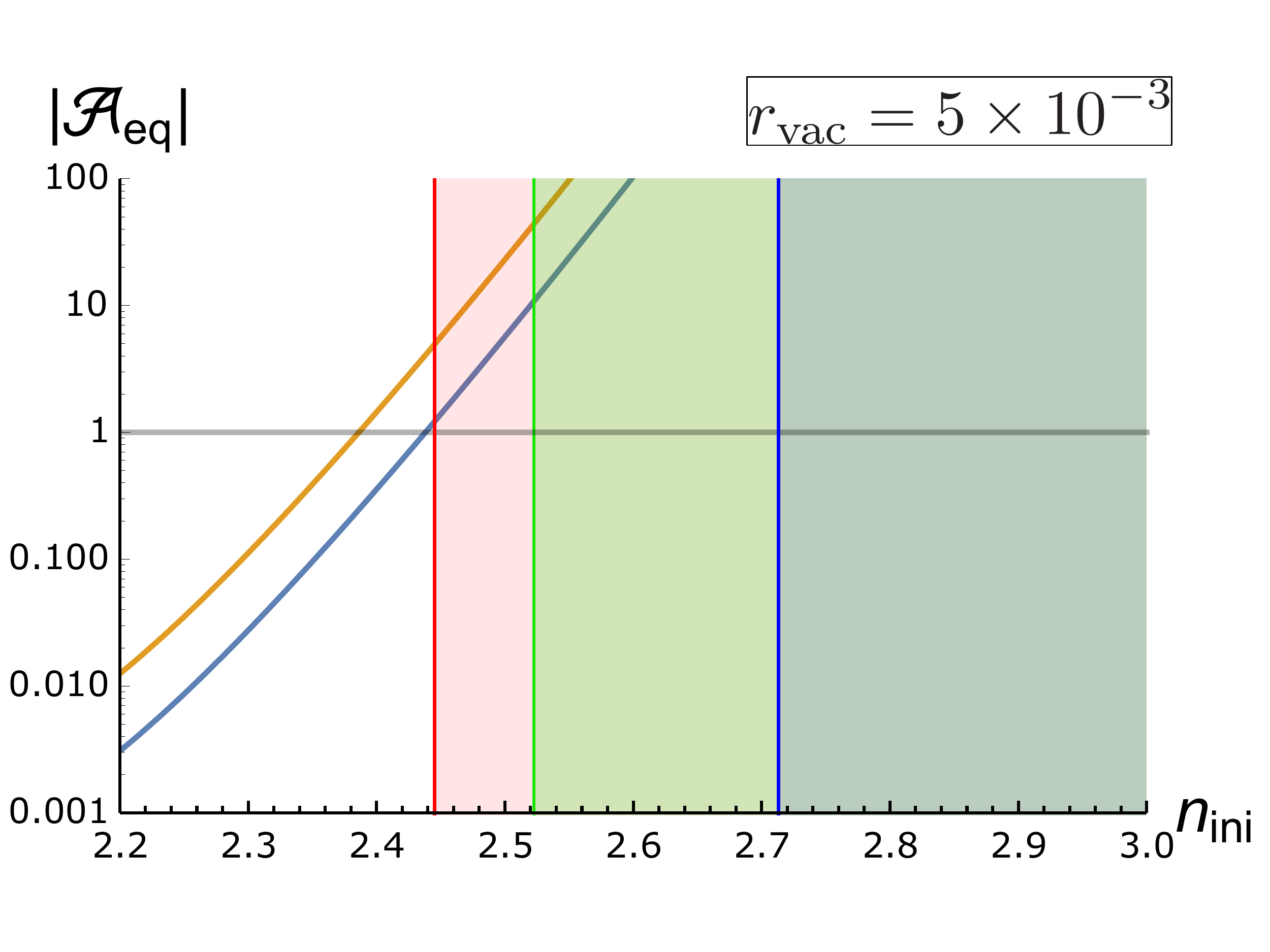}
  \end{center}
 \end{minipage}
\caption{The amplitude of tensor non-Gaussianity in the squeezed limit (left column) and equilateral limit (right column) with respect to $n_{\rm ini}$ and $r_{\rm vac} = 5\times10^{-4}$ (top panels), $10^{-3}$ (middle panels), $5\times10^{-3}$ (bottom panels).
In each panel, the blue (orange) line represents $\mathcal{A}^{+++}_{\text{sq/eq}}$ ($\mathcal{A}^{+\times\times}_{\text{sq/eq}})$.
At the green dotted line, the total tensor-to-scalar ratio becomes $r_{0.002}(n_{\rm ini}) = 10^{-3}$ evaluated at $k = 0.002 ~\text{Mpc}^{-1}$, while the green shaded region represents $r_{0.002}(n_{\rm ini}) \gtrsim 0.06$, which are excluded by the current Planck observation.
The red shaded region represents either severer constraint in Eq.~\eqref{eq: ns}, where the generation of scalar spectral tilt is non-negligible.
The blue shaded region represents $N_G \gtrsim N_G^{\rm max}$, where the perturbative treatment $\bar{A}_i \gg \delta A_i$ comes to be suspicious.
We also show the line $|\mathcal{A}_{\text{sq}/\text{eq}}| = 1$ which could be a target parameter space of the upcoming CMB B-mode measurements \cite{Shiraishi:2019yux}.
In these plots, we use the set of model parameters \eqref{eq: parameters} and pivot scales \eqref{eq: pivots}.
}
\label{fig: fnl}
\end{figure}

Taking the above constraints into account, we plot the amplitudes of non-linear parameters $\mathcal{A}^{+++/+\times\times}_{\text{sq}}$ and $\mathcal{A}^{+++/+\times\times}_{\text{eq}}$ in Figure \ref{fig: fnl} with the following model parameters and the pivot scales in each limit
\begin{align}
&\Lambda =10^{-2}\Mpl, \quad  N_A = 15 \ ,\label{eq: parameters} \\ 
&(\text{squeezed}): k_{\rm long} = 0.0002~\text{Mpc}^{-1} \ (l \sim 3) = e^{-7}k_A \quad \ll \quad k_{\rm short} = 0.05~\text{Mpc}^{-1} \ (l \sim 690) \ , \notag \\
&(\text{equilateral}): k = 0.0002~\text{Mpc}^{-1} = e^{-7}k_A \ ,
\label{eq: pivots}
\end{align}
where $l$ represents the angular multipole moment for the CMB angular power spectrum.
Owing to the strongly red-tilted scale dependence, the non-Gaussianity rapidly increases as $n_{\rm ini}$ goes up, and the equilateral limit can be greater than the squeezed limit since we can take the largest CMB scale for all momenta in the former limit.
In Figure \ref{fig: fnl}, we also depict the allowed parameter regions for three different choices of $r_{\rm vac}$.
We note that we have taken the average of $\theta_{\hat{\bm{k}}}$ in the estimation of sourced curvature power spectrum  \eqref{sourced Pz} and tensor power spectrum \eqref{sourced Ph}.
As $r_{\rm vac}$ increases, generations of sourced perturbations controlled by $\Delta n$ become unable to be large due to the CMB constraints, and consequently the allowed parameter space shrinks.
Among these constraints, we can see that the generation of spectral tilt in curvature perturbation is most sensitive to the increase of $r_{\rm vac}$.
Even so, there remains the allowed regions where $\mathcal{A}$ can take large values $\mathcal{A} \gtrsim \mathcal{O}(1)$.

%
%

In addition to $\mathcal{A}$, the angular structures $\mathcal{G}_{\rm sq/eq}$ in each limit also contribute the overall amplitude of bispectrum.
For the isotropic CMB measurement, we will actually need to take an average of the angular function in CMB analysis, namely, averaging $\mathcal{G}$ over all directions of background vector field $\hat{\bar{\bm{E}}}$.
This procedure makes the total amplitude of non-Gaussianity smaller by an order of magnitude (in particular, crucial to the squeezed configuration).
However,
the averaged squeezed bispectrum still leaves a statistical anisotropy characterized by the multipole moments with respect to angles between long-wavelength and short-wavelength modes  \cite{Shiraishi:2013vja,Franciolini:2018eno}.
In order to find such an expression, the angle average should be taken in two-dimensional $4\pi$ sky surface.
Our result is, however, valid only for the special case that $\hat{\bar{\bm{E}}}$ lies on the momentum triangle, which means that the domain of isotropic average is restricted to only one-dimensional $2\pi$ circle in the sky.
Therefore, we will need to find the general functional form of tensor bispectrum regardless of whether $\hat{\bar{\bm{E}}}$ is in parallel with the triangle plane or not.
We leave a construction of such a template as a future issue.

\section{Summary and Outlook}
\label{sum}

Primordial gravitational waves are the smoking gun of inflationary universe.
Although no significant signal has been detected from the CMB temperature and E-mode polarization data, the upcoming
B-mode polarization surveys are expected to give us substantial information about the origin of primordial tensor perturbations.
Especially, testing the non-Gaussianity of primordial gravitational waves is a powerful way to clarify the underlying high-energy physics in the early universe.
In this study, we explored the generation of tensor non-Gaussianity 
in a kind of anisotropic inflation \cite{Fujita:2018zbr}, where a classical $U(1)$ electric field is provided by a rolling spectator field via the kinetic coupling.
There are three stages in the background evolution.
(i) Initially, the background electric field is negligibly small but increases due to the energy transfer from the spectator field.
(ii) At the intermediate stage called attractor phase, the enhanced electric energy backreacts to the kinetic energy of the spectator field and settles in a constant value.
(iii) The spectator field becomes stabilized and accordingly the gauge field decays on the last stage of inflation.
In perturbations, due to the presence of background vector field, the fluctuations of gauge field and spectator field are coupled to the tensor perturbations at linear level and the amplified gauge modes source the statistically anisotropic tensor perturbations on the super-horizon scale.
The key point is that the spectator mode is also sourced by the one polarization mode of the gauge field and the enhanced spectator mode contributes to the generation of tensor mode.
We employed the in-in formalism and computed the bispectrum of tensor mode,
and found that $\langle h^+h^+h^+\rangle$ and $\langle h^+h^\times h^\times \rangle$ are the dominant contribution in the tensor bispectrum.
This seems to reflect the fact that the scenario is parity-conserving.
The resultant bispectrum is enhanced at small momentum scales
due to the highly red-tilted scale dependence of tensor perturbations.
For the comparison of spectral shapes, we evaluated the non-linear parameters in the squeezed limit $f^{+++/+\times\times}_{\rm NL,sq}$ and the equilateral limit $f^{+++/+\times\times}_{\rm NL,eq}$.
We found that both limits exhibit a rich dependence on angles between the preferred direction and the wave vectors respected by the production of polarization tensors, including higher multipole moments.
We showed that the amplitudes in each limit are potentially testable with the future B-mode observations such as LiteBIRD \cite{Matsumura:2013aja} or CMB-S4 \cite{Abazajian:2016yjj} missions.

For obtaining more precise forecast constraints, we should develop the CMB angular bispectrum derived from this model and examine the preferable estimator to limit the statistical anisotropies in the primordial bispectrum templates.
In the case of scalar bispectrum, the nonzero signals appear in the outside of trianglar condition $|l_1 - l_2| \leq l_3 \leq l_1 + l_2$, where the statistically isotropic bispectra are exactly zero \cite{Shiraishi:2011ph, Bartolo:2011ee}.
We expect that a similar result is obtained in CMB tensor bispectrum and becomes a novel information about the primordial gravitational waves.
And, it is also interesting to study the tensor non-Gaussianity in the model of anisotropic inflation with two-form field \cite{Obata:2018ilf}.
We leave these issues in our future work.


\section*{Acknowledgement}
\label{ack}

We would like to thank Takahiro Tanaka for fruitful discussion
and also Tomohiro Fujita for the collaboration at the early stage.
In this work, IO was supported by JSPS Overseas Research Fellowship. 
KM was supported by World Premier International Research Center Initiative (WPI Initiative), MEXT, Japan, the Program of Excellence in Photon Science, and the JSPS Research Fellowships for Young Scientists Grant No. JP20J20248.
SY is supported by JSPS Grant-in-Aid for Scientific Research(B) No. JP20H01932 and (C) No. JP20K03968.
\appendix

\section{Commutations of \texorpdfstring{$\delta \dot{\hat{A}}^X_{\rm int}$}{} and \texorpdfstring{$\delta\hat\sigma_{\rm src}$}{}}
\label{appendix negligible}

In this appendix, we describe the evolution of the perturbations $\delta \dot{\hat{A}}^{X/Y}_{\rm int}$ and $\delta\hat\sigma_{\rm src}$ and show that their commutators become negligible on the super-horizon scale.
During the growing phase of the gauge field, the equation of motion for the mode function $\delta A_{\rm int}$ is approximately given by
\begin{align}
  \left[
    \partial_x^2 +1 -\dfrac{n(n-1)}{x^2}
  \right]
  \left( \bar{I}\delta A_{\text{int}, k} \right)
  &\simeq 0 \ ,
\end{align}
where we have used the dimensionless time variable $x = -k\tau$ and $\bar{I}(x)\propto x^n$.
With the Bunch-Davies boundary condition, the solution of $\delta A_{\rm int, k}$ is 
\begin{equation}
  \bar{I}\delta A_{\text{int}, k}(x)
  = \dfrac{1}{2} \sqrt{\dfrac{\pi}{k}}e^{i\tfrac{n}{2}\pi}\sqrt{x}H_{n-1/2}^{(1)}(x),
\end{equation}
where $H^{(1)}_{n-1/2}$ is the Hankel function of the first kind with the order $n-1/2$.
Since we consider the case $n>2$, in the limit of $x \ll 1$,
\begin{align}
  \lim_{x \to 0} \bar{I}\delta A_{\text{int}, k}
  &= \sqrt{\dfrac{\pi}{2k}}e^{i\tfrac{n}{2}\pi}
  \left[
  \dfrac{1}{\Gamma(n+1/2)}\left( \dfrac{x}{2} \right)^{n}
  -i \dfrac{\Gamma\left( n-1/2\right)}{\pi}\left( \dfrac{x}{2}\right)^{1-n}
  \right].
\label{Aint growing solution}
\end{align}
Hence, the fluctuation of the electric field $\Bar{I}\delta \dot{A}^X_{\rm int}/a$ in the super-horizon limit is given by
\begin{align}
  \lim_{x \to 0} i\dfrac{\bar{I}\delta \dot{A}^{X/Y}_{\text{int}, k}}{a}
  &= e^{i\tfrac{n}{2}\pi}
  \dfrac{4H^2\Gamma\left( n+1/2\right)}{\sqrt{2\pi k^3}}\left( \dfrac{x}{2}\right)^{2-n} \ ,
\label{Eint growing solution}
\end{align}
where the first term in the bracket of Eq.~\eqref{Aint growing solution} makes no contribution.
On the other hand, $\delta\sigma_{\rm src}$ in the growing phase obeys
\begin{equation}
  \left[
  \partial_x^2 +1 -\dfrac{2}{x^2}
  \right]\left( a\delta \sigma_{\text{src},\bm{k}} \right)
  =
  2i\sin \theta_{\hat{\bm{k}}} \dfrac{\sqrt{2\bar{\rho}_E}}{H\Lambda x}\bar{I}\partial_x \delta A^X_{\text{int}, k} \ ,
\end{equation}
which is solved as
\begin{equation}
  a\delta\sigma_{\text{src}, \bm{k}}(x)
  = 2 i \sin \theta_{\hat{\bm{k}}}
  \int_0^{\infty} \mathrm{d}y ~G_R(x,y) \dfrac{\sqrt{2\bar{\rho}_E(y)}}{H\Lambda y}\bar{I}\partial_y \delta A^X_{\text{int}, k}(y) \ ,
\label{sigmasrc growing solution}
\end{equation}
where the retarded Green's function $G_R(x,y) = -\Theta(y-x)(x^3-y^3)/(3xy)$ satisfies $[\partial_x^2+1-2/x^2]G_R(x,y) = \delta(x-y)$.
Since $\bar{\rho}_E(x) \simeq \bar{\rho}_{E,\text{att}}x^{-2\Delta n}$ and $\bar{I}(x) \propto x^n$ in the growing phase, the time integral on the super-horizon regime is performed as
\begin{equation}
  a\delta \sigma_{\text{src}, \bm{k}} \simeq -e^{i\tfrac{n\pi}{2}}\sin \theta_{\hat{\bm{k}}} \dfrac{1}{\sqrt{2k}x} \sqrt{\dfrac{3}{\pi \Delta n}} \dfrac{2^n\Gamma(n+1/2)}{2n-1} \dfrac{x^{4-2n}}{x_A^{2-n}} \label{sigmasrc growing solution k} \ .
\end{equation}
Therefore, dropping the irrelevant phase factor $e^{in\pi/2}$, Eqs.~\eqref{Eint growing solution} and \eqref{sigmasrc growing solution k} are real numbers on the super-horizon scale in the growing phase.
This condition also persists even on the attractor phase since their complex phases do not rotate due to their equational forms.
Therefore, $\hat{\mathcal{F}}^+(\delta \hat{A}^X_{\rm int}, \ \delta\hat{\sigma}_{\rm src})$ and $\hat{\mathcal{F}}^\times(\delta \hat{A}^Y_{\rm int})$ in Eqs.~\eqref{eq: Fp} and \eqref{eq: Fc} become classical values on the super-horizon scale
and we can reasonably assume that their commutators are negligible in the computation of correlation functions.

\section{Contributions from other diagrams}
\label{con}

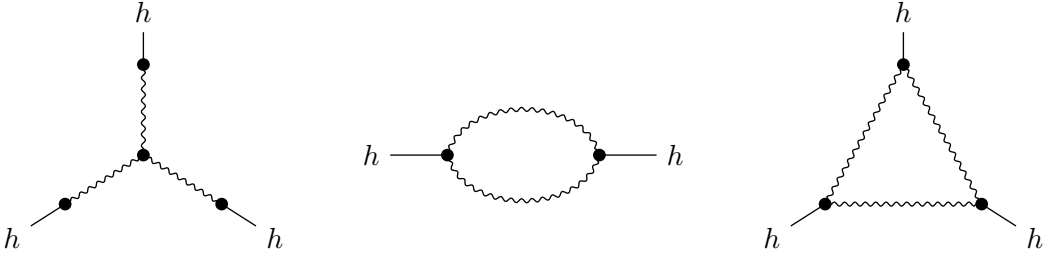
\begin{figure}[thpb]    
\begin{center}     
\begin{tikzpicture} 
\begin{feynhand}
    \vertex [particle] (p1) at (-7-1.73,-1) {$h$};
    \vertex [particle] (p2) at (-7+1.73,-1) {$h$};
    \vertex [particle] (p3) at (-7,2) {$h$};
    \vertex [dot] (wp1) at (-7-1.73+0.7,-1+0.8-0.35) {};
    \vertex [dot] (wp2) at (-7+1.73-0.7,-1+0.8-0.35) {};
    \vertex [dot] (wp3) at (-7,1.30) {};
    \vertex [dot] (wp4) at (-7,0.10) {};
    \propag [plain] (p1) to (wp1);
    \propag [plain] (p2) to (wp2);
    \propag [plain] (p3) to (wp3);
    \propag [photon] (wp1) to (wp4);
    \propag [photon] (wp2) to (wp4);
    \propag [photon] (wp4) to (wp3);
    \vertex [particle] (i1) at (-4,0.1) {$h$};
    \vertex [particle] (f1) at (0,0.1) {$h$};
    \vertex [dot] (w1) at (-3,0.1) {};
    \vertex [dot] (w2) at (-1,0.1) {};
    \propag [plain] (i1) to (w1);
    \propag [photon] (w1) to [out = 70, in= 110] (w2);
    \propag [photon] (w1) to [out = -70, in= -110] (w2);
    \propag [plain] (w2) to (f1);
    \vertex [particle] (p31) at (3-1.73,-1) {$h$};
    \vertex [particle] (p32) at (3+1.73,-1) {$h$};
    \vertex [particle] (p33) at (3,2) {$h$};
    \vertex [dot] (wp31) at (3-1.73+0.7,-1+0.8-0.35) {};
    \vertex [dot] (wp32) at (3+1.73-0.7,-1+0.8-0.35) {};
    \vertex [dot] (wp33) at (3,1.30) {};
    \propag [plain] (p31) to (wp31);
    \propag [plain] (p32) to (wp32);
    \propag [plain] (p33) to (wp33);
    \propag [photon] (wp31) to (wp32);
    \propag [photon] (wp32) to (wp33);
    \propag [photon] (wp33) to (wp31);
    \end{feynhand}
\end{tikzpicture}
\caption{Other diagrams which contribute to the tensor bispectrum in our model. The wavy line represents $\delta A^{X/Y}_{\rm int}$ or $\delta\sigma_{\rm src}$ including their mixing effects (the crossed cycle is omitted).}
\label{fig: hhhdiagrams2}
\end{center}
\end{figure}

Here, we discuss the possibility of changing our result from other diagrams.
Especially, we compute other tree-level contributions and the one-loop effect to the correlation functions in Fig.~\ref{fig: hhhdiagrams2}, and show that these contributions are negligible in our result.

First, we compute the tree-level contribution (the left diagram in Fig.~\ref{fig: hhhdiagrams2}).
The main interaction Hamiltonian of this diagram is as follows:
\begin{align}
H_{3m} &= H_{AA\sigma} + H_{A\sigma\sigma} + H_{\sigma\sigma\sigma} \ , \\
H_{AA\sigma} &= a^3\dfrac{1}{2\Lambda}\int \dfrac{d\bm{k}~d\bm{p}~d\bm{q}}{(2\pi)^6}\delta^{(3)}(\bm{k}+\bm{p}+\bm{q})\delta\sigma_{\bm{k}}\left[ -\dfrac{\bar{I}^2}{a^2}\delta\dot{A}^X_{\bm{p}}\delta\dot{A}^X_{\bm{q}}e^X_i(\hat{\bm{p}})e^X_i(\hat{\bm{q}}) \right. \notag \\
 &\left. + \dfrac{\bar{I}^2}{a^2}\delta\dot{A}^Y_{\bm{p}}\delta\dot{A}^Y_{\bm{q}}e^Y_i(\hat{\bm{p}})e^Y_i(\hat{\bm{q}}) + 2i\dfrac{\bar{I}^2}{a^2}\delta\dot{A}^X_{\bm{p}}\delta\dot{A}^Y_{\bm{q}}e^X_i(\hat{\bm{p}})e^Y_i(\hat{\bm{q}}) \right] \ , \\
H_{A\sigma\sigma} &= a^3 \dfrac{2\sqrt{2\bar{\rho}_E}}{\Lambda^2}\int \dfrac{d\bm{k}~d\bm{p}~d\bm{q}}{(2\pi)^6}\delta^{(3)}(\bm{k}+\bm{p}+\bm{q})i\sin\theta_{\hat{\bm{k}}}\dfrac{\bar{I}\delta\dot{\hat{A}}^X_{\bm{k}}}{a}\delta\hat{\sigma}_{\bm{p}}\delta\hat{\sigma}_{\bm{q}} \ , \\
H_{\sigma\sigma\sigma} &= -a^3\dfrac{4}{3}\dfrac{\bar{\rho}_E}{\Lambda^3}\int \dfrac{d\bm{k}~d\bm{p}~d\bm{q}}{(2\pi)^6}\delta^{(3)}(\bm{k}+\bm{p}+\bm{q})~\delta\hat{\sigma}_{\bm{k}}\delta\hat{\sigma}_{\bm{p}}\delta\hat{\sigma}_{\bm{q}} \ .
\end{align}
Using Eq.~\eqref{eq: xz}, we obtain
 \begin{align}
H_{3m} &= \dfrac{a^3}{\Lambda}\int\dfrac{d\bm{k}d\bm{p}d\bm{q}}{(2\pi)^6}\delta^{(3)}(\bm{k}+\bm{p}+\bm{q})\hat{\mathcal{F}}^{m}_{\bm{k} \ \bm{p} \ \bm{q}}(\delta\hat{A}_{\rm int}, \delta\hat{\sigma}_{\rm osc}, t) \ , \\
\hat{\mathcal{F}}^{m}_{\bm{k} \ \bm{p} \ \bm{q}} &\equiv -\dfrac{\bar{I}^2}{2a^2}\delta\hat{\sigma}_{\text{src},\bm{k}}\delta\dot{\hat{A}}^X_{\text{int},\bm{p}}\delta\dot{\hat{A}}^X_{\text{int},\bm{q}}\cos\theta_{\hat{\bm{p}}\cdot\hat{\bm{q}}} + \dfrac{\bar{I}^2}{2a^2}\delta\hat{\sigma}_{\text{src},\bm{k}}\delta\dot{\hat{A}}^Y_{\text{int},\bm{p}}\delta\dot{\hat{A}}^Y_{\text{int},\bm{q}} \notag \\
&+ \dfrac{2\sqrt{2\bar{\rho}_E}}{\Lambda}i\sin\theta_{\hat{\bm{k}}}\dfrac{\bar{I}\delta\dot{\hat{A}}^X_{\text{int},\bm{k}}}{a}\delta\hat{\sigma}_{\text{src},\bm{p}}\delta\hat{\sigma}_{\text{src},\bm{q}} - \dfrac{4}{3}\dfrac{\bar{\rho}_E}{\Lambda^2}\delta\hat{\sigma}_{\text{src},\bm{k}}\delta\hat{\sigma}_{\text{src},\bm{p}}\delta\hat{\sigma}_{\text{src},\bm{q}} \ .
\end{align}
%
%
%
Then, the three-point function includes the following expectation value
 \begin{align}
&\left\langle\left[ H_{3m}(t_1), \left[ H_{2+}(t_2), \left[H_{2+}(t_3), \left[ H_{2+}(t_4), \hat{\psi}^+_{\bm{k_1}}(t)\hat{\psi}^{+}_{\bm{k_2}}(t)\hat{\psi}^{+}_{\bm{k_3}}(t)\right] \right] \right] \right] \right\rangle + (\text{permutation of $H_{3m}$}) \notag \\
&= \Lambda^2\Mpl^{-3}a(t_1)^3a(t_2)^2a(t_3)^2a(t_4)^2\int\dfrac{d\bm{k}d\bm{p}d\bm{q}}{(2\pi)^6}\delta^{(3)}(\bm{k}+\bm{p}+\bm{q})~(2i)^3 \times \notag \\
 &\left\langle\left[ \hat{\mathcal{F}}^{m}_{\bm{k} \ \bm{p} \ \bm{q}}(t_1), \ \hat{\mathcal{F}}^+_{\bm{k}_3}(t_2)\hat{\mathcal{F}}^+_{\bm{k}_2}(t_3)\hat{\mathcal{F}}^+_{\bm{k}_1}(t_4) \right]\right\rangle \text{Im}[\psi^{+}_{k_1}(t_4)\psi^{+*}_{k_1}(t)]\text{Im}[\psi^{+}_{k_2}(t_3)\psi^{+*}_{k_2}(t)]\text{Im}[\psi^{+}_{k_3}(t_2)\psi^{+*}_{k_3}(t)] \notag \\
 &+ ... \ .
\end{align}
We find that the commutator of $\delta\dot{\hat{A}}_{\rm int}$ or $\delta\hat{\sigma}_{\rm src}$ is always included, which is negligible from the discussion in Appendix \ref{appendix negligible}.
Therefore, these contributions are irrelevant in the generation of tensor-non-Gaussianity.

Next, we evaluate the contribution of one-loop diagrams in Fig.~\ref{fig: hhhdiagrams2}.
In addition to the enhancement of tensor perturbations on the super-horizon regime, these diagrams include another enhancement coming from the momentum integral in the IR limit of one of the two gauge modes.
These contributions might be serious since the amplified gauge mode is highly red-tilted in our scenario.
In order to evaluate the magnitude of loop effects, we only compute the tensor power spectrum of cross mode $\langle h^\times_{\bm{k}}h^\times_{\bm{k'}}\rangle$ for our analytical convenience.
Regarding the two-point function, we calculate
\begin{align}
\langle \hat{\psi}^{\times}_{\bm{k}}(t)\hat{\psi}^{\times}_{\bm{k}'}(t) \rangle^{\rm (loop)} &= -\int_{-\infty}^tdt_2\int_{-\infty}^{t_2}dt_{1}\left\langle \left[ H_{hAA}(t_1), \left[H_{hAA}(t_2), \hat{\psi}^{\times}_{\bm{k}}(t)\hat{\psi}^{\times}_{\bm{k}'}(t) \right] \right] \right\rangle \notag \\
&= (2\pi)^3\delta^{(3)}(\bm{k}+\bm{k'})\dfrac{2k^2}{9x^2\Mpl^2H^6}\int\dfrac{d\bm{p}}{(2\pi)^3}\cos^2\theta_{\hat{\bm{k}}\cdot(-\hat{\bm{p}})}\int_{x_{\rm min}}^{x_2} \dfrac{dx_1}{x_1}\int_{x_{\rm min}}^x \dfrac{dx_2}{x_2}\prod_{i=1}^2\dfrac{x_i^3-x^3}{x_i^3} \notag \\
 &\times \dfrac{\bar{I}\delta\dot{A}^X_{\bm{p}}(x_1)}{a}\dfrac{\bar{I}\delta\dot{A}^{X*}_{\bm{p}}(x_2)}{a}\dfrac{\bar{I}\delta\dot{A}^Y_{\bm{k}-\bm{p}}(x_1)}{a}\dfrac{\bar{I}\delta\dot{A}^{Y*}_{\bm{k}-\bm{p}}(x_2)}{a} + (x_1 \leftrightarrow x_2) \ , 
\end{align}
where the time integration is restricted to times between $x$ and $x_{\rm min} \equiv \text{min}\left[k/p, \ k/|\bm{k}-\bm{p}|\right]$, when both sourcing modes $\delta A_{\bm{p}}$ and $\delta A_{\bm{k}-\bm{p}}$ have exited the horizon.
Then, defining the dimensionless momentum variables $p_* \equiv p/k \ , |\bm{k}-\bm{p}|_* = |\bm{k}-\bm{p}|/k$, we obtain
\begin{align}
\langle \hat{h}^{\times}_{\bm{k}}(t)\hat{h}^{\times}_{\bm{k}'}(t) \rangle^{\rm (loop)} &= (2\pi)^3\delta^{(3)}(\bm{k}+\bm{k'})\dfrac{2H^4}{9k^3\Mpl^4}\int\dfrac{d\bm{p_*}}{(2\pi)^3}\dfrac{\cos^2\theta_{\hat{\bm{k}}\cdot(-\hat{\bm{p}})}\cos^2 2\theta_{\hat{\bm{p}}}}{p_*^{3}|\bm{k}-\bm{p}|_*^3}\left(\dfrac{k_A}{p}\right)^{2\Delta n}\left(\dfrac{k_A}{|\bm{k}-\bm{p}|}\right)^{2\Delta n} \notag \\
 &\times \int_{x_{\rm min}}^{x} \dfrac{dx_1}{x_1}\int_{x_{\rm min}}^{x} \dfrac{dx_2}{x_2}\prod_{i=1}^2\dfrac{x_i^3-x^3}{x_i^3}\gamma(n)^4\dfrac{2\bar{\rho}_E(x_1)}{H^2\Lambda^2}\dfrac{2\bar{\rho}_E(x_2)}{H^2\Lambda^2} \notag \\
&\sim (2\pi)^3\delta^{(3)}(\bm{k}+\bm{k'})\dfrac{2\pi^2}{k^3}\dfrac{H^4}{2^3\pi^4\Mpl^4}N_A^2\gamma(n)^4\Delta n\left(\dfrac{k_A}{k_{\rm min}}\right)^{2\Delta n}\left(\dfrac{k_A}{k}\right)^{2\Delta n} \ ,
\end{align}
where we have introduced the lowest momentum scale $k_{\rm min}$ corresponding to the horizon size when the growing phase starts.
Therefore, the ratio of 1-loop spectrum to the tree-level spectrum is evaluated as
\begin{align}
\dfrac{\mathcal{P}_h^{\rm (loop)}}{\mathcal{P}^{\rm (tree)}_h} \sim \dfrac{2H^2}{\pi^2\Lambda^2}\dfrac{\gamma(n)^2}{\Delta n}\left(\dfrac{k_A}{k_{\rm min}}\right)^{2\Delta n} \ , \label{eq: loopratio}
\end{align}
which is the order of $(k_{\rm min}^{3/2}\delta\dot{A}_{\bm{k}_{\rm min}}/\dot{\bar{A}})^2$.
As well, the ratio of one-loop diagram to tree-diagram of bispectrum $B^{\rm (loop)}_h/B^{\rm (tree)}_h$ is found to be roughly the same as in Eq.~\eqref{eq: loopratio}.
We find that our used model parameters \eqref{eq: parameters} with $k_{\rm min} = e^{-7}k_A$ (Eq.~\eqref{eq: pivots}) satisfy the condition that the above ratio is sufficiently smaller than unity.

\bibliographystyle{apsrev4-1}
\bibliography{Ref}

\begin{thebibliography}{62}%
\makeatletter
\providecommand \@ifxundefined [1]{%
 \@ifx{#1\undefined}
}%
\providecommand \@ifnum [1]{%
 \ifnum #1\expandafter \@firstoftwo
 \else \expandafter \@secondoftwo
 \fi
}%
\providecommand \@ifx [1]{%
 \ifx #1\expandafter \@firstoftwo
 \else \expandafter \@secondoftwo
 \fi
}%
\providecommand \natexlab [1]{#1}%
\providecommand \enquote  [1]{``#1''}%
\providecommand \bibnamefont  [1]{#1}%
\providecommand \bibfnamefont [1]{#1}%
\providecommand \citenamefont [1]{#1}%
\providecommand \href@noop [0]{\@secondoftwo}%
\providecommand \href [0]{\begingroup \@sanitize@url \@href}%
\providecommand \@href[1]{\@@startlink{#1}\@@href}%
\providecommand \@@href[1]{\endgroup#1\@@endlink}%
\providecommand \@sanitize@url [0]{\catcode `\\12\catcode `\$12\catcode
  `\&12\catcode `\#12\catcode `\^12\catcode `\_12\catcode `\%12\relax}%
\providecommand \@@startlink[1]{}%
\providecommand \@@endlink[0]{}%
\providecommand \url  [0]{\begingroup\@sanitize@url \@url }%
\providecommand \@url [1]{\endgroup\@href {#1}{\urlprefix }}%
\providecommand \urlprefix  [0]{URL }%
\providecommand \Eprint [0]{\href }%
\providecommand \doibase [0]{http://dx.doi.org/}%
\providecommand \selectlanguage [0]{\@gobble}%
\providecommand \bibinfo  [0]{\@secondoftwo}%
\providecommand \bibfield  [0]{\@secondoftwo}%
\providecommand \translation [1]{[#1]}%
\providecommand \BibitemOpen [0]{}%
\providecommand \bibitemStop [0]{}%
\providecommand \bibitemNoStop [0]{.\EOS\space}%
\providecommand \EOS [0]{\spacefactor3000\relax}%
\providecommand \BibitemShut  [1]{\csname bibitem#1\endcsname}%
\let\auto@bib@innerbib\@empty
\bibitem [{\citenamefont {Sorbo}(2011)}]{Sorbo:2011rz}%
  \BibitemOpen
  \bibfield  {author} {\bibinfo {author} {\bibfnamefont {L.}~\bibnamefont
  {Sorbo}},\ }\href {\doibase 10.1088/1475-7516/2011/06/003} {\bibfield
  {journal} {\bibinfo  {journal} {JCAP}\ }\textbf {\bibinfo {volume} {06}},\
  \bibinfo {pages} {003} (\bibinfo {year} {2011})},\ \Eprint
  {http://arxiv.org/abs/1101.1525} {arXiv:1101.1525 [astro-ph.CO]} \BibitemShut
  {NoStop}%
\bibitem [{\citenamefont {Barnaby}\ \emph {et~al.}(2011)\citenamefont
  {Barnaby}, \citenamefont {Namba},\ and\ \citenamefont
  {Peloso}}]{Barnaby:2011vw}%
  \BibitemOpen
  \bibfield  {author} {\bibinfo {author} {\bibfnamefont {N.}~\bibnamefont
  {Barnaby}}, \bibinfo {author} {\bibfnamefont {R.}~\bibnamefont {Namba}}, \
  and\ \bibinfo {author} {\bibfnamefont {M.}~\bibnamefont {Peloso}},\ }\href
  {\doibase 10.1088/1475-7516/2011/04/009} {\bibfield  {journal} {\bibinfo
  {journal} {JCAP}\ }\textbf {\bibinfo {volume} {04}},\ \bibinfo {pages} {009}
  (\bibinfo {year} {2011})},\ \Eprint {http://arxiv.org/abs/1102.4333}
  {arXiv:1102.4333 [astro-ph.CO]} \BibitemShut {NoStop}%
\bibitem [{\citenamefont {Cook}\ and\ \citenamefont
  {Sorbo}(2012)}]{Cook:2011hg}%
  \BibitemOpen
  \bibfield  {author} {\bibinfo {author} {\bibfnamefont {J.~L.}\ \bibnamefont
  {Cook}}\ and\ \bibinfo {author} {\bibfnamefont {L.}~\bibnamefont {Sorbo}},\
  }\href {\doibase 10.1103/PhysRevD.85.023534} {\bibfield  {journal} {\bibinfo
  {journal} {Phys. Rev. D}\ }\textbf {\bibinfo {volume} {85}},\ \bibinfo
  {pages} {023534} (\bibinfo {year} {2012})},\ \bibinfo {note} {[Erratum:
  Phys.Rev.D 86, 069901 (2012)]},\ \Eprint {http://arxiv.org/abs/1109.0022}
  {arXiv:1109.0022 [astro-ph.CO]} \BibitemShut {NoStop}%
\bibitem [{\citenamefont {Mukohyama}\ \emph {et~al.}(2014)\citenamefont
  {Mukohyama}, \citenamefont {Namba}, \citenamefont {Peloso},\ and\
  \citenamefont {Shiu}}]{Mukohyama:2014gba}%
  \BibitemOpen
  \bibfield  {author} {\bibinfo {author} {\bibfnamefont {S.}~\bibnamefont
  {Mukohyama}}, \bibinfo {author} {\bibfnamefont {R.}~\bibnamefont {Namba}},
  \bibinfo {author} {\bibfnamefont {M.}~\bibnamefont {Peloso}}, \ and\ \bibinfo
  {author} {\bibfnamefont {G.}~\bibnamefont {Shiu}},\ }\href {\doibase
  10.1088/1475-7516/2014/08/036} {\bibfield  {journal} {\bibinfo  {journal}
  {JCAP}\ }\textbf {\bibinfo {volume} {08}},\ \bibinfo {pages} {036} (\bibinfo
  {year} {2014})},\ \Eprint {http://arxiv.org/abs/1405.0346} {arXiv:1405.0346
  [astro-ph.CO]} \BibitemShut {NoStop}%
\bibitem [{\citenamefont {Domcke}\ \emph {et~al.}(2016)\citenamefont {Domcke},
  \citenamefont {Pieroni},\ and\ \citenamefont {Binétruy}}]{Domcke:2016bkh}%
  \BibitemOpen
  \bibfield  {author} {\bibinfo {author} {\bibfnamefont {V.}~\bibnamefont
  {Domcke}}, \bibinfo {author} {\bibfnamefont {M.}~\bibnamefont {Pieroni}}, \
  and\ \bibinfo {author} {\bibfnamefont {P.}~\bibnamefont {Binétruy}},\ }\href
  {\doibase 10.1088/1475-7516/2016/06/031} {\bibfield  {journal} {\bibinfo
  {journal} {JCAP}\ }\textbf {\bibinfo {volume} {06}},\ \bibinfo {pages} {031}
  (\bibinfo {year} {2016})},\ \Eprint {http://arxiv.org/abs/1603.01287}
  {arXiv:1603.01287 [astro-ph.CO]} \BibitemShut {NoStop}%
\bibitem [{\citenamefont {Garcia-Bellido}\ \emph {et~al.}(2017)\citenamefont
  {Garcia-Bellido}, \citenamefont {Peloso},\ and\ \citenamefont
  {Unal}}]{Garcia-Bellido:2017aan}%
  \BibitemOpen
  \bibfield  {author} {\bibinfo {author} {\bibfnamefont {J.}~\bibnamefont
  {Garcia-Bellido}}, \bibinfo {author} {\bibfnamefont {M.}~\bibnamefont
  {Peloso}}, \ and\ \bibinfo {author} {\bibfnamefont {C.}~\bibnamefont
  {Unal}},\ }\href {\doibase 10.1088/1475-7516/2017/09/013} {\bibfield
  {journal} {\bibinfo  {journal} {JCAP}\ }\textbf {\bibinfo {volume} {09}},\
  \bibinfo {pages} {013} (\bibinfo {year} {2017})},\ \Eprint
  {http://arxiv.org/abs/1707.02441} {arXiv:1707.02441 [astro-ph.CO]}
  \BibitemShut {NoStop}%
\bibitem [{\citenamefont {{\"O}zsoy}(2018)}]{Ozsoy:2017blg}%
  \BibitemOpen
  \bibfield  {author} {\bibinfo {author} {\bibfnamefont {O.}~\bibnamefont
  {{\"O}zsoy}},\ }\href {\doibase 10.1088/1475-7516/2018/04/062} {\bibfield
  {journal} {\bibinfo  {journal} {JCAP}\ }\textbf {\bibinfo {volume} {04}},\
  \bibinfo {pages} {062} (\bibinfo {year} {2018})},\ \Eprint
  {http://arxiv.org/abs/1712.01991} {arXiv:1712.01991 [astro-ph.CO]}
  \BibitemShut {NoStop}%
\bibitem [{\citenamefont {{\"O}zsoy}(2020)}]{Ozsoy:2020ccy}%
  \BibitemOpen
  \bibfield  {author} {\bibinfo {author} {\bibfnamefont {O.}~\bibnamefont
  {{\"O}zsoy}},\ }\href@noop {} {\  (\bibinfo {year} {2020})},\ \Eprint
  {http://arxiv.org/abs/2005.10280} {arXiv:2005.10280 [astro-ph.CO]}
  \BibitemShut {NoStop}%
\bibitem [{\citenamefont {Dimastrogiovanni}\ and\ \citenamefont
  {Peloso}(2013)}]{Dimastrogiovanni:2012ew}%
  \BibitemOpen
  \bibfield  {author} {\bibinfo {author} {\bibfnamefont {E.}~\bibnamefont
  {Dimastrogiovanni}}\ and\ \bibinfo {author} {\bibfnamefont {M.}~\bibnamefont
  {Peloso}},\ }\href {\doibase 10.1103/PhysRevD.87.103501} {\bibfield
  {journal} {\bibinfo  {journal} {Phys. Rev. D}\ }\textbf {\bibinfo {volume}
  {87}},\ \bibinfo {pages} {103501} (\bibinfo {year} {2013})},\ \Eprint
  {http://arxiv.org/abs/1212.5184} {arXiv:1212.5184 [astro-ph.CO]} \BibitemShut
  {NoStop}%
\bibitem [{\citenamefont {Adshead}\ \emph {et~al.}(2013)\citenamefont
  {Adshead}, \citenamefont {Martinec},\ and\ \citenamefont
  {Wyman}}]{Adshead:2013qp}%
  \BibitemOpen
  \bibfield  {author} {\bibinfo {author} {\bibfnamefont {P.}~\bibnamefont
  {Adshead}}, \bibinfo {author} {\bibfnamefont {E.}~\bibnamefont {Martinec}}, \
  and\ \bibinfo {author} {\bibfnamefont {M.}~\bibnamefont {Wyman}},\ }\href
  {\doibase 10.1103/PhysRevD.88.021302} {\bibfield  {journal} {\bibinfo
  {journal} {Phys. Rev. D}\ }\textbf {\bibinfo {volume} {88}},\ \bibinfo
  {pages} {021302} (\bibinfo {year} {2013})},\ \Eprint
  {http://arxiv.org/abs/1301.2598} {arXiv:1301.2598 [hep-th]} \BibitemShut
  {NoStop}%
\bibitem [{\citenamefont {Obata}\ and\ \citenamefont
  {Soda}(2016{\natexlab{a}})}]{Obata:2016tmo}%
  \BibitemOpen
  \bibfield  {author} {\bibinfo {author} {\bibfnamefont {I.}~\bibnamefont
  {Obata}}\ and\ \bibinfo {author} {\bibfnamefont {J.}~\bibnamefont {Soda}},\
  }\href {\doibase 10.1103/PhysRevD.93.123502} {\bibfield  {journal} {\bibinfo
  {journal} {Phys. Rev. D}\ }\textbf {\bibinfo {volume} {93}},\ \bibinfo
  {pages} {123502} (\bibinfo {year} {2016}{\natexlab{a}})},\ \bibinfo {note}
  {[Addendum: Phys.Rev.D 95, 109903 (2017)]},\ \Eprint
  {http://arxiv.org/abs/1602.06024} {arXiv:1602.06024 [hep-th]} \BibitemShut
  {NoStop}%
\bibitem [{\citenamefont {Maleknejad}(2016)}]{Maleknejad:2016qjz}%
  \BibitemOpen
  \bibfield  {author} {\bibinfo {author} {\bibfnamefont {A.}~\bibnamefont
  {Maleknejad}},\ }\href {\doibase 10.1007/JHEP07(2016)104} {\bibfield
  {journal} {\bibinfo  {journal} {JHEP}\ }\textbf {\bibinfo {volume} {07}},\
  \bibinfo {pages} {104} (\bibinfo {year} {2016})},\ \Eprint
  {http://arxiv.org/abs/1604.03327} {arXiv:1604.03327 [hep-ph]} \BibitemShut
  {NoStop}%
\bibitem [{\citenamefont {Obata}\ and\ \citenamefont
  {Soda}(2016{\natexlab{b}})}]{Obata:2016xcr}%
  \BibitemOpen
  \bibfield  {author} {\bibinfo {author} {\bibfnamefont {I.}~\bibnamefont
  {Obata}}\ and\ \bibinfo {author} {\bibfnamefont {J.}~\bibnamefont {Soda}},\
  }\href {\doibase 10.1103/PhysRevD.94.044062} {\bibfield  {journal} {\bibinfo
  {journal} {Phys. Rev. D}\ }\textbf {\bibinfo {volume} {94}},\ \bibinfo
  {pages} {044062} (\bibinfo {year} {2016}{\natexlab{b}})},\ \Eprint
  {http://arxiv.org/abs/1607.01847} {arXiv:1607.01847 [astro-ph.CO]}
  \BibitemShut {NoStop}%
\bibitem [{\citenamefont {Dimastrogiovanni}\ \emph {et~al.}(2017)\citenamefont
  {Dimastrogiovanni}, \citenamefont {Fasiello},\ and\ \citenamefont
  {Fujita}}]{Dimastrogiovanni:2016fuu}%
  \BibitemOpen
  \bibfield  {author} {\bibinfo {author} {\bibfnamefont {E.}~\bibnamefont
  {Dimastrogiovanni}}, \bibinfo {author} {\bibfnamefont {M.}~\bibnamefont
  {Fasiello}}, \ and\ \bibinfo {author} {\bibfnamefont {T.}~\bibnamefont
  {Fujita}},\ }\href {\doibase 10.1088/1475-7516/2017/01/019} {\bibfield
  {journal} {\bibinfo  {journal} {JCAP}\ }\textbf {\bibinfo {volume} {01}},\
  \bibinfo {pages} {019} (\bibinfo {year} {2017})},\ \Eprint
  {http://arxiv.org/abs/1608.04216} {arXiv:1608.04216 [astro-ph.CO]}
  \BibitemShut {NoStop}%
\bibitem [{\citenamefont {Fujita}\ \emph
  {et~al.}(2018{\natexlab{a}})\citenamefont {Fujita}, \citenamefont {Namba},\
  and\ \citenamefont {Tada}}]{Fujita:2017jwq}%
  \BibitemOpen
  \bibfield  {author} {\bibinfo {author} {\bibfnamefont {T.}~\bibnamefont
  {Fujita}}, \bibinfo {author} {\bibfnamefont {R.}~\bibnamefont {Namba}}, \
  and\ \bibinfo {author} {\bibfnamefont {Y.}~\bibnamefont {Tada}},\ }\href
  {\doibase 10.1016/j.physletb.2017.12.014} {\bibfield  {journal} {\bibinfo
  {journal} {Phys. Lett. B}\ }\textbf {\bibinfo {volume} {778}},\ \bibinfo
  {pages} {17} (\bibinfo {year} {2018}{\natexlab{a}})},\ \Eprint
  {http://arxiv.org/abs/1705.01533} {arXiv:1705.01533 [astro-ph.CO]}
  \BibitemShut {NoStop}%
\bibitem [{\citenamefont {Thorne}\ \emph {et~al.}(2018)\citenamefont {Thorne},
  \citenamefont {Fujita}, \citenamefont {Hazumi}, \citenamefont {Katayama},
  \citenamefont {Komatsu},\ and\ \citenamefont {Shiraishi}}]{Thorne:2017jft}%
  \BibitemOpen
  \bibfield  {author} {\bibinfo {author} {\bibfnamefont {B.}~\bibnamefont
  {Thorne}}, \bibinfo {author} {\bibfnamefont {T.}~\bibnamefont {Fujita}},
  \bibinfo {author} {\bibfnamefont {M.}~\bibnamefont {Hazumi}}, \bibinfo
  {author} {\bibfnamefont {N.}~\bibnamefont {Katayama}}, \bibinfo {author}
  {\bibfnamefont {E.}~\bibnamefont {Komatsu}}, \ and\ \bibinfo {author}
  {\bibfnamefont {M.}~\bibnamefont {Shiraishi}},\ }\href {\doibase
  10.1103/PhysRevD.97.043506} {\bibfield  {journal} {\bibinfo  {journal} {Phys.
  Rev. D}\ }\textbf {\bibinfo {volume} {97}},\ \bibinfo {pages} {043506}
  (\bibinfo {year} {2018})},\ \Eprint {http://arxiv.org/abs/1707.03240}
  {arXiv:1707.03240 [astro-ph.CO]} \BibitemShut {NoStop}%
\bibitem [{\citenamefont {Domcke}\ \emph {et~al.}(2019)\citenamefont {Domcke},
  \citenamefont {Mares}, \citenamefont {Muia},\ and\ \citenamefont
  {Pieroni}}]{Domcke:2018rvv}%
  \BibitemOpen
  \bibfield  {author} {\bibinfo {author} {\bibfnamefont {V.}~\bibnamefont
  {Domcke}}, \bibinfo {author} {\bibfnamefont {B.}~\bibnamefont {Mares}},
  \bibinfo {author} {\bibfnamefont {F.}~\bibnamefont {Muia}}, \ and\ \bibinfo
  {author} {\bibfnamefont {M.}~\bibnamefont {Pieroni}},\ }\href {\doibase
  10.1088/1475-7516/2019/04/034} {\bibfield  {journal} {\bibinfo  {journal}
  {JCAP}\ }\textbf {\bibinfo {volume} {04}},\ \bibinfo {pages} {034} (\bibinfo
  {year} {2019})},\ \Eprint {http://arxiv.org/abs/1807.03358} {arXiv:1807.03358
  [hep-ph]} \BibitemShut {NoStop}%
\bibitem [{\citenamefont {Maleknejad}\ and\ \citenamefont
  {Komatsu}(2019)}]{Maleknejad:2018nxz}%
  \BibitemOpen
  \bibfield  {author} {\bibinfo {author} {\bibfnamefont {A.}~\bibnamefont
  {Maleknejad}}\ and\ \bibinfo {author} {\bibfnamefont {E.}~\bibnamefont
  {Komatsu}},\ }\href {\doibase 10.1007/JHEP05(2019)174} {\bibfield  {journal}
  {\bibinfo  {journal} {JHEP}\ }\textbf {\bibinfo {volume} {05}},\ \bibinfo
  {pages} {174} (\bibinfo {year} {2019})},\ \Eprint
  {http://arxiv.org/abs/1808.09076} {arXiv:1808.09076 [hep-ph]} \BibitemShut
  {NoStop}%
\bibitem [{\citenamefont {Mirzagholi}\ \emph {et~al.}(2020)\citenamefont
  {Mirzagholi}, \citenamefont {Komatsu}, \citenamefont {Lozanov},\ and\
  \citenamefont {Watanabe}}]{Mirzagholi:2020irt}%
  \BibitemOpen
  \bibfield  {author} {\bibinfo {author} {\bibfnamefont {L.}~\bibnamefont
  {Mirzagholi}}, \bibinfo {author} {\bibfnamefont {E.}~\bibnamefont {Komatsu}},
  \bibinfo {author} {\bibfnamefont {K.~D.}\ \bibnamefont {Lozanov}}, \ and\
  \bibinfo {author} {\bibfnamefont {Y.}~\bibnamefont {Watanabe}},\ }\href@noop
  {} {\  (\bibinfo {year} {2020})},\ \Eprint {http://arxiv.org/abs/2003.05931}
  {arXiv:2003.05931 [gr-qc]} \BibitemShut {NoStop}%
\bibitem [{\citenamefont {Watanabe}\ and\ \citenamefont
  {Komatsu}(2020)}]{Watanabe:2020ctz}%
  \BibitemOpen
  \bibfield  {author} {\bibinfo {author} {\bibfnamefont {Y.}~\bibnamefont
  {Watanabe}}\ and\ \bibinfo {author} {\bibfnamefont {E.}~\bibnamefont
  {Komatsu}},\ }\href@noop {} {\  (\bibinfo {year} {2020})},\ \Eprint
  {http://arxiv.org/abs/2004.04350} {arXiv:2004.04350 [hep-th]} \BibitemShut
  {NoStop}%
\bibitem [{\citenamefont {Cook}\ and\ \citenamefont
  {Sorbo}(2013)}]{Cook:2013xea}%
  \BibitemOpen
  \bibfield  {author} {\bibinfo {author} {\bibfnamefont {J.~L.}\ \bibnamefont
  {Cook}}\ and\ \bibinfo {author} {\bibfnamefont {L.}~\bibnamefont {Sorbo}},\
  }\href {\doibase 10.1088/1475-7516/2013/11/047} {\bibfield  {journal}
  {\bibinfo  {journal} {JCAP}\ }\textbf {\bibinfo {volume} {11}},\ \bibinfo
  {pages} {047} (\bibinfo {year} {2013})},\ \Eprint
  {http://arxiv.org/abs/1307.7077} {arXiv:1307.7077 [astro-ph.CO]} \BibitemShut
  {NoStop}%
\bibitem [{\citenamefont {Namba}\ \emph {et~al.}(2016)\citenamefont {Namba},
  \citenamefont {Peloso}, \citenamefont {Shiraishi}, \citenamefont {Sorbo},\
  and\ \citenamefont {Unal}}]{Namba:2015gja}%
  \BibitemOpen
  \bibfield  {author} {\bibinfo {author} {\bibfnamefont {R.}~\bibnamefont
  {Namba}}, \bibinfo {author} {\bibfnamefont {M.}~\bibnamefont {Peloso}},
  \bibinfo {author} {\bibfnamefont {M.}~\bibnamefont {Shiraishi}}, \bibinfo
  {author} {\bibfnamefont {L.}~\bibnamefont {Sorbo}}, \ and\ \bibinfo {author}
  {\bibfnamefont {C.}~\bibnamefont {Unal}},\ }\href {\doibase
  10.1088/1475-7516/2016/01/041} {\bibfield  {journal} {\bibinfo  {journal}
  {JCAP}\ }\textbf {\bibinfo {volume} {01}},\ \bibinfo {pages} {041} (\bibinfo
  {year} {2016})},\ \Eprint {http://arxiv.org/abs/1509.07521} {arXiv:1509.07521
  [astro-ph.CO]} \BibitemShut {NoStop}%
\bibitem [{\citenamefont {Agrawal}\ \emph
  {et~al.}(2018{\natexlab{a}})\citenamefont {Agrawal}, \citenamefont {Fujita},\
  and\ \citenamefont {Komatsu}}]{Agrawal:2017awz}%
  \BibitemOpen
  \bibfield  {author} {\bibinfo {author} {\bibfnamefont {A.}~\bibnamefont
  {Agrawal}}, \bibinfo {author} {\bibfnamefont {T.}~\bibnamefont {Fujita}}, \
  and\ \bibinfo {author} {\bibfnamefont {E.}~\bibnamefont {Komatsu}},\ }\href
  {\doibase 10.1103/PhysRevD.97.103526} {\bibfield  {journal} {\bibinfo
  {journal} {Phys. Rev. D}\ }\textbf {\bibinfo {volume} {97}},\ \bibinfo
  {pages} {103526} (\bibinfo {year} {2018}{\natexlab{a}})},\ \Eprint
  {http://arxiv.org/abs/1707.03023} {arXiv:1707.03023 [astro-ph.CO]}
  \BibitemShut {NoStop}%
\bibitem [{\citenamefont {Agrawal}\ \emph
  {et~al.}(2018{\natexlab{b}})\citenamefont {Agrawal}, \citenamefont {Fujita},\
  and\ \citenamefont {Komatsu}}]{Agrawal:2018mrg}%
  \BibitemOpen
  \bibfield  {author} {\bibinfo {author} {\bibfnamefont {A.}~\bibnamefont
  {Agrawal}}, \bibinfo {author} {\bibfnamefont {T.}~\bibnamefont {Fujita}}, \
  and\ \bibinfo {author} {\bibfnamefont {E.}~\bibnamefont {Komatsu}},\ }\href
  {\doibase 10.1088/1475-7516/2018/06/027} {\bibfield  {journal} {\bibinfo
  {journal} {JCAP}\ }\textbf {\bibinfo {volume} {06}},\ \bibinfo {pages} {027}
  (\bibinfo {year} {2018}{\natexlab{b}})},\ \Eprint
  {http://arxiv.org/abs/1802.09284} {arXiv:1802.09284 [astro-ph.CO]}
  \BibitemShut {NoStop}%
\bibitem [{\citenamefont {Dimastrogiovanni}\ \emph {et~al.}(2018)\citenamefont
  {Dimastrogiovanni}, \citenamefont {Fasiello}, \citenamefont {Hardwick},
  \citenamefont {Assadullahi}, \citenamefont {Koyama},\ and\ \citenamefont
  {Wands}}]{Dimastrogiovanni:2018xnn}%
  \BibitemOpen
  \bibfield  {author} {\bibinfo {author} {\bibfnamefont {E.}~\bibnamefont
  {Dimastrogiovanni}}, \bibinfo {author} {\bibfnamefont {M.}~\bibnamefont
  {Fasiello}}, \bibinfo {author} {\bibfnamefont {R.~J.}\ \bibnamefont
  {Hardwick}}, \bibinfo {author} {\bibfnamefont {H.}~\bibnamefont
  {Assadullahi}}, \bibinfo {author} {\bibfnamefont {K.}~\bibnamefont {Koyama}},
  \ and\ \bibinfo {author} {\bibfnamefont {D.}~\bibnamefont {Wands}},\ }\href
  {\doibase 10.1088/1475-7516/2018/11/029} {\bibfield  {journal} {\bibinfo
  {journal} {JCAP}\ }\textbf {\bibinfo {volume} {11}},\ \bibinfo {pages} {029}
  (\bibinfo {year} {2018})},\ \Eprint {http://arxiv.org/abs/1806.05474}
  {arXiv:1806.05474 [astro-ph.CO]} \BibitemShut {NoStop}%
\bibitem [{\citenamefont {Fujita}\ \emph {et~al.}(2019)\citenamefont {Fujita},
  \citenamefont {Namba},\ and\ \citenamefont {Obata}}]{Fujita:2018vmv}%
  \BibitemOpen
  \bibfield  {author} {\bibinfo {author} {\bibfnamefont {T.}~\bibnamefont
  {Fujita}}, \bibinfo {author} {\bibfnamefont {R.}~\bibnamefont {Namba}}, \
  and\ \bibinfo {author} {\bibfnamefont {I.}~\bibnamefont {Obata}},\ }\href
  {\doibase 10.1088/1475-7516/2019/04/044} {\bibfield  {journal} {\bibinfo
  {journal} {JCAP}\ }\textbf {\bibinfo {volume} {04}},\ \bibinfo {pages} {044}
  (\bibinfo {year} {2019})},\ \Eprint {http://arxiv.org/abs/1811.12371}
  {arXiv:1811.12371 [astro-ph.CO]} \BibitemShut {NoStop}%
\bibitem [{\citenamefont {Kamionkowski}\ and\ \citenamefont
  {Souradeep}(2011)}]{Kamionkowski:2010rb}%
  \BibitemOpen
  \bibfield  {author} {\bibinfo {author} {\bibfnamefont {M.}~\bibnamefont
  {Kamionkowski}}\ and\ \bibinfo {author} {\bibfnamefont {T.}~\bibnamefont
  {Souradeep}},\ }\href {\doibase 10.1103/PhysRevD.83.027301} {\bibfield
  {journal} {\bibinfo  {journal} {Phys. Rev. D}\ }\textbf {\bibinfo {volume}
  {83}},\ \bibinfo {pages} {027301} (\bibinfo {year} {2011})},\ \Eprint
  {http://arxiv.org/abs/1010.4304} {arXiv:1010.4304 [astro-ph.CO]} \BibitemShut
  {NoStop}%
\bibitem [{\citenamefont {Shiraishi}\ \emph
  {et~al.}(2013{\natexlab{a}})\citenamefont {Shiraishi}, \citenamefont
  {Ricciardone},\ and\ \citenamefont {Saga}}]{Shiraishi:2013kxa}%
  \BibitemOpen
  \bibfield  {author} {\bibinfo {author} {\bibfnamefont {M.}~\bibnamefont
  {Shiraishi}}, \bibinfo {author} {\bibfnamefont {A.}~\bibnamefont
  {Ricciardone}}, \ and\ \bibinfo {author} {\bibfnamefont {S.}~\bibnamefont
  {Saga}},\ }\href {\doibase 10.1088/1475-7516/2013/11/051} {\bibfield
  {journal} {\bibinfo  {journal} {JCAP}\ }\textbf {\bibinfo {volume} {11}},\
  \bibinfo {pages} {051} (\bibinfo {year} {2013}{\natexlab{a}})},\ \Eprint
  {http://arxiv.org/abs/1308.6769} {arXiv:1308.6769 [astro-ph.CO]} \BibitemShut
  {NoStop}%
\bibitem [{\citenamefont {Shiraishi}\ \emph {et~al.}(2016)\citenamefont
  {Shiraishi}, \citenamefont {Hikage}, \citenamefont {Namba}, \citenamefont
  {Namikawa},\ and\ \citenamefont {Hazumi}}]{Shiraishi:2016yun}%
  \BibitemOpen
  \bibfield  {author} {\bibinfo {author} {\bibfnamefont {M.}~\bibnamefont
  {Shiraishi}}, \bibinfo {author} {\bibfnamefont {C.}~\bibnamefont {Hikage}},
  \bibinfo {author} {\bibfnamefont {R.}~\bibnamefont {Namba}}, \bibinfo
  {author} {\bibfnamefont {T.}~\bibnamefont {Namikawa}}, \ and\ \bibinfo
  {author} {\bibfnamefont {M.}~\bibnamefont {Hazumi}},\ }\href {\doibase
  10.1103/PhysRevD.94.043506} {\bibfield  {journal} {\bibinfo  {journal} {Phys.
  Rev. D}\ }\textbf {\bibinfo {volume} {94}},\ \bibinfo {pages} {043506}
  (\bibinfo {year} {2016})},\ \Eprint {http://arxiv.org/abs/1606.06082}
  {arXiv:1606.06082 [astro-ph.CO]} \BibitemShut {NoStop}%
\bibitem [{\citenamefont {Shiraishi}(2019)}]{Shiraishi:2019yux}%
  \BibitemOpen
  \bibfield  {author} {\bibinfo {author} {\bibfnamefont {M.}~\bibnamefont
  {Shiraishi}},\ }\href {\doibase 10.3389/fspas.2019.00049} {\bibfield
  {journal} {\bibinfo  {journal} {Front. Astron. Space Sci.}\ }\textbf
  {\bibinfo {volume} {6}},\ \bibinfo {pages} {49} (\bibinfo {year} {2019})},\
  \Eprint {http://arxiv.org/abs/1905.12485} {arXiv:1905.12485 [astro-ph.CO]}
  \BibitemShut {NoStop}%
\bibitem [{\citenamefont {Watanabe}\ \emph {et~al.}(2009)\citenamefont
  {Watanabe}, \citenamefont {Kanno},\ and\ \citenamefont
  {Soda}}]{Watanabe:2009ct}%
  \BibitemOpen
  \bibfield  {author} {\bibinfo {author} {\bibfnamefont {M.-a.}\ \bibnamefont
  {Watanabe}}, \bibinfo {author} {\bibfnamefont {S.}~\bibnamefont {Kanno}}, \
  and\ \bibinfo {author} {\bibfnamefont {J.}~\bibnamefont {Soda}},\ }\href
  {\doibase 10.1103/PhysRevLett.102.191302} {\bibfield  {journal} {\bibinfo
  {journal} {Phys. Rev. Lett.}\ }\textbf {\bibinfo {volume} {102}},\ \bibinfo
  {pages} {191302} (\bibinfo {year} {2009})},\ \Eprint
  {http://arxiv.org/abs/0902.2833} {arXiv:0902.2833 [hep-th]} \BibitemShut
  {NoStop}%
\bibitem [{\citenamefont {Kanno}\ \emph {et~al.}(2010)\citenamefont {Kanno},
  \citenamefont {Soda},\ and\ \citenamefont {Watanabe}}]{Kanno:2010nr}%
  \BibitemOpen
  \bibfield  {author} {\bibinfo {author} {\bibfnamefont {S.}~\bibnamefont
  {Kanno}}, \bibinfo {author} {\bibfnamefont {J.}~\bibnamefont {Soda}}, \ and\
  \bibinfo {author} {\bibfnamefont {M.-a.}\ \bibnamefont {Watanabe}},\ }\href
  {\doibase 10.1088/1475-7516/2010/12/024} {\bibfield  {journal} {\bibinfo
  {journal} {JCAP}\ }\textbf {\bibinfo {volume} {12}},\ \bibinfo {pages} {024}
  (\bibinfo {year} {2010})},\ \Eprint {http://arxiv.org/abs/1010.5307}
  {arXiv:1010.5307 [hep-th]} \BibitemShut {NoStop}%
\bibitem [{\citenamefont {Do}\ \emph {et~al.}(2011)\citenamefont {Do},
  \citenamefont {Kao},\ and\ \citenamefont {Lin}}]{Do:2011zza}%
  \BibitemOpen
  \bibfield  {author} {\bibinfo {author} {\bibfnamefont {T.~Q.}\ \bibnamefont
  {Do}}, \bibinfo {author} {\bibfnamefont {W.}~\bibnamefont {Kao}}, \ and\
  \bibinfo {author} {\bibfnamefont {I.-C.}\ \bibnamefont {Lin}},\ }\href
  {\doibase 10.1103/PhysRevD.83.123002} {\bibfield  {journal} {\bibinfo
  {journal} {Phys. Rev. D}\ }\textbf {\bibinfo {volume} {83}},\ \bibinfo
  {pages} {123002} (\bibinfo {year} {2011})}\BibitemShut {NoStop}%
\bibitem [{\citenamefont {Ohashi}\ \emph
  {et~al.}(2013{\natexlab{a}})\citenamefont {Ohashi}, \citenamefont {Soda},\
  and\ \citenamefont {Tsujikawa}}]{Ohashi:2013pca}%
  \BibitemOpen
  \bibfield  {author} {\bibinfo {author} {\bibfnamefont {J.}~\bibnamefont
  {Ohashi}}, \bibinfo {author} {\bibfnamefont {J.}~\bibnamefont {Soda}}, \ and\
  \bibinfo {author} {\bibfnamefont {S.}~\bibnamefont {Tsujikawa}},\ }\href
  {\doibase 10.1103/PhysRevD.88.103517} {\bibfield  {journal} {\bibinfo
  {journal} {Phys. Rev. D}\ }\textbf {\bibinfo {volume} {88}},\ \bibinfo
  {pages} {103517} (\bibinfo {year} {2013}{\natexlab{a}})},\ \Eprint
  {http://arxiv.org/abs/1310.3053} {arXiv:1310.3053 [hep-th]} \BibitemShut
  {NoStop}%
\bibitem [{\citenamefont {Naruko}\ \emph {et~al.}(2015)\citenamefont {Naruko},
  \citenamefont {Komatsu},\ and\ \citenamefont {Yamaguchi}}]{Naruko:2014bxa}%
  \BibitemOpen
  \bibfield  {author} {\bibinfo {author} {\bibfnamefont {A.}~\bibnamefont
  {Naruko}}, \bibinfo {author} {\bibfnamefont {E.}~\bibnamefont {Komatsu}}, \
  and\ \bibinfo {author} {\bibfnamefont {M.}~\bibnamefont {Yamaguchi}},\ }\href
  {\doibase 10.1088/1475-7516/2015/04/045} {\bibfield  {journal} {\bibinfo
  {journal} {JCAP}\ }\textbf {\bibinfo {volume} {04}},\ \bibinfo {pages} {045}
  (\bibinfo {year} {2015})},\ \Eprint {http://arxiv.org/abs/1411.5489}
  {arXiv:1411.5489 [astro-ph.CO]} \BibitemShut {NoStop}%
\bibitem [{\citenamefont {Ito}\ and\ \citenamefont {Soda}(2018)}]{Ito:2017bnn}%
  \BibitemOpen
  \bibfield  {author} {\bibinfo {author} {\bibfnamefont {A.}~\bibnamefont
  {Ito}}\ and\ \bibinfo {author} {\bibfnamefont {J.}~\bibnamefont {Soda}},\
  }\href {\doibase 10.1140/epjc/s10052-018-5534-5} {\bibfield  {journal}
  {\bibinfo  {journal} {Eur. Phys. J. C}\ }\textbf {\bibinfo {volume} {78}},\
  \bibinfo {pages} {55} (\bibinfo {year} {2018})},\ \Eprint
  {http://arxiv.org/abs/1710.09701} {arXiv:1710.09701 [hep-th]} \BibitemShut
  {NoStop}%
\bibitem [{\citenamefont {Watanabe}\ \emph {et~al.}(2010)\citenamefont
  {Watanabe}, \citenamefont {Kanno},\ and\ \citenamefont
  {Soda}}]{Watanabe:2010fh}%
  \BibitemOpen
  \bibfield  {author} {\bibinfo {author} {\bibfnamefont {M.-a.}\ \bibnamefont
  {Watanabe}}, \bibinfo {author} {\bibfnamefont {S.}~\bibnamefont {Kanno}}, \
  and\ \bibinfo {author} {\bibfnamefont {J.}~\bibnamefont {Soda}},\ }\href
  {\doibase 10.1143/PTP.123.1041} {\bibfield  {journal} {\bibinfo  {journal}
  {Prog. Theor. Phys.}\ }\textbf {\bibinfo {volume} {123}},\ \bibinfo {pages}
  {1041} (\bibinfo {year} {2010})},\ \Eprint {http://arxiv.org/abs/1003.0056}
  {arXiv:1003.0056 [astro-ph.CO]} \BibitemShut {NoStop}%
\bibitem [{\citenamefont {Watanabe}\ \emph {et~al.}(2011)\citenamefont
  {Watanabe}, \citenamefont {Kanno},\ and\ \citenamefont
  {Soda}}]{Watanabe:2010bu}%
  \BibitemOpen
  \bibfield  {author} {\bibinfo {author} {\bibfnamefont {M.-a.}\ \bibnamefont
  {Watanabe}}, \bibinfo {author} {\bibfnamefont {S.}~\bibnamefont {Kanno}}, \
  and\ \bibinfo {author} {\bibfnamefont {J.}~\bibnamefont {Soda}},\ }\href
  {\doibase 10.1111/j.1745-3933.2011.01010.x} {\bibfield  {journal} {\bibinfo
  {journal} {Mon. Not. Roy. Astron. Soc.}\ }\textbf {\bibinfo {volume} {412}},\
  \bibinfo {pages} {L83} (\bibinfo {year} {2011})},\ \Eprint
  {http://arxiv.org/abs/1011.3604} {arXiv:1011.3604 [astro-ph.CO]} \BibitemShut
  {NoStop}%
\bibitem [{\citenamefont {Ohashi}\ \emph
  {et~al.}(2013{\natexlab{b}})\citenamefont {Ohashi}, \citenamefont {Soda},\
  and\ \citenamefont {Tsujikawa}}]{Ohashi:2013qba}%
  \BibitemOpen
  \bibfield  {author} {\bibinfo {author} {\bibfnamefont {J.}~\bibnamefont
  {Ohashi}}, \bibinfo {author} {\bibfnamefont {J.}~\bibnamefont {Soda}}, \ and\
  \bibinfo {author} {\bibfnamefont {S.}~\bibnamefont {Tsujikawa}},\ }\href
  {\doibase 10.1088/1475-7516/2013/12/009} {\bibfield  {journal} {\bibinfo
  {journal} {JCAP}\ }\textbf {\bibinfo {volume} {12}},\ \bibinfo {pages} {009}
  (\bibinfo {year} {2013}{\natexlab{b}})},\ \Eprint
  {http://arxiv.org/abs/1308.4488} {arXiv:1308.4488 [astro-ph.CO]} \BibitemShut
  {NoStop}%
\bibitem [{\citenamefont {Groeneboom}\ and\ \citenamefont
  {Eriksen}(2009)}]{Groeneboom:2008fz}%
  \BibitemOpen
  \bibfield  {author} {\bibinfo {author} {\bibfnamefont {N.~E.}\ \bibnamefont
  {Groeneboom}}\ and\ \bibinfo {author} {\bibfnamefont {H.~K.}\ \bibnamefont
  {Eriksen}},\ }\href {\doibase 10.1088/0004-637X/690/2/1807} {\bibfield
  {journal} {\bibinfo  {journal} {Astrophys. J.}\ }\textbf {\bibinfo {volume}
  {690}},\ \bibinfo {pages} {1807} (\bibinfo {year} {2009})},\ \Eprint
  {http://arxiv.org/abs/0807.2242} {arXiv:0807.2242 [astro-ph]} \BibitemShut
  {NoStop}%
\bibitem [{\citenamefont {Groeneboom}\ \emph {et~al.}(2010)\citenamefont
  {Groeneboom}, \citenamefont {Ackerman}, \citenamefont {Wehus},\ and\
  \citenamefont {Eriksen}}]{Groeneboom:2009cb}%
  \BibitemOpen
  \bibfield  {author} {\bibinfo {author} {\bibfnamefont {N.~E.}\ \bibnamefont
  {Groeneboom}}, \bibinfo {author} {\bibfnamefont {L.}~\bibnamefont
  {Ackerman}}, \bibinfo {author} {\bibfnamefont {I.~K.}\ \bibnamefont {Wehus}},
  \ and\ \bibinfo {author} {\bibfnamefont {H.~K.}\ \bibnamefont {Eriksen}},\
  }\href {\doibase 10.1088/0004-637X/722/1/452} {\bibfield  {journal} {\bibinfo
   {journal} {Astrophys. J.}\ }\textbf {\bibinfo {volume} {722}},\ \bibinfo
  {pages} {452} (\bibinfo {year} {2010})},\ \Eprint
  {http://arxiv.org/abs/0911.0150} {arXiv:0911.0150 [astro-ph.CO]} \BibitemShut
  {NoStop}%
\bibitem [{\citenamefont {{Hanson}}\ \emph {et~al.}(2010)\citenamefont
  {{Hanson}}, \citenamefont {{Lewis}},\ and\ \citenamefont
  {{Challinor}}}]{2010PhRvD..81j3003H}%
  \BibitemOpen
  \bibfield  {author} {\bibinfo {author} {\bibfnamefont {D.}~\bibnamefont
  {{Hanson}}}, \bibinfo {author} {\bibfnamefont {A.}~\bibnamefont {{Lewis}}}, \
  and\ \bibinfo {author} {\bibfnamefont {A.}~\bibnamefont {{Challinor}}},\
  }\href {\doibase 10.1103/PhysRevD.81.103003} {\bibfield  {journal} {\bibinfo
  {journal} {Phys. Rev. D}\ }\textbf {\bibinfo {volume} {81}},\ \bibinfo {eid}
  {103003} (\bibinfo {year} {2010})},\ \Eprint {http://arxiv.org/abs/1003.0198}
  {arXiv:1003.0198 [astro-ph.CO]} \BibitemShut {NoStop}%
\bibitem [{\citenamefont {Bennett}\ \emph {et~al.}(2013)\citenamefont {Bennett}
  \emph {et~al.}}]{Bennett:2012zja}%
  \BibitemOpen
  \bibfield  {author} {\bibinfo {author} {\bibfnamefont {C.}~\bibnamefont
  {Bennett}} \emph {et~al.} (\bibinfo {collaboration} {WMAP}),\ }\href
  {\doibase 10.1088/0067-0049/208/2/20} {\bibfield  {journal} {\bibinfo
  {journal} {Astrophys. J. Suppl.}\ }\textbf {\bibinfo {volume} {208}},\
  \bibinfo {pages} {20} (\bibinfo {year} {2013})},\ \Eprint
  {http://arxiv.org/abs/1212.5225} {arXiv:1212.5225 [astro-ph.CO]} \BibitemShut
  {NoStop}%
\bibitem [{\citenamefont {Kim}\ and\ \citenamefont
  {Komatsu}(2013)}]{Kim:2013gka}%
  \BibitemOpen
  \bibfield  {author} {\bibinfo {author} {\bibfnamefont {J.}~\bibnamefont
  {Kim}}\ and\ \bibinfo {author} {\bibfnamefont {E.}~\bibnamefont {Komatsu}},\
  }\href {\doibase 10.1103/PhysRevD.88.101301} {\bibfield  {journal} {\bibinfo
  {journal} {Phys. Rev. D}\ }\textbf {\bibinfo {volume} {88}},\ \bibinfo
  {pages} {101301} (\bibinfo {year} {2013})},\ \Eprint
  {http://arxiv.org/abs/1310.1605} {arXiv:1310.1605 [astro-ph.CO]} \BibitemShut
  {NoStop}%
\bibitem [{\citenamefont {Akrami}\ \emph {et~al.}(2018)\citenamefont {Akrami}
  \emph {et~al.}}]{Akrami:2018odb}%
  \BibitemOpen
  \bibfield  {author} {\bibinfo {author} {\bibfnamefont {Y.}~\bibnamefont
  {Akrami}} \emph {et~al.} (\bibinfo {collaboration} {Planck}),\ }\href@noop {}
  {\  (\bibinfo {year} {2018})},\ \Eprint {http://arxiv.org/abs/1807.06211}
  {arXiv:1807.06211 [astro-ph.CO]} \BibitemShut {NoStop}%
\bibitem [{\citenamefont {Bartolo}\ \emph {et~al.}(2013)\citenamefont
  {Bartolo}, \citenamefont {Matarrese}, \citenamefont {Peloso},\ and\
  \citenamefont {Ricciardone}}]{Bartolo:2012sd}%
  \BibitemOpen
  \bibfield  {author} {\bibinfo {author} {\bibfnamefont {N.}~\bibnamefont
  {Bartolo}}, \bibinfo {author} {\bibfnamefont {S.}~\bibnamefont {Matarrese}},
  \bibinfo {author} {\bibfnamefont {M.}~\bibnamefont {Peloso}}, \ and\ \bibinfo
  {author} {\bibfnamefont {A.}~\bibnamefont {Ricciardone}},\ }\href {\doibase
  10.1103/PhysRevD.87.023504} {\bibfield  {journal} {\bibinfo  {journal} {Phys.
  Rev. D}\ }\textbf {\bibinfo {volume} {87}},\ \bibinfo {pages} {023504}
  (\bibinfo {year} {2013})},\ \Eprint {http://arxiv.org/abs/1210.3257}
  {arXiv:1210.3257 [astro-ph.CO]} \BibitemShut {NoStop}%
\bibitem [{\citenamefont {Fujita}\ and\ \citenamefont
  {Obata}(2018)}]{Fujita:2017lfu}%
  \BibitemOpen
  \bibfield  {author} {\bibinfo {author} {\bibfnamefont {T.}~\bibnamefont
  {Fujita}}\ and\ \bibinfo {author} {\bibfnamefont {I.}~\bibnamefont {Obata}},\
  }\href {\doibase 10.1088/1475-7516/2018/01/049} {\bibfield  {journal}
  {\bibinfo  {journal} {JCAP}\ }\textbf {\bibinfo {volume} {01}},\ \bibinfo
  {pages} {049} (\bibinfo {year} {2018})},\ \Eprint
  {http://arxiv.org/abs/1711.11539} {arXiv:1711.11539 [astro-ph.CO]}
  \BibitemShut {NoStop}%
\bibitem [{\citenamefont {Fujita}\ \emph
  {et~al.}(2018{\natexlab{b}})\citenamefont {Fujita}, \citenamefont {Obata},
  \citenamefont {Tanaka},\ and\ \citenamefont {Yokoyama}}]{Fujita:2018zbr}%
  \BibitemOpen
  \bibfield  {author} {\bibinfo {author} {\bibfnamefont {T.}~\bibnamefont
  {Fujita}}, \bibinfo {author} {\bibfnamefont {I.}~\bibnamefont {Obata}},
  \bibinfo {author} {\bibfnamefont {T.}~\bibnamefont {Tanaka}}, \ and\ \bibinfo
  {author} {\bibfnamefont {S.}~\bibnamefont {Yokoyama}},\ }\href {\doibase
  10.1088/1475-7516/2018/07/023} {\bibfield  {journal} {\bibinfo  {journal}
  {JCAP}\ }\textbf {\bibinfo {volume} {07}},\ \bibinfo {pages} {023} (\bibinfo
  {year} {2018}{\natexlab{b}})},\ \Eprint {http://arxiv.org/abs/1801.02778}
  {arXiv:1801.02778 [astro-ph.CO]} \BibitemShut {NoStop}%
\bibitem [{\citenamefont {Hiramatsu}\ \emph {et~al.}(2018)\citenamefont
  {Hiramatsu}, \citenamefont {Yokoyama}, \citenamefont {Fujita},\ and\
  \citenamefont {Obata}}]{Hiramatsu:2018vfw}%
  \BibitemOpen
  \bibfield  {author} {\bibinfo {author} {\bibfnamefont {T.}~\bibnamefont
  {Hiramatsu}}, \bibinfo {author} {\bibfnamefont {S.}~\bibnamefont {Yokoyama}},
  \bibinfo {author} {\bibfnamefont {T.}~\bibnamefont {Fujita}}, \ and\ \bibinfo
  {author} {\bibfnamefont {I.}~\bibnamefont {Obata}},\ }\href {\doibase
  10.1103/PhysRevD.98.083522} {\bibfield  {journal} {\bibinfo  {journal} {Phys.
  Rev. D}\ }\textbf {\bibinfo {volume} {98}},\ \bibinfo {pages} {083522}
  (\bibinfo {year} {2018})},\ \Eprint {http://arxiv.org/abs/1808.08044}
  {arXiv:1808.08044 [astro-ph.CO]} \BibitemShut {NoStop}%
\bibitem [{\citenamefont {Nilles}\ \emph {et~al.}(2001)\citenamefont {Nilles},
  \citenamefont {Peloso},\ and\ \citenamefont {Sorbo}}]{Nilles:2001fg}%
  \BibitemOpen
  \bibfield  {author} {\bibinfo {author} {\bibfnamefont {H.~P.}\ \bibnamefont
  {Nilles}}, \bibinfo {author} {\bibfnamefont {M.}~\bibnamefont {Peloso}}, \
  and\ \bibinfo {author} {\bibfnamefont {L.}~\bibnamefont {Sorbo}},\ }\href
  {\doibase 10.1088/1126-6708/2001/04/004} {\bibfield  {journal} {\bibinfo
  {journal} {JHEP}\ }\textbf {\bibinfo {volume} {04}},\ \bibinfo {pages} {004}
  (\bibinfo {year} {2001})},\ \Eprint {http://arxiv.org/abs/hep-th/0103202}
  {arXiv:hep-th/0103202} \BibitemShut {NoStop}%
\bibitem [{\citenamefont {Gumrukcuoglu}\ \emph {et~al.}(2010)\citenamefont
  {Gumrukcuoglu}, \citenamefont {Himmetoglu},\ and\ \citenamefont
  {Peloso}}]{Gumrukcuoglu:2010yc}%
  \BibitemOpen
  \bibfield  {author} {\bibinfo {author} {\bibfnamefont {A.}~\bibnamefont
  {Gumrukcuoglu}}, \bibinfo {author} {\bibfnamefont {B.}~\bibnamefont
  {Himmetoglu}}, \ and\ \bibinfo {author} {\bibfnamefont {M.}~\bibnamefont
  {Peloso}},\ }\href {\doibase 10.1103/PhysRevD.81.063528} {\bibfield
  {journal} {\bibinfo  {journal} {Phys. Rev. D}\ }\textbf {\bibinfo {volume}
  {81}},\ \bibinfo {pages} {063528} (\bibinfo {year} {2010})},\ \Eprint
  {http://arxiv.org/abs/1001.4088} {arXiv:1001.4088 [astro-ph.CO]} \BibitemShut
  {NoStop}%
\bibitem [{\citenamefont {Weinberg}(2005)}]{Weinberg:2005vy}%
  \BibitemOpen
  \bibfield  {author} {\bibinfo {author} {\bibfnamefont {S.}~\bibnamefont
  {Weinberg}},\ }\href {\doibase 10.1103/PhysRevD.72.043514} {\bibfield
  {journal} {\bibinfo  {journal} {Phys. Rev. D}\ }\textbf {\bibinfo {volume}
  {72}},\ \bibinfo {pages} {043514} (\bibinfo {year} {2005})},\ \Eprint
  {http://arxiv.org/abs/hep-th/0506236} {arXiv:hep-th/0506236} \BibitemShut
  {NoStop}%
\bibitem [{\citenamefont {Yokoyama}\ and\ \citenamefont
  {Soda}(2008)}]{Yokoyama:2008xw}%
  \BibitemOpen
  \bibfield  {author} {\bibinfo {author} {\bibfnamefont {S.}~\bibnamefont
  {Yokoyama}}\ and\ \bibinfo {author} {\bibfnamefont {J.}~\bibnamefont
  {Soda}},\ }\href {\doibase 10.1088/1475-7516/2008/08/005} {\bibfield
  {journal} {\bibinfo  {journal} {JCAP}\ }\textbf {\bibinfo {volume} {08}},\
  \bibinfo {pages} {005} (\bibinfo {year} {2008})},\ \Eprint
  {http://arxiv.org/abs/0805.4265} {arXiv:0805.4265 [astro-ph]} \BibitemShut
  {NoStop}%
\bibitem [{\citenamefont {Karciauskas}\ \emph {et~al.}(2009)\citenamefont
  {Karciauskas}, \citenamefont {Dimopoulos},\ and\ \citenamefont
  {Lyth}}]{Karciauskas:2008bc}%
  \BibitemOpen
  \bibfield  {author} {\bibinfo {author} {\bibfnamefont {M.}~\bibnamefont
  {Karciauskas}}, \bibinfo {author} {\bibfnamefont {K.}~\bibnamefont
  {Dimopoulos}}, \ and\ \bibinfo {author} {\bibfnamefont {D.~H.}\ \bibnamefont
  {Lyth}},\ }\href {\doibase 10.1103/PhysRevD.80.023509} {\bibfield  {journal}
  {\bibinfo  {journal} {Phys. Rev. D}\ }\textbf {\bibinfo {volume} {80}},\
  \bibinfo {pages} {023509} (\bibinfo {year} {2009})},\ \bibinfo {note}
  {[Erratum: Phys.Rev.D 85, 069905 (2012)]},\ \Eprint
  {http://arxiv.org/abs/0812.0264} {arXiv:0812.0264 [astro-ph]} \BibitemShut
  {NoStop}%
\bibitem [{\citenamefont {Bartolo}\ \emph {et~al.}(2009)\citenamefont
  {Bartolo}, \citenamefont {Dimastrogiovanni}, \citenamefont {Matarrese},\ and\
  \citenamefont {Riotto}}]{Bartolo:2009pa}%
  \BibitemOpen
  \bibfield  {author} {\bibinfo {author} {\bibfnamefont {N.}~\bibnamefont
  {Bartolo}}, \bibinfo {author} {\bibfnamefont {E.}~\bibnamefont
  {Dimastrogiovanni}}, \bibinfo {author} {\bibfnamefont {S.}~\bibnamefont
  {Matarrese}}, \ and\ \bibinfo {author} {\bibfnamefont {A.}~\bibnamefont
  {Riotto}},\ }\href {\doibase 10.1088/1475-7516/2009/10/015} {\bibfield
  {journal} {\bibinfo  {journal} {JCAP}\ }\textbf {\bibinfo {volume} {10}},\
  \bibinfo {pages} {015} (\bibinfo {year} {2009})},\ \Eprint
  {http://arxiv.org/abs/0906.4944} {arXiv:0906.4944 [astro-ph.CO]} \BibitemShut
  {NoStop}%
\bibitem [{\citenamefont {Bartolo}\ \emph {et~al.}(2012)\citenamefont
  {Bartolo}, \citenamefont {Dimastrogiovanni}, \citenamefont {Liguori},
  \citenamefont {Matarrese},\ and\ \citenamefont {Riotto}}]{Bartolo:2011ee}%
  \BibitemOpen
  \bibfield  {author} {\bibinfo {author} {\bibfnamefont {N.}~\bibnamefont
  {Bartolo}}, \bibinfo {author} {\bibfnamefont {E.}~\bibnamefont
  {Dimastrogiovanni}}, \bibinfo {author} {\bibfnamefont {M.}~\bibnamefont
  {Liguori}}, \bibinfo {author} {\bibfnamefont {S.}~\bibnamefont {Matarrese}},
  \ and\ \bibinfo {author} {\bibfnamefont {A.}~\bibnamefont {Riotto}},\ }\href
  {\doibase 10.1088/1475-7516/2012/01/029} {\bibfield  {journal} {\bibinfo
  {journal} {JCAP}\ }\textbf {\bibinfo {volume} {01}},\ \bibinfo {pages} {029}
  (\bibinfo {year} {2012})},\ \Eprint {http://arxiv.org/abs/1107.4304}
  {arXiv:1107.4304 [astro-ph.CO]} \BibitemShut {NoStop}%
\bibitem [{\citenamefont {Shiraishi}\ \emph
  {et~al.}(2013{\natexlab{b}})\citenamefont {Shiraishi}, \citenamefont
  {Komatsu}, \citenamefont {Peloso},\ and\ \citenamefont
  {Barnaby}}]{Shiraishi:2013vja}%
  \BibitemOpen
  \bibfield  {author} {\bibinfo {author} {\bibfnamefont {M.}~\bibnamefont
  {Shiraishi}}, \bibinfo {author} {\bibfnamefont {E.}~\bibnamefont {Komatsu}},
  \bibinfo {author} {\bibfnamefont {M.}~\bibnamefont {Peloso}}, \ and\ \bibinfo
  {author} {\bibfnamefont {N.}~\bibnamefont {Barnaby}},\ }\href {\doibase
  10.1088/1475-7516/2013/05/002} {\bibfield  {journal} {\bibinfo  {journal}
  {JCAP}\ }\textbf {\bibinfo {volume} {05}},\ \bibinfo {pages} {002} (\bibinfo
  {year} {2013}{\natexlab{b}})},\ \Eprint {http://arxiv.org/abs/1302.3056}
  {arXiv:1302.3056 [astro-ph.CO]} \BibitemShut {NoStop}%
\bibitem [{\citenamefont {Franciolini}\ \emph {et~al.}(2018)\citenamefont
  {Franciolini}, \citenamefont {Kehagias}, \citenamefont {Riotto},\ and\
  \citenamefont {Shiraishi}}]{Franciolini:2018eno}%
  \BibitemOpen
  \bibfield  {author} {\bibinfo {author} {\bibfnamefont {G.}~\bibnamefont
  {Franciolini}}, \bibinfo {author} {\bibfnamefont {A.}~\bibnamefont
  {Kehagias}}, \bibinfo {author} {\bibfnamefont {A.}~\bibnamefont {Riotto}}, \
  and\ \bibinfo {author} {\bibfnamefont {M.}~\bibnamefont {Shiraishi}},\ }\href
  {\doibase 10.1103/PhysRevD.98.043533} {\bibfield  {journal} {\bibinfo
  {journal} {Phys. Rev. D}\ }\textbf {\bibinfo {volume} {98}},\ \bibinfo
  {pages} {043533} (\bibinfo {year} {2018})},\ \Eprint
  {http://arxiv.org/abs/1803.03814} {arXiv:1803.03814 [astro-ph.CO]}
  \BibitemShut {NoStop}%
\bibitem [{\citenamefont {Matsumura}\ \emph {et~al.}(2014)\citenamefont
  {Matsumura} \emph {et~al.}}]{Matsumura:2013aja}%
  \BibitemOpen
  \bibfield  {author} {\bibinfo {author} {\bibfnamefont {T.}~\bibnamefont
  {Matsumura}} \emph {et~al.},\ }\href {\doibase 10.1007/s10909-013-0996-1}
  {\bibfield  {journal} {\bibinfo  {journal} {J. Low Temp. Phys.}\ }\textbf
  {\bibinfo {volume} {176}},\ \bibinfo {pages} {733} (\bibinfo {year}
  {2014})},\ \Eprint {http://arxiv.org/abs/1311.2847} {arXiv:1311.2847
  [astro-ph.IM]} \BibitemShut {NoStop}%
\bibitem [{\citenamefont {Abazajian}\ \emph {et~al.}(2016)\citenamefont
  {Abazajian} \emph {et~al.}}]{Abazajian:2016yjj}%
  \BibitemOpen
  \bibfield  {author} {\bibinfo {author} {\bibfnamefont {K.~N.}\ \bibnamefont
  {Abazajian}} \emph {et~al.} (\bibinfo {collaboration} {CMB-S4}),\ }\href@noop
  {} {\  (\bibinfo {year} {2016})},\ \Eprint {http://arxiv.org/abs/1610.02743}
  {arXiv:1610.02743 [astro-ph.CO]} \BibitemShut {NoStop}%
\bibitem [{\citenamefont {Shiraishi}\ and\ \citenamefont
  {Yokoyama}(2011)}]{Shiraishi:2011ph}%
  \BibitemOpen
  \bibfield  {author} {\bibinfo {author} {\bibfnamefont {M.}~\bibnamefont
  {Shiraishi}}\ and\ \bibinfo {author} {\bibfnamefont {S.}~\bibnamefont
  {Yokoyama}},\ }\href {\doibase 10.1143/PTP.126.923} {\bibfield  {journal}
  {\bibinfo  {journal} {Prog. Theor. Phys.}\ }\textbf {\bibinfo {volume}
  {126}},\ \bibinfo {pages} {923} (\bibinfo {year} {2011})},\ \Eprint
  {http://arxiv.org/abs/1107.0682} {arXiv:1107.0682 [astro-ph.CO]} \BibitemShut
  {NoStop}%
\bibitem [{\citenamefont {Obata}\ and\ \citenamefont
  {Fujita}(2019)}]{Obata:2018ilf}%
  \BibitemOpen
  \bibfield  {author} {\bibinfo {author} {\bibfnamefont {I.}~\bibnamefont
  {Obata}}\ and\ \bibinfo {author} {\bibfnamefont {T.}~\bibnamefont {Fujita}},\
  }\href {\doibase 10.1103/PhysRevD.99.023513} {\bibfield  {journal} {\bibinfo
  {journal} {Phys. Rev. D}\ }\textbf {\bibinfo {volume} {99}},\ \bibinfo
  {pages} {023513} (\bibinfo {year} {2019})},\ \Eprint
  {http://arxiv.org/abs/1808.00548} {arXiv:1808.00548 [astro-ph.CO]}
  \BibitemShut {NoStop}%
\end{thebibliography}%

\end{document}